\documentclass{ws-ijmpa-arXiv} 
\usepackage{amsmath,amssymb}
\usepackage[compress]{cite}
\usepackage{titlesec}
\usepackage{subfigure}
\usepackage{hyperref,url}
\usepackage{endnotes}
\usepackage{lineno}

\def\frac#1#2{\textstyle{{{#1} \over {#2}}}}

\def\lsim{\mathrel{\rlap{\lower4pt\hbox{\hskip1pt$\sim$}}
    \raise1pt\hbox{$<$}}}
\def\gsim{\mathrel{\rlap{\lower4pt\hbox{\hskip1pt$\sim$}}
    \raise1pt\hbox{$>$}}}

\def\Z{\hat Z}

\newcommand{\beq}{\begin{equation}}
\newcommand{\eeq}{\end{equation}}
\newcommand{\bea}{\begin{eqnarray}}
\newcommand{\eea}{\end{eqnarray}}

\begin{document}

\title{THE PHYSICS PROGRAMME OF THE MoEDAL EXPERIMENT AT THE LHC}

\author{B.~ACHARYA$^{1,2}, $ J. ALEXANDRE$^{1}$, J.~BERNAB\'{E}U$^{3}$, M.~CAMPBELL$^{4}$, S.~CECCHINI$^{5}$, J.~CHWASTOWSKI$^{6}$,  M.~DE~MONTIGNY$^{7}$, D. DERENDARZ$^{6}$, A.~DE~ROECK$^{4}$, J.~R.~ELLIS$^{1,4}$, M.~FAIRBAIRN$^{1}$,  D.~FELEA$^{8}$, M.~FRANK$^{9}$, D.~FREKERS$^{10}$, 
C. GARCIA$^{3}$,  G.~GIACOMELLI$^{5,5a}\dagger$,  J.~JAK\r{U}BEK$^{11}$,  A.~KATRE$^{12}$,   D-W KIM$^{13}$, M.G.L.~KING$^{3}$, K.~KINOSHITA$^{14}$, 
D.~LACARRERE$^{4}$, S. C. LEE$^{11}$, C. LEROY$^{15}$, A.~MARGIOTTA$^{5}$, N. MAURI$^{5,5a}$, N.~E.~MAVROMATOS$^{1,4}$, 
P.~MERMOD$^{12}$, V.~A.~MITSOU$^{3}$, R.~ORAVA$^{16}$,   L. PASQUALINI$^{5,5a}$,  L.~PATRIZII$^{5}$, G.~E.~P\u{A}V\u{A}LA\c{S}$^{8}$, J.~L.~PINFOLD$^{7}$\footnote{Communicating author.}\hspace{0.5em},   M.~PLATKEVI\v{C}$^{11}$, V.~POPA$^{8}$,  M. POZZATO$^{5}$, S.~POSPISIL$^{11}$,  A.~RAJANTIE$^{17}$,
 Z. SAHNOUN$^{5, 5b}$, M.~SAKELLARIADOU$^{1}$, S.~SARKAR$^{1}$, G.~SEMENOFF$^{18}$, G.~SIRRI$^{5}$, K.~SLIWA$^{19}$, R.~SOLUK$^{7}$, M.~SPURIO$^{5, 5a}$, 
    Y.N. SRIVASTAVA$^{20}$, R. STASZEWSKI$^{6}$,  J.~SWAIN$^{20}$, M. TENTI$^{5, 5a}$, V.~TOGO$^{5}$,  M. TRZEBINSKI$^{6}$, J.~A.~TUSZY\'{N}SKI$^{7}$, V.~VENTO$^{3}$, O.~VIVES$^{3}$, Z.~VYKYDAL$^{11}$, \and A.~WIDOM$^{20}$, J. H. YOON$^{21}$.  \\ 
(for the MoEDAL Collaboration) \\ [0.2cm]}

\address{
$^{1}$Theoretical Particle Physics and Cosmology Group, Physics Department, King's~College~London, UK\\
$^{2}$International Centre for Theoretical Physics, Trieste, Italy\\ 
$^{3}$IFIC, Universitat de Val\`{e}ncia - CSIC, Valencia, Spain\\
$^{4}$Physics Department, CERN, Geneva, Switzerland\\
$^{5}$ INFN, Section of Bologna, 40127, Bologna , Italy\\ 
$^{5a}$ Department of Physics \& Astronomy, University of Bologna, Italy\\  
$^{5b}$ Centre on Astronomy, Astrophysics and Geophysics, Algiers, Algeria\\
$^{6}$ Institute of Nuclear Physics Polish Academy of Sciences, Cracow, Poland \\
$^{7}$Physics Department, University of Alberta, Edmonton Alberta, Canada\\
$^{8}$Institute of Space Science, M\u{a}gurele,  Romania\\
$^{9}$Department of Physics, Concordia University, Montreal, Quebec,  Canada\\
$^{10}$Physics Department, University of Muenster, Muenster, Germany\\
$^{11}$IEAP, Czech Technical University in Prague, Czech~Republic\\
$^{12}$Section de Physique, Universit\'{e} de Gen\`{e}ve, Switzerland\\
$^{13}$Physics Department, Gangneung-Wonju National University, Gangneung, Korea\\
$^{14}$Physics Department, University of Cincinnati, Cincinnati OH, USA\\
$^{15}$ Physics Department, University de Montr\'{e}al, Montr\'{e}al, Qu\'{e}bec, Canada\\
$^{16}$Physics Department, University of Helsinki, Helsinki, Finland\\
$^{17}$Physics Department, Imperial College London, UK\\
$^{18}$Department of Physics, University of British Columbia, Vancouver BC, Canada\\
$^{19}$Department of Physics and Astronomy, Tufts University, Medford MA, USA\\
$^{20}$Department of Physics,  Northeastern University, Boston, USA\\
$^{21}$Physics Department, Konkuk University, Seoul, Korea\\}

\maketitle  

\vspace*{-0.7cm}

\begin{center}
{\footnotesize $^\dagger$Deceased. We dedicate this paper to Giorgio's memory. We will strive to make this experiment a great success and a tribute to  his memory. He will be sorely missed.\par }
\end{center}

\begin{center}
\footnotesize{KCL-PH-TH/2014-02, LCTS/2014-02, CERN-PH-TH/2014-021, IFIC/14-16,  Imperial/TP/2014/AR/1}
\end{center}

\newpage
\vskip 2cm

\begin{center}
{\bf Abstract}
\end{center}

\begin{abstract}
The MoEDAL experiment at Point~8 of the LHC ring is the seventh and newest LHC experiment. It is dedicated to the  search for highly ionizing particle 
avatars of physics beyond the Standard Model, extending significantly the discovery horizon of the LHC. A MoEDAL discovery would have revolutionary 
implications for our fundamental  understanding of the Microcosm.
 MoEDAL is an unconventional and largely passive LHC  detector comprised of the largest array of Nuclear Track Detector stacks ever deployed at an 
 accelerator, surrounding the intersection  region at Point~8 on the LHC ring.  Another novel feature is the use of paramagnetic trapping volumes to 
 capture both electrically and magnetically charged highly-ionizing particles predicted in new physics scenarios. It includes an array of TimePix pixel
  devices for monitoring highly-ionizing particle backgrounds. The main passive elements of the MoEDAL detector do not require a trigger system, 
  electronic readout, or online computerized data acquisition. The aim of this paper is to give an  overview of the MoEDAL physics reach,  which is 
  largely complementary to the programs of the large multi-purpose LHC detectors ATLAS and CMS.
 \end{abstract}


\vskip 1.5cm
\tableofcontents
\clearpage

\section{Introduction}

In 2010 the CERN (European Laboratory for Particle Physics) Research Board unanimously approved MoEDAL (Monopole and Exotics Detector at the LHC)~\cite{moedal-web,moedal-tdr}, the $7^{\rm th}$ international experiment at the Large Hadron Collider (LHC)~\cite{LHC}, which is designed to seek out avatars of new physics with highly-ionizing particle signatures. The most important motivation for the MoEDAL experiment is to continue the quest for the magnetic monopole and dyons~\cite{Dirac1931kp,Diracs_idea,tHooft1974qc,Polyakov1974ek,Julia1975ff,N,Witten1979ey,Lazarides1980cc,Sorkin1983ns,Gross1983hb,Cho1996qd,
Schwinger1969ib,Preskill1984gd,AV,Kephart2006zd,CHO-KIM-YOON,ZHANG,Rajantie2005hi} to LHC energies.  However, the experiment is also 
designed to search for  massive, stable or long-lived, slow-moving particles~\cite{Fairbairn07} with single or multiple electric charges that arise in 
many scenarios of physics beyond the Standard Model (BSM)~\cite{SUSY, MSSM,CMSSM,CMSSM1, CMSSM2, CMSSM3, CMSSM4, CMSSM5,CMSSM6,CMSSM7, RV, FATHIGGS,
xyon,ADD,ADD1,ADD2, Randall,TEV-1,TEV-1a,TEV-1b, UED,shiu, westmuckett, westmuckett1, westmuckett2, westmuckett3, mitsou,vectorlike,KILIC,Coleman:1985ki,SUSYQballs}.

Magnetic monopoles that carry a non-zero magnetic charge and dyons possessing both magnetic and electric charge are among the most fascinating  hypothetical particles. Even though there is no generally acknowledged empirical  evidence for their existence, there are strong theoretical reasons to believe that they do exist, and they are predicted by many theories including grand unified theories and superstring theory. 

The laws of electrodynamics guarantee that the lightest magnetic monopole would be a stable particle, and because monopoles interact strongly with the electromagnetic field they are straightforward to detect experimentally. The scattering processes of electrically charged fermions with magnetic monopoles are dependent on their microscopic properties even at low energies,\cite{Rubakov1981rg,Callan1982ah} and therefore they would provide a unique window to physics at high energies beyond the Standard Model. In particular they would elucidate some of the most fundamental aspects of electrodynamics (or, more precisely, the electroweak hypercharge) such as its relation to other elementary particle interactions.

Likewise, the existence of massive, stable or long-lived, slow-moving particles with single or multiple electric charges would also have drastic implications for models of particle physics and cosmology. Therefore, a MoEDAL discovery  would have revolutionary implications for our understanding of the microcosm, potentially providing insights into such  fundamental questions as: Does magnetic charge exist? Are there extra dimensions or new symmetries of nature?  What is the mechanism for the generation of mass? What is the nature of dark matter? How did the Big Bang unfurl at the earliest times?

MoEDAL is an unconventional and largely passive LHC detector comprised of the largest array ($\sim$ 100 m$^{2}$)  of Nuclear Track Detector (NTD) stacks ever deployed at an accelerator, surrounding the intersection region at Point~8 on the LHC ring. Essentially, MoEDAL is like a giant camera ready to reveal ``photographic" evidence for new physics and also to trap long-lived new particles for further study. 

In this paper we describe the physics goals of a five-year programme of work that combines interdisciplinary experimental techniques and facilities --- from collider and astroparticle physics research --- to expand significantly the horizon for discovery at the LHC, in a way complementary to the other LHC detectors. The official start to MoEDAL data  taking is currently Spring 2015, after the long LHC shutdown. 

The structure of the paper is as follows. First we describe the MoEDAL detector and then present the mechanisms by which the MoEDAL detector will register the signatures of new physics. The bulk of the paper is devoted to describing the physics program of the MoEDAL experiment in various scenarios for BSM physics. In this arena we start with the consideration of the MoEDAL's  main physics motivation --- the search for the magnetic monopole and other manifestations of magnetic charge. We then move to the consideration of singly electrically-charged Massive (pseudo)Stable Particles (SMPs) in a number of scenarios. We consider as a separate case the search for particles with double or multiple electrical charges in a number of models.   

A detailed comparison of MoEDAL's sensitivity\footnote{Defined to be a convolution of  the efficiency and acceptance} to those of other experiments in this arena, for each new physics scenario presented below, is beyond the scope of the paper. Rather, we outline here a programme of studies at the discovery frontier where MoEDAL would provide a unique and complementary physics coverage to the existing LHC experiments.  We emphasize that, even in those cases where we can expect a considerable overlap between the new physics reach of MoEDAL and the other LHC experiments, MoEDAL's contribution would be invaluable, particularly in the event of a discovery. This is because MoEDAL is an entirely different type of LHC detector. It is immune to fake signals from Standard Model backgrounds and has totally different systematics from other LHC detectors. 

MoEDAL is a passive detector  unfettered by the requirement to trigger and is not subject to the
limitations imposed by real-time electronic readout systems. Importantly, its NTD system can be directly calibrated for the detection of highly-ionizing particles using heavy-ion beams, which is not possible for the other LHC detectors. Also, it is the only LHC detector than can detect directly magnetic charge. Last but not least, MoEDAL will have a permanent record of any new particle that it detects and may even be able to capture that new particle for further study.

We  note that in this work we consider only collider-based searches for new physics scenarios signalled by highly-ionizing phenomena. 
Cosmological and astrophysical constraints on such scenarios, along with the various assumptions required, are not explored here. Comprehensive reviews of
non-accelerator searches are presented elsewhere  \cite{Fairbairn07}\cite{Fairbairn14}.

\section{The MoEDAL Detector}

\begin{figure}[htb]
\begin{center}
\includegraphics[width=25pc]{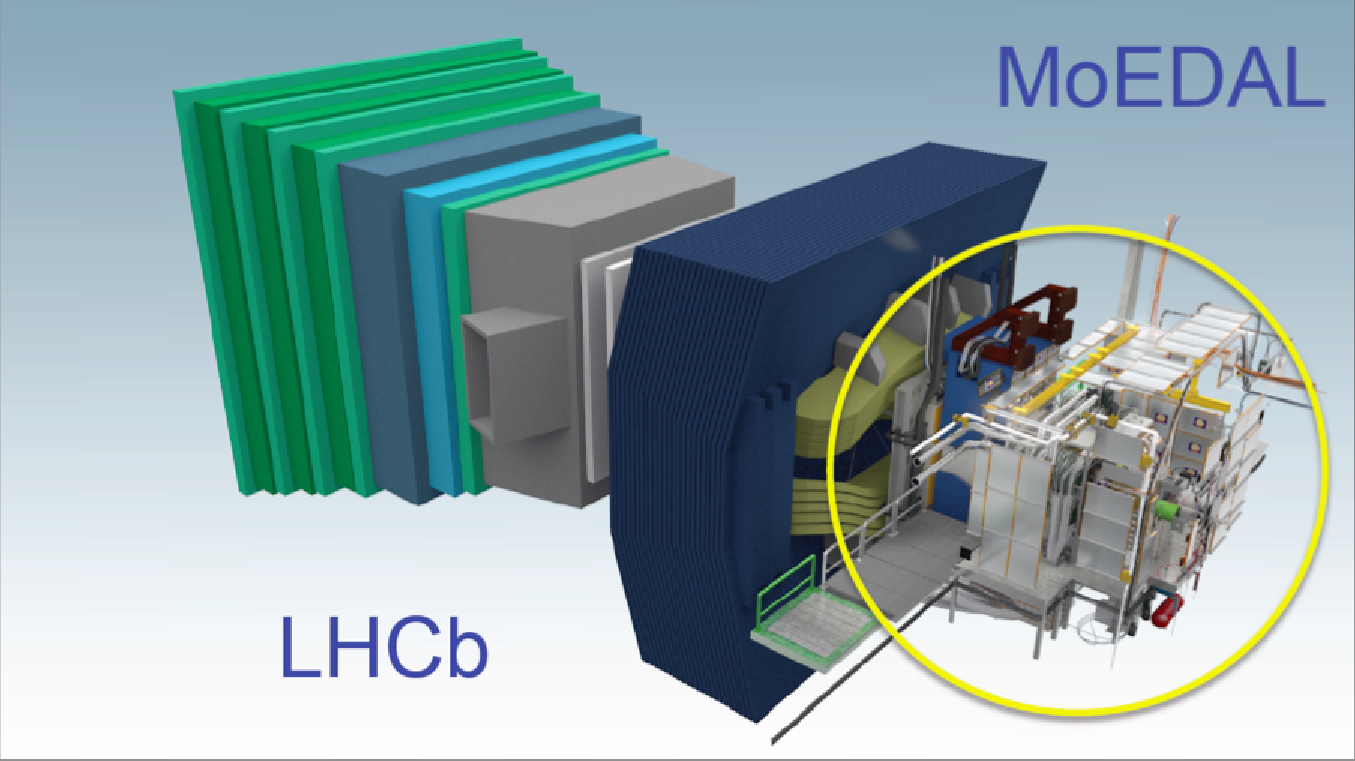}
\caption{ A three-dimensional schematic depiction of the deployment of the MoEDAL
detector around the LHCb VELO region at Point~8 of the LHC.}
\label{Fig:moedal-lhcb}
\end{center}
\end{figure}

The MoEDAL detector is deployed around the intersection region at Point~8 of the LHC in the LHCb experiment's VELO (VErtex LOcator)~\cite{LHCb-detector} cavern. A three-dimensional depiction of the MoEDAL experiment is presented in Fig.~\ref{Fig:moedal-lhcb}. It is a unique and largely passive LHC detector comprised of four sub-detector systems. 

The main subdetector system comprises a large array (100~m$^{2}$) of CR39\textregistered~\cite{CR39},  Makrofol\textregistered~\cite{MAKROFOL} and lexan\textregistered~\cite{lexan} NTD stacks surrounding the intersection region. In p-p running the only source of Standard Model particles that are highly ionizing enough to leave a track in MoEDAL's  NTDs are spallation products with range that is in the vast majority of cases  much less than the thickness of one sheet of the NTD  stack. Even then the ionizing signature will be that of a very low energy electrically charged stopping particle. This signature is  very different to that of a  penetrating electrically  or magnetically  charged particle that will usually  traverse every sheet in a MoEDAL NTD stack,  accurately defining a track that points back to the MoEDAL intersection region. In the case of heavy ion running one might expect a background from high ionizing fragments. However, such heavy-ion fragments  are produced in the  far forward direction and do not enter the acceptance of the MoEDAL detector.

A unique feature of the MoEDAL detector  is the use of paramagnetic trapping volumes (MMTs) to capture electrically- and magnetically-charged highly-ionizing particles for subsequent analysis at a remote detector facility. Magnetically-charged particles will be monitored at a remote Magnetometer Facility. The search for the decays of long-lived electrically charged particles  that are stopped in the trapping detectors will subsequently be carried out at a remote underground facility such as SNOLAB.

The only non-passive MoEDAL sub-detector system is comprised of an array of around ten TimePix pixel devices  
forming  a real-time radiation monitoring system devoted to the monitoring of highly-ionizing backgrounds
in the MoEDAL cavern.

\subsection{The Nuclear Track Detector system}

\begin{figure}[htb]
\begin{center}
\includegraphics[width=24pc]{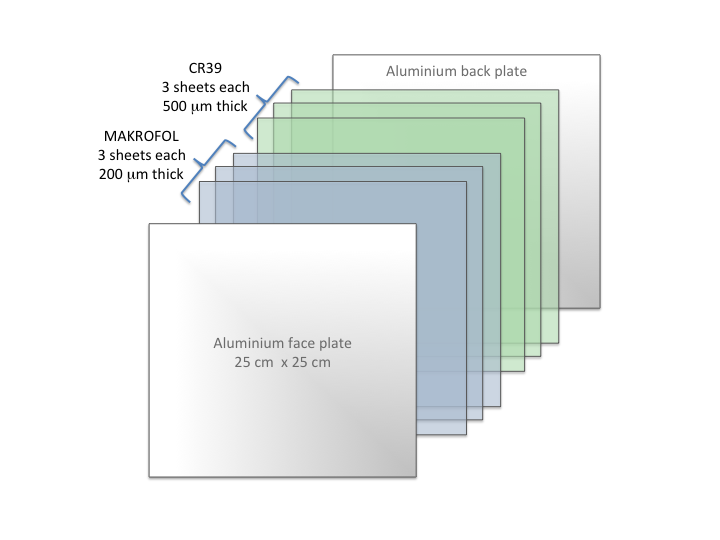}
\caption{The composition of a MoEDAL LT-NTD  stack, the lexan sheets are not shown.}
\label{Fig:NTD}
\end{center}
\end{figure}

The main subsystem referred to as the low threshold NTD (LT-NTD) array,  the array originally defined in the TDR~\cite{moedal-tdr}, is the largest array  of plastic NTD stacks  ($\sim$ 250) ever deployed at an accelerator. Each stack, 25 cm$^{2}$ $\times$ 25 cm$^{2}$ in size,  consists of three sheets of CR39 polymer, three of Makrofol  and three of Lexan . A depiction of a MoEDAL TDR NTD stack is shown in Fig.~\ref{Fig:NTD}. CR39  is the NTD with the lowest detection threshold, it can detect particles with ionization equivalent to $Z/\beta \sim 5$, where $Z$ is electric charge of the impinging particle and $\beta$ its velocity, expressed as a fraction of the speed of light.  A standard minimum-ionizing particle produced in an interaction at the LHC would   have a Z/$\beta \simeq$  1. The charge  resolution of CR39  detectors is better than $0.1e$ \cite{STEFANO}, where $e$ is the electric charge.

The LTD-NTD array has been enhanced by the  High Charge Catcher (HCC) sub-detector with threshold $Z/\beta \gtrsim 50$ comprising three Makrofol  sheets in an aluminium foil envelope. These lightweight low-mass detector stacks can be applied directly to the outside of the VELO detector housings and on other accessible surfaces in the region, for example  the front face of the LHCb RICH detector, rather than on the walls and ceilings of the VELO cavern. In this way we can increase the geometrical acceptance for magnetic monopoles to over $\sim 60\%$.

The exposed NTD stacks will be replaced each year, or as required. The removed plastic will be etched at the INFN Bologna etching Lab and  in Cincinnati and subsequently scanned using the manual and semi-automatic systems of the Bologna Lab. As soon as available a  fully automatic  state-of-the-art high-throughput optical scanning-microscopy for high-resolution and large-area surface analysis based on the AMBIS (Anti Motion-Blurring Imaging System) device will be deployed at the MoEDAL plastic analysis centres. This apparatus  is capable of searching quickly large areas of NTD material ($\sim 100$~cm$^{2}$ in 40~minutes) for extremely small features (${\cal O}(10) \mu$m).  In this way the usual scanning process is turned into an image enhancement and pattern recognition process that can be handled with specialized software. In this way  the low $Z/\beta$ threshold  can be maintained.
 
   
In a multi-sheet stack detector, the position and direction information from individual pits can be combined to track the  particle back to the interaction region. Also,  the actual or effective  $Z/\beta$ values can be used  to determine if  the change in ionization energy loss is   consistent with the scenario under investigation.
    
The signal for a magnetic monopole would be a set of 20~etch pits aligned with a precision of  $\sim 10~\mu$m, pointing towards the intersection point, with etch pits remaining the same size or decreasing slightly in size, since the monopole energy loss decreases with falling $\beta$, making it distinct from an electrically-charged particle.      

\subsection{The magnetic monopole trapper detector system}
 
The Magnetic Monopole Trapper (MMT) is the third and newest sub-detector system to be added to the MoEDAL detector. This detector consists of passive stacks of aluminium  trapping volumes placed adjacent to the VELO detector. These stacks - intended to trap magnetic monopoles and highly-ionizing \mbox{(quasi-)stable} massive  charged particles that stop within their volume -  will be replaced once a year.  After removal the exposed trapping volumes  will be sent first to the SQUID facility at ETH Zurich to be scanned for trapped magnetic charge. The signal for a magnetic monopole in the MMT trapping detectors at the ETH facility would be a sustained current resulting from the passage of a monopole though the SQUID detector.

\begin{figure}[htb]
\begin{center}
\includegraphics[width=20pc]{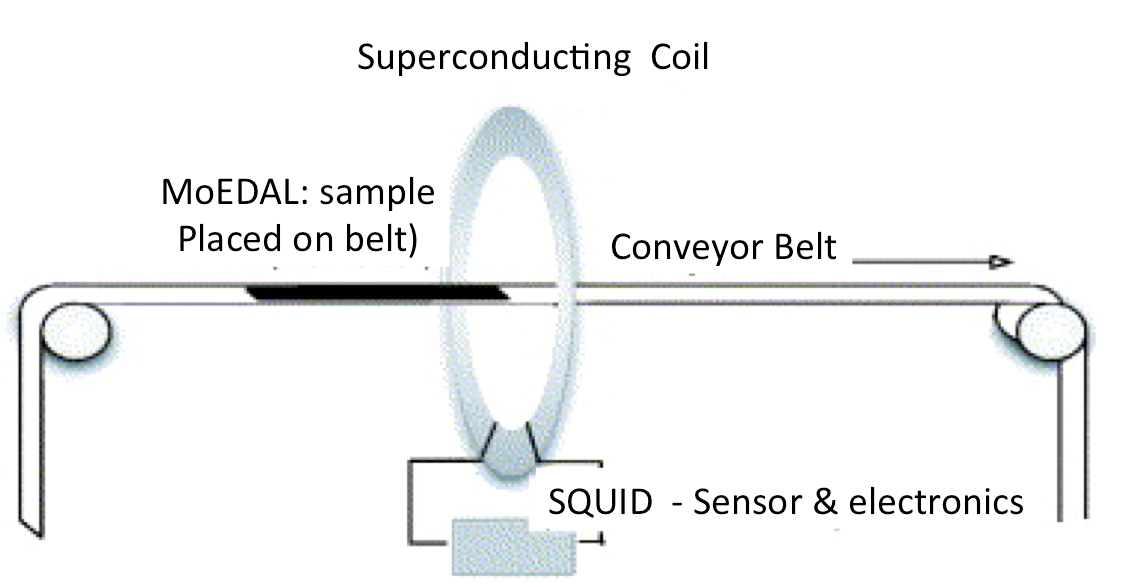}
\caption{ A schematic diagram showing how the SQUID apparatus is used.
The sample  travels in steps  completely through the 
coil. The current in the superconducting coil is read out after each step.}
\label{Fig:SQUID}
\end{center}
\end{figure}

\begin{figure}[htb]
\begin{center}
\includegraphics[width=20pc]{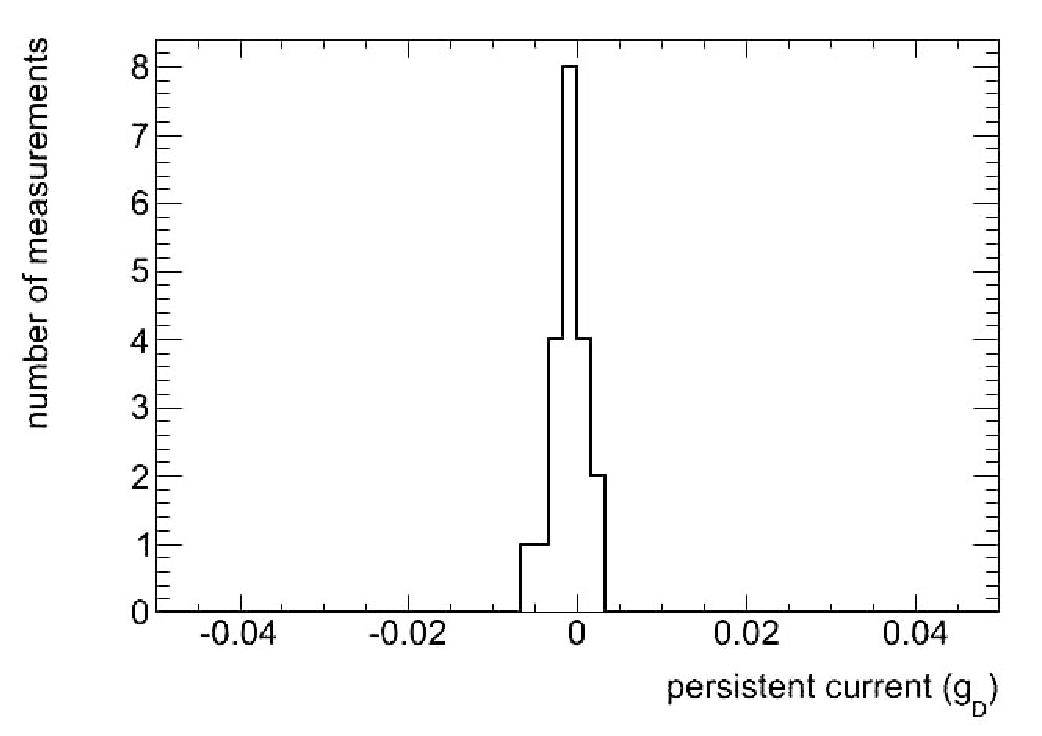}
\caption{ A histogram showing the measurements of a number of MoEDAL MMT detector blanks, showing
that the monitoring SQUID magnetometer is sensitive to monopoles with charge as low as one hundredth
of a Dirac charge.}
\label{Fig:SQUID-sense}
\end{center}
\end{figure}

A schematic diagram describing the  SQUID magnetometer scanning process is shown in Fig.~\ref{Fig:SQUID}. The sensitivity of the SQUID magnetometer housed at ETH Zurich, as determined by a ``pseudo-monopole'' formed from a very long conventional magnet, is shown in Fig.~\ref{Fig:SQUID-sense}.  The solenoid method is not the only method by which the SQUID magnetometer can be 
calibrated. Further details of this technique can be found elsewhere \cite{SQUIDS}. 

After the SQUID scan has been performed, the MMT trapping volumes will be sent to an underground laboratory - SNOLAB is currently the favored site -  to be monitored for the  decay of  trapped very long-lived electrically charged particles. The rate of decay observed  can be used to find the particle's lifetime.

The MMT sub-detector has the two attractive advantages of speed and complementarity. It will result in the publication of the first monopole search in 14-TeV $pp$ collisions, and has the potential to procure a robust and independent cross-check of a discovery as well as a unique measurement of the magnetic properties of a monopole.

\subsection{The TimePix radiation monitoring system}

The fourth and only non-passive sub-detector system is an array of TimePix pixel devices (NTPX) \cite{TimePix}. It is used to monitor possible highly-ionizing beam-related backgrounds. Each pixel of the innovative TimePix chip contains a preamplifier, a discriminator with hysteresis and 4-bit DAC for threshold adjustment, synchronization  logic and a 14-bit counter with overflow control. The TimePix chip has an active area of $\sim$2cm$^{2}$ segmented into a 256 $\times$ 256 square pixel matrix, where each pixel is 55$\mu$ on the side.
  
\begin{figure}[htb]
\begin{center}
\includegraphics[width=20pc]{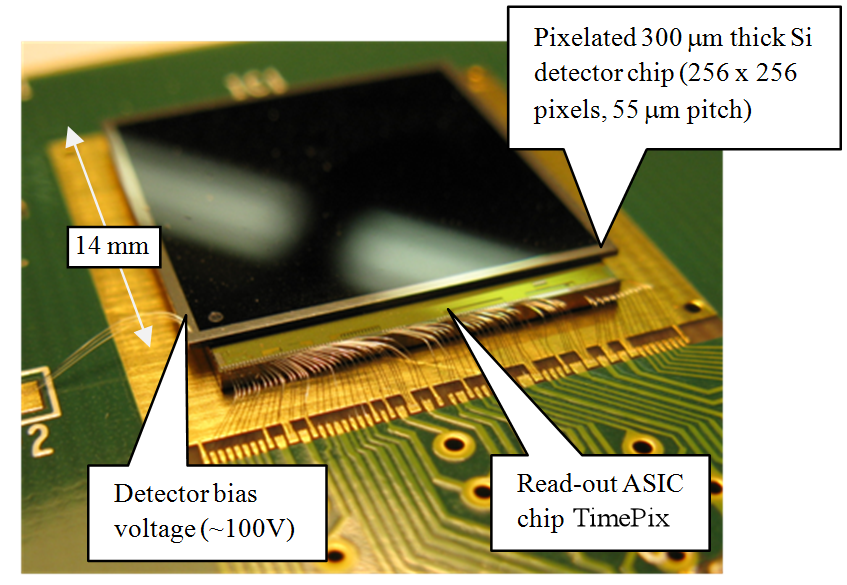}
\caption{ A photograph of a Timepix chip with its main features indicated.}
\label{Fig:TimePixChip}
\end{center}
\end{figure}

MoEDAL uses TimePix in ``Time-over-Threshold'' mode, so that each pixel can act as an ADC that can supply an energy measurement. A photograph of a TimePix pixel chip is shown in Fig.~\ref{Fig:TimePixChip}. The online TimePix radiation monitoring system will be accessed via the web. Thus, it is not necessary to run shifts even though the TimePix array is a real-time MoEDAL detector system.

Some $5-10$ TimePix1 silicon pixel imaging devices will be deployed to sample the radiation levels around the MoEDA/LHCb cavern.
 Each Timepix detector is essentially a small electronic bubble chamber capable of imaging complete spallation events in its $300~\mu$m thick silicon sensitive volume. Data readout and  event display produced is provided by the ``PixelMan''  software developed by the CTU-IEAP group.

\section{Interactions of Electrically- and Magnetically-Charged Particles in MoEDAL}

In this Section we describe the expected interactions of SMPs 
with electrical and/or magnetic charge in the MoEDAL detector. In addition, a description is given
 of the theory of electromagnetic energy loss for electrically- and magnetically-charged particles. 
 We also consider the physics of monopole trapping.  Specific scenarios for highly-ionizing particles 
 with magnetic and/or electric charge will be discussed in subsequent sections.

\subsection{Ionization energy loss in matter}

A main detection mode for SMPs in MoEDAL is  via the measurement of continuous ionization 
energy loss $dE/dx$. Both electrically and magnetically charged SMPs lose energy principally 
through ionization energy loss as they propagate through matter. The theory of electromagnetic 
energy loss for both electrically and magnetically charged particles is well established~\cite{PDG}.

\subsubsection{Ionization energy loss for electrically-charged particles}

As an electrically charged particle moves through matter it loses energy either by interactions 
with atomic electrons or by collisions with atomic nuclei in the material. The first of these results 
in the freeing of electrons from the atoms in the material (i.e., ionization) while the second results 
in the displacement of atoms from the lattice. The energy loss due to the second process is called 
the Non-Ionizing Energy Loss (NIEL). Except at very low $\beta$,  the energy loss ($dE/dx$), due to the ionization energy 
is much larger than the NIEL~\cite{NIEL} in most particle detectors.

The mean rate of energy loss (or stopping power) for moderately relativistic charged particles 
other than electrons is given by the Bethe-Bloch formula~\cite{PDG},
\begin{equation}\label{eq:Bethe-Bloch_mod}
\dfrac{dE}{dx} = 4\pi N_{A} r^{2}_{e}m_{e}c^{2}\dfrac{Z}{A}\dfrac{z^{2}}{\beta^{2}}
\left[\dfrac{1}{2}\ln\dfrac{2m_{e}c^{2}\beta^{2}\gamma^{2}T_{\rm max.}}{I_{e}^{2}}
-\beta^{2} -\dfrac{\delta(\beta\gamma)}{2}\right],
\end{equation}
where: $Z$($A$) is the atomic number (mass) of the medium; $z$ is the charge of the incident 
particle; $m_{e}$ and $r_{e}$ are the mass and classical radius of the electron,
respectively; $N_{A}$ is Avogadro's number;
$\beta$ is the velocity of the incident particle as a fraction of the speed of light 
($c$); $\gamma = 1/\sqrt{1 -\beta^{2}}$;  and,  $I_{e}$ is the mean ionization potential of the 
medium. The ionization potential can be parameterized by~\cite{lewin}, $I_{e}(Z) = 
(12Z + 7)$~eV for $Z  \le 13$ and  $(9.76Z +  58.8Z^{-0.19})$~eV for $Z > 13$. The 
quantity $T_{\rm max.}$ is the maximum kinetic energy that can be imparted to a free 
electron in a single collision, and is given, for a particle with mass $M$,  by:
\begin{equation}
T_{\rm max.} = \dfrac{2m_{e}c^{2}\beta^{2}\gamma^{2}}{1 + \dfrac{2\gamma m_{e}}{M} + \left(\dfrac{m_{e}}{M}\right)^{2}}.
\end{equation}
The $\delta$ term represents the density effect correction to ionization energy loss. As the
 particle energy increases, its electric field flattens and extends, so that the distant-collision
  contribution increases as $\ln\beta\gamma$. However, real media become polarized, 
  diminishing the extension of the field and thereby limiting the relativistic rise at high energy.
   Thus the $\delta$ term is important for massive particles --- where $T_{\rm max.} 
   \approx 2m_{e}c^{2}\beta^{2}\gamma^{2}$ --- with $\beta\gamma >3$. Slight differences
    in $\delta$  occur for different charges moving with low velocity. The density effect correction
     is usually computed using Sternheimer parameterization~\cite{sternheimer}. 
 
\begin{figure}[ht]
\centering
\includegraphics[width=0.99\textwidth]{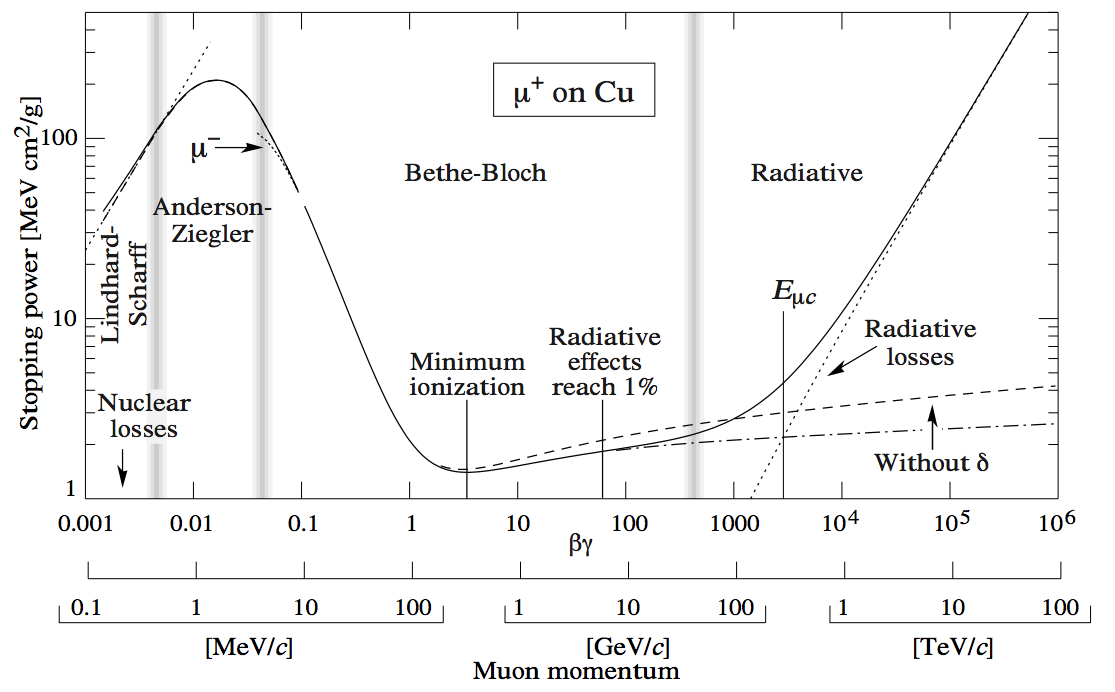}
\caption{\label{fig:vector-like}The stopping power $dE/dx$ for positively-charged muons in 
copper as a function of $\beta\gamma = p/M$, taken from the Particle Data Group book~\cite{PDG}. 
The solid curve indicates the total stopping power. Data below the break at $\beta\gamma \approx 1$ 
are taken from Ref.~\cite{ICRU49} and data at higher energies are from Ref.~\cite{PDGHE, PDGHE1}. The
 radiative effects apparent at very high energies are much less relevant for particles heavier than 
 muons. The different regions indicated by the vertical bands are described in the Particle Data 
 Group book~\cite{PDG}, as are the small difference between positive and negative charges at
low values of velocity (the Barkas effect~\cite{barkas}) shown by the short dotted lines labelled $\mu^{-}$. }
\end{figure}

The Bethe-Bloch formula~(\ref{eq:Bethe-Bloch_mod}) is based on a first-order Born approximation.
 However, for lower energy, higher-order corrections are important. These corrections are usually 
 included by adding the ``Bloch correction'' $z^{2}L_{2}(\beta)$ inside the square bracket of 
 Eq.~(\ref{eq:Bethe-Bloch}). An additional correction term, $zL_{1}\beta$, makes the stopping 
 power for a positive particle somewhat larger than for a negative particle, all other factors being 
 equal~\cite{barkas}. This is indicated by the short dotted lines labelled $\mu^{-}$ in Fig.~\ref{fig:vector-like}.
 
At  low energies, in the so-called Lindhard region~\cite{lindhard}, where particles are moving 
with speed less than around $0.01c$, the velocity of the incident particle is comparable to, or 
less than, the velocity of the outer atomic electrons, and the Bethe-Bloch formula is no longer 
valid. In this region Lindhard has introduced a successful formula for the stopping power that 
is proportional to the particle's velocity, $\beta$~\cite{lindhard}:
\begin{equation}
\dfrac{dE}{dx} = N\xi_{e}8\pi e^{2}a_{0} \dfrac{Z z}{(Z^{\frac{2}{3}} + z^{\frac{2}{3}})} \dfrac{\beta}{\beta_{0}},
\end{equation}
\noindent
where $N$ is the number of atoms per unit volume, $\xi \approx z^{1/6}$, and $a_{0}$ is the
 Bohr radius of the hydrogen atom. The formula holds for $\beta < z^{2/3}\beta_{0}$, where 
 $\beta_{0} = e^{2}/(2\epsilon_{0}hc)$ is the electron velocity in the classical lowest Bohr 
 orbit of the hydrogen atom.

The intermediate region,  roughly $0.01 <  \beta < 0.05$, described by Anderson and 
Ziegler \cite{AANDZ} and defined by the relation:
\begin{equation}
\max\left[ \dfrac{\alpha z^{1/3}}{(1+\alpha z^{1/3})}, ~~\dfrac{(2Z^{0.5} + 1)}{400} \right]  \le  \beta \le  \dfrac{\alpha z} {(1 + \alpha z)} ,
\end{equation}
\noindent
does not have a satisfactory theoretical description. However, Lewin~\cite{lewin} described a useful phenomenological
 polynomial interpolation over the intermediate region, of the form:
\begin{equation}
\dfrac{dE}{dx} = A\beta^{3} + B\beta^{2} + C\beta + D,
\end{equation}
\noindent
where the coefficients $A$, $B$, $C$ and $D$ are discussed in Ref.~\cite{lewin}.

\subsubsection{Ionization energy loss of magnetically-charged particles }

A fast ($\beta >$ 10$^{-2}$) magnetic monopole with a single Dirac charge ($g_{D}$) \footnote{
The concept of  Dirac (magnetic) charge is presented in Section 5.}   has an  equivalent  electric 
charge of  Z$_{e}$q = $\beta$(137e/2). Thus  for a relativistic monopole the energy loss is around 
$4,700$ times ($68.5^2$) that of a Minimum-Ionizing electrically-charged Particle (MIP). 
Thus, one would expect a monopole to leave a unique and striking ionization signature.

\begin{figure}[htb]
\begin{center}
\includegraphics[width=29pc]{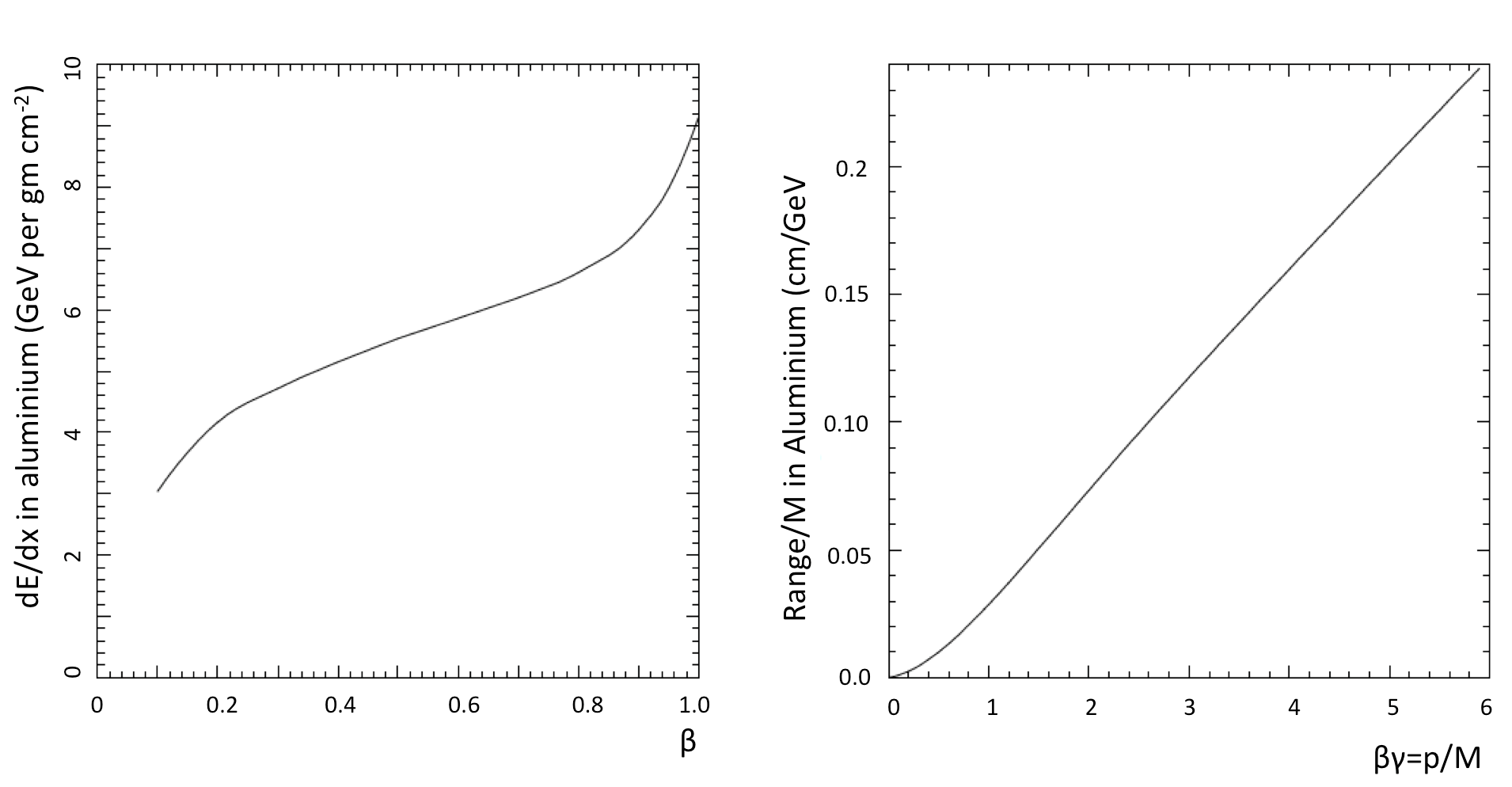}
\caption{Left: The $dE/dx$ for a Dirac monopole in aluminium as a function of the velocity of 
the monopole, obtained from~\cite{ahlen} and adjusted for the electron density in aluminium. 
Right: The ratio of range to mass for a Dirac monopole in aluminium versus $\beta\gamma$, 
calculated from the stopping power $dE/dx$.}
\label{Fig:monopole-stop}
\end{center}
\end{figure}

The ionization energy loss by magnetic monopoles can be described by a formula very similar
 to the Bethe-Bloch equation, 
  with the electric charge term 
 replaced by the electric charge equivalent of the magnetic charge to the monopole. 
 At relativistic velocities the energy loss is therefore constant, independent of $\beta$.
The detailed formula for the stopping power
    of a magnetic monopole with magnetic charge $ng$ ($n=1,2,3...$) is given by~\cite{ahlen}:
\begin{equation}\label{eq:Bethe-Bloch}
\dfrac{dE}{ dx} = \dfrac{4\pi e^{2}(ng)^{2}}{m_{e} c^{2}}n_{e}\left[\dfrac{1}{2}\ln\dfrac{2m_{e}c^{2}\beta^{2}\gamma^{2}T_{\rm max.}}{I^{2}_{m}}
\dfrac{1}{2} - \dfrac{\delta}{2}  + \dfrac{K(|g|)}{2} -B(|g|)        \right],
\end{equation}
\noindent
where $n_{e}$ is the number of electrons per unit volume in the medium, $I_{m}$ is the mean
 ionization potential for magnetic monopoles, which is close in value to $I_{e}$, and $K(|g|) = 0.406, 0.346)$ 
and  $B(|g|) ( =  0.248, 0.672)$  are correction terms for  $g = 1g_{D}/2g_{D}$, respectively~\cite{ahlen}.
The relationship between $I_{m}$ and $I_{e}$ is given by $I_{m} = I_{e}e^{-D/2}$, where the 
power $D$ has been calculated by Sternheimer~\cite{sternheimerD} for various elements, for 
example $D(Al) = 0.056$, $D(Fe)=0.14$ and $D(C) = 0.22$. 
The stopping power for a unit Dirac magnetic monopole in aluminium as a function of
 the velocity of the monopole is shown in Fig.~\ref{Fig:monopole-stop} . The 
 above formalism allows accurate estimates of the stopping 
 power of magnetic monopoles for $\beta \ge 0.1$ and  $\gamma \le 100$. 

The computed range ~\cite{monopole-range} is shown in Fig.~\ref{Fig:monopole-stop} (right) for a Dirac monopole 
as a function of $p/M$ = $\beta\gamma$, where $p$ and $M$ are the momentum and the
mass of the monopole, respectively. The monopole energy loss calculation is implemented 
as part of a {\tt GEANT} package for the simulation of monopole trajectories in a detector~\cite{GEANT}.
 
The inspection of the Bethe-Bloch formulae for electronically and magnetically charged 
particles for a singly-charged Dirac monopole and a unit electric charge moving with 
velocity $\beta$ shows that the ratio of their  stopping powers  is $\sim 4700\beta^{2}$.
 It can be seen from Fig.~\ref{Fig:monopole-stop}  that as the monopole slows the ionization
  decreases as compared to electrically-charged particles where the opposite is true.   
 

One of the most successful approaches for calculating the stopping powers of materials
 has been to treat them approximately as a free (degenerate) gas of electrons - a ``Fermi gas''. 
 This is clearly appropriate for interactions with the conduction electrons of metallic absorbers.
  For nonmetallic absorbers it represents a reasonable approximation for heavy atoms ($Z \ge10$),
   for which the Thomas-Fermi description is valid. Ahlen and Kinoshita~\cite{AHLENKINOSH} 
   have computed the energy loss of  slow monopoles in Fermi gases:
\begin{equation}\label{eq:ahlenandkinosh}
\left(\dfrac{dE}{dx}\right)_{m} = \dfrac{2\pi N_{e} (ng)^{2}e^{2}v}{m_{e}c^{2}v_{F}}\left[\ln\dfrac{1}{Z_{min}} - \dfrac{1}{2}\right] \, ,
\end{equation}
where $v_{F}$ is the Fermi velocity, $v$ is the projectile velocity, $Z_{min} =
\hbar/(2m_{e}v_{F}a_{0})$,   $N_{e}$ is the density of non-conducting electrons in the 
medium, and $a_{0} = 0.5 \times 10^{-8}$~cm (roughly the ``mean free path'' of an electron
 bound in an atom).  This equation is expected to be valid for non-conductors for
  $10^{-4} \le \beta \le 10^{-2}$. 
 
For conductors one should add another term that depends on the conduction electrons. 
However, Eq.~(\ref{eq:ahlenandkinosh}) should actually provide a good description of 
energy loss in a conductor for $\beta \le 10^{-2}$, with the parameters $Z_{min} 
=\hbar/(2mv_{F}\Lambda$), and $\Lambda \approx 50aT_{m}/T$, where $a$ is a lattice 
parameter; $T_{m}$ is the melting point of the metal, $T$ is the temperature and $N_{e}$ 
is the density of conduction electrons.

\begin{figure}[htb]
\begin{center}
\includegraphics[width=20pc]{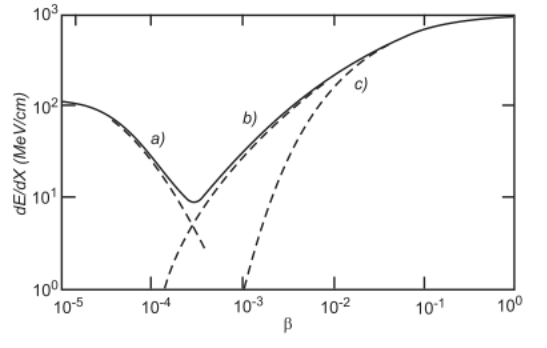}
\caption{ The energy losses in MeV/cm of a $g = g_{D}$ monopole in liquid 
hydrogen as a function of $\beta$. Curve a) monopole with $\beta $ 10$^{-4}$ cannot
excite atoms energy loss here i due to elastic monopole-hydrogen
 atom/nucleus  scattering. in the region covered by curve (b)   energy loss via the
  ionization or excitation of atoms and molecules   of the medium (``electronic'' energy loss) 
  dominates for $\beta >$ 10$^{-2}$. The dE/dx of magnetic monopoles with 10$^{-4}$ $ < \beta <$ 10$^{-3}$
  is mainly due to  excitations in atoms. Curve c) shows the ionization energy loss.
  The figure was obtained from Ref. \protect  \cite{GIORGIO}.}
\label{Fig:vlemel}
\end{center}
\end{figure}

For velocities $\beta \le 10^{-4}$ magnetic monopoles cannot excite atoms, but they can
 lose energy in elastic collisions with atoms or nuclei. In the case of monopole-atom elastic
  scattering this process is dominated by the coupling of the atomic electron magnetic 
  moments with the magnetic monopole field. An estimate of the energy loss can be achieved
   by  considering the elastic interaction of a monopole and an atom characterized only  by
    its magnetic moment~\cite{APPROXLEEL}:
\begin{equation}\label{eq:MONOPOLEATOM}
\dfrac{dE}{dx} \approx N_{a}E_{c.m.}\sigma \approx \dfrac{N_{a}\hbar^{2}}{m_{e}} ,
\end{equation} 
\noindent
where $N_{a}$ is the number of atoms/cm$^{2}$. The results of a more precise 
calculation~\cite{BRACCI} are shown in Fig.~\ref{Fig:vlemel}. The energy is released to the 
medium in the form of thermal and acoustic energy. Monopole-nucleus elastic scattering is 
expected to be dominated by the interaction of the monopole magnetic charge with the 
magnetic moment of the nucleus. Thus we obtain a formula rather like Eq.~(\ref{eq:MONOPOLEATOM}).

In a second approach~\cite{AHLENKINOSH} it was assumed that the atoms of the absorbing 
material have no magnetic moment and that the interaction is dominated by the transverse
 electric field. In this case, at large impact parameters the atom will respond to an applied
  electric field only through its induced electric dipole moment. This approach results in the 
  following expression for energy loss:
\begin{equation}
\dfrac{dE}{dx}_{\rm atomic} = \dfrac{4\pi
 N_{a}g^{2}Z^{2}r^{2}}{M_{\rm nuc}c^{2}}\ln\left(\dfrac{M_{\rm nuc}vca_{0}}{gZe}\right),
\end{equation}
\noindent
where $M_{\rm nuc}$ is the mass of the nucleus, $a_{0}$ is the Bohr radius, $v$ is the
 monopole velocity as a fraction of the speed of light and $Z$ is the atomic number of the 
 medium.  Taking silicon as an example, we see that $dE/dx_{\rm nuc} = 
 1.4~{\rm MeV/(g/cm}^{2})$ at $\beta = 10^{-3}$, which is about 7\% of the electronic 
 stopping power. For $\beta = 10^{-2}$, $dE/dx_{\rm nuc}$ is only about 1\% of the 
 electronic stopping power.
 

\subsection{Ionization energy loss in plastic Nuclear Track Detectors}



\begin{figure}[htb]
\begin{center}
\includegraphics[width=19pc]{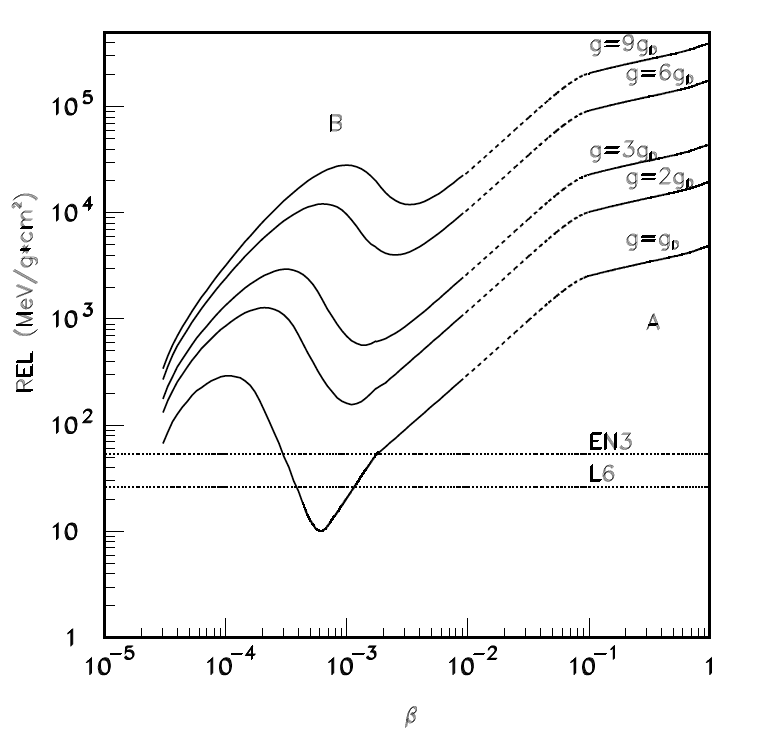}
\caption{The restricted energy loss of a magnetic monopole as a function of $\beta$ 
in the nuclear track detector CR39 ($\rho$ = I .31 g/cm$^{3}$). The detection 
thresholds of two types of CR39 used by MACRO experiment  (T1 and T2) 
are also shown. Notice that the solid curves represent $\beta$  regions where
 the calculations are more reliable, the dashed lines are interpolations. 
 Explanations about the curves in regions A
and B are given in the text.  The figure  is  obtained from Ref. \protect \cite{DERK}. }
\label{Fig:REL}
\end{center}
\end{figure}

A key quantity  for  assessing NTDs is the Restricted Energy Loss (REL), 
which is the energy deposited within $\sim$ 10 nm from the track. For computation 
of the REL, say for CR39, only energy transfers to atoms above 12~eV are considered, 
because it is estimated that at least 12~eV is required to break the molecular bonds~\cite{DERK}.
The REL is mainly due to the monopole itself with some contribution 
from short-range $\delta$-rays.
At high velocities ($\beta > 0.05$) the REL for the monopole tracks  are obtained excluding
 the energy transfers that result in $\delta$-ray  production and thus in energy deposited 
 far away from the track \cite{AHLEN2}.
 The REL for high-velocity monopoles is shown in the (A) region of 
 Fig.~\ref{Fig:REL} \cite{DERK}. 

At lower velocities ($\beta < 10^{-2}$) there are two contributions to REL. First there 
is the contribution due to ionization 
that  can be computed assuming that the medium is a degenerate electron gas \cite{AHLENKINOSH}.
However, another estimate  \cite{RITSON} gives a smaller estimate of ionization energy loss.
The second contribution to REL for slow monopoles is due to the elastic recoil of 
atoms.
 The procedure to compute this contribution
 can be found in Ref.~\cite{NAKAMURA} and~\cite{DERK}. The elastic recoil is due
  to the diamagnetic interaction between the magnetic monopole and the carbon and
   oxygen atoms of the medium. 
 
   The approach of Ref.~\cite{NAKAMURA}
    is used to estimate the atomic elastic recoil contribution, which gives rise to a 
    bump in REL around $\beta g/g_{D}$, shown in Fig.~\ref{Fig:REL}. The ionization 
    contribution dominates at $\beta$ higher than the value at which the minimum 
    REL occurs. For $ 10^{-2} < \beta < 10^{-1}$ a smooth interpolation has been 
    performed. As is shown in Fig.~\ref{Fig:REL}, monopoles with $ g > 2g_{D}$ can be
     detected by the CR39 detector for $\beta > 3 \times 10^{-5}$. 

The practical threshold of CR39 was determined by direct measurements with 
heavy ions to be $Z/ \beta \gtrsim 5$ \cite{GIORGIO2,GIORGIO2a,GIORGIO2b}. This  corresponds to an REL of
 $\sim 25~{\rm MeV~cm}^{2}{\rm g}^{-1}$. The threshold of Makrofol is 
 approximately ten times higher.

\subsubsection{Track formation in NTDs}

When a particle passes through a NTD, it leaves a trail of damage along its path called the
 latent track (LT) with diameter of the order of 10 nm \cite{LATENT}.
  The LT can be enlarged and thus made visible to an optical microscope
  by chemically etching the detector. The etchants are highly basic or acid solutions, 
  e.g., aqueous solutions of NaOH or KOH of different concentrations and temperatures, 
  usually in the range $40 \rightarrow 70^o$C. Etching takes place via the  dissolution 
  of the disordered region of the LT, which is in a state of higher free energy than the 
  undamaged bulk material.
 
The track etch rate $v_T$ is defined as the rate at which the detector material is 
chemically etched along the LT, whereas the bulk etch rate $v_B$ is defined as 
the rate at which the undamaged material of the NTD is etched away. The track 
etch rate $v_{T}$ depends on the energy loss of the incident particles and on
 the chemical etching conditions (concentration, temperature, etc.) whereas 
 $v_{B}$ depends only on the etching conditions.  The reduced etch rate $p$ 
 is defined as $p = v_{T}/v_{B}$. If  $v_{T} > v_{B}$  ($p>1$) etch-pit cones are
  formed.
 
\begin{figure}[htb]
\begin{center}
\includegraphics[width=22pc]{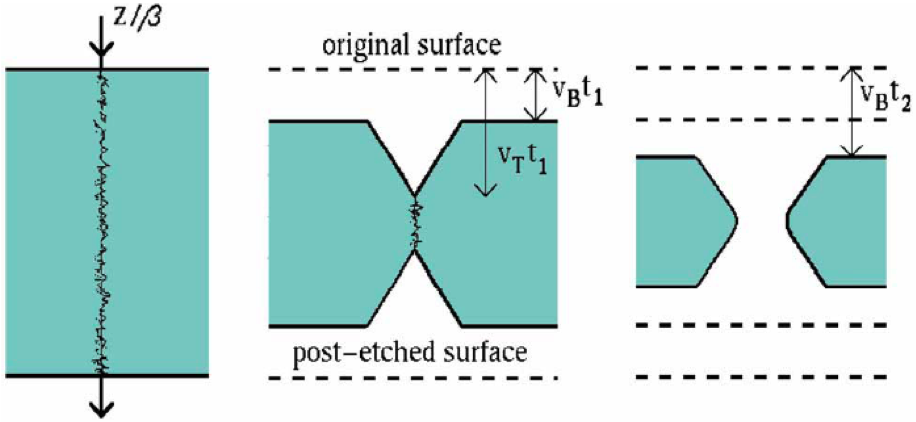}
\caption{Depiction of the latent track left by the passage of an ionizing particle through
 a NTD and the subsequent formation of etch-pit cones.}
\label{Fig:etching}
\end{center}
\end{figure}

The formation of etch pit cones for a particle  impinging normally on the NTD is depicted
in Fig.~\ref{Fig:etching}. Etching a layer for a short time yields two 
etched cones on each side of the sheet. The primary ionization rate may be 
determined from the geometry of the etched cones. For CR39 this technique
 yields  measurements of the electric charge of heavy nuclei to a precision of
  $0.1e$ if one uses several layers of plastic sheets, placed perpendicular
   to the incoming ions \cite{CHARGERES, CHARGERES1}.

For an electrically-charged particle slowing down appreciably within the 
NTD stack, the opening angle of the etch-pit cone would become smaller. 
For a particle stopping inside the detector, if the chemical etching continues
 beyond the track range, the cone end develops into a spherical shape (end
  of range tracks). For a monopole slowing down appreciably within an NTD 
  stack its ionization would fall rather than rise and thus the opening angle of the etch pits would
   become larger rather than smaller. For relativistic monopoles at the LHC the 
   energy loss across an NTD stack would usually be essentially constant, 
   yielding a trail of equal diameter etch-pit pairs, assuming that all NTD 
   sheets were etched in the same manner and for the same time.

\section{Bound States Between Monopoles and Matter}

The study of the interactions of magnetic monopoles in matter is of critical importance in the understanding of the energy loss of monopoles as well as
the formation of bound systems of monopoles and atomic nuclei.   The consideration of the formation of bound states between monopoles and matter was 
necessary for  the design of the MoEDAL trapping detector subsystem (MMT).

\subsection{Trapping monopoles}

Once the monopoles are produced in a collision at the LHC, they will travel through the surrounding material losing energy as described above. In some cases they will lose enough energy to slow down and become trapped in the material, presumably by binding to a nucleus of an atom forming the material.  We assume that this binding is due to the interaction between the magnetic charge of the monopole and the magnetic moment of the nucleus. In the case of dyons the situation is simple, the correct sign of the electric charge will always bind electrically to  nuclei. Thus, we consider here only the binding of a monopole to a nucleus. 

The magnetic moment of a nucleus with charge $Ze$ and mass $M =AM_{p}$ is:
\begin{equation}
 \boldsymbol{\mu}= \dfrac{Ze}{2M}g_{N}\bold{S},
\end{equation}
where $ \bold{S}$ is the spin of the nucleus and $g_{N}$ is the gyromagnetic ratio of the nucleus, and $M$ is the reduced mass of the monopole-nucleus system. If the monopole is heavy compared to the nucleus, which is a good assumption for searches at the LHC,  the reduced mass is essentially equal to the mass of the monopole. If the  monopole has a magnetic charge $g=n/2e$ (using the units $\hbar = c = 1$), where $n = \pm 1, \pm 2, ...$, the electromagnetic field around the monopole-nucleus system carries an angular momentum  $|q| = |nZ|/2$, which can combine with the orbital angular momentum to give the values: $l = |q|, |q| + 1, ....$. This in turn can be coupled with the nuclear spin $S$ to give a total angular momentum: $j = |q| + 1 - S, |q| + 1,....$, provided $ |q| \ge S$~\footnote{If $|n|  = 1$, this is only true for magnetic charge coupled to  $^{2}H  (S=1, |q|=1/2)$, $^{8}Li (S = 2, |q| = 3/2)$ and $ ^{10}B  (S = 3, |q|=5/2)$.}.

We consider first the non-relativistic binding for single nucleons with  $S=1/2$~\cite{BRACCIBOUND,MALKUS,GAMBERG}. Binding to neutrons (with $Z=0$) will occur in the lowest angular momentum state, $J =1/2$, if $| g_{N}|/2 > 3/(2n)$. Since $g_{N}/2 = -1.91$, this condition is satisfied for all $n$. Defining a ``reduced'' gyromagnetic ratio of the nucleus, $g'_{N} = g_{N}(A/Z)$, binding will occur in the special lowest angular momentum state $J = l- 1/2$, if $ g'_{N}/2 > 1 + 1/4l$. The proton has a gyromagnetic ratio of $5.586$, thus the proton will bind to a monopole.

Binding can occur in higher angular momentum states $J$ if and only if: 
\begin{equation}
|g'_{N}|/2  > g_{c} = \dfrac{2}{l}\left| J^{2} + J - l^{2} \right| + 2 \, .
\end{equation}
In order to calculate the binding energy, one must regulate the potential at $r = 0$. The results shown in Table~\ref{tab:trapping}~\cite{GAMBERG} assume a hard core.

\begin{table}[htbp]
  \tbl{Weakly-bound states of nuclei to a magnetic monopole. The angular momentum
quantum number $J$ of the lowest bound state is indicated. A single  Dirac charge was assumed in all
   calculations. }
   {\begin{tabular}{|c|crrrc|c|} \toprule
    \hline 
   Nucleus &  Spin(S) & $ g_{N}/2$  & $g'_{N}/2$  &  $J$                    & $E_{bind}$ & Ref. \\ \hline
 $  n $                 &  1/2         &      -1.91       &                      &         1/2          &  350 keV     &  \cite{sivers} \\
 $^{1}_{1}$H   &  1/2         &        2.79      &   2.79           &   $l$ -1/2 =0   & 15.1 keV     & \cite{BRACCIBOUND} \\
  $^{1}_{1}$H  &   1/2       &         2.79              &     2.79                 &   $l$ -1/2 =0     & 320 keV      & \cite{sivers} \\
 $^{1}_{1}$H   &   1/2       &      2.79                &   2.79                   &   $l$ -1/2 =0     & 50-1000 keV & \cite{OLAUSSEN} \\
 $^{3}_{2}$He &   1/2       &     0.857       &    1.71        & $l$ +1/2 =3/2   &  13.4 keV     & \cite{BRACCIBOUND} \\
 $^{27}_{13}$Al & 5/2   &     3.63         &     7.56         & $l$ - 5/2 = 4    &   2.6 MeV      &\cite{OLAUSSEN} \\ 
 $^{27}_{13}$Al & 5/2    &    3.63           &   7.56          &  $l$- 5/2 = 4   &  560 keV       & \cite{GOEBEL} \\
  $^{113}_{48}$Be &1/2   &   -0.62          &  -1.46         & $l$ +1/2 = 49/2 & 6.3 keV      &  \cite{BRACCIBOUND} \\ \hline
        \end{tabular}
    \label{tab:trapping}}
\end{table}

The more general case for non-relativistic binding for a general $S$ was considered, for example, in Ref~\cite{sivers}. We assume that $l \ge S$, the only exceptions being $^{2}$H, $^{7}$Li and $^{10}$B. The binding in the lowest angular momentum state $J= l - S$ is given by the same criteria as in the spin-$1/2$ case considered above. Binding in the next state, with $ J = l - S + 1$, occurs if $\lambda_{\pm} > 1/4$ where, 
\begin{equation}
\lambda_{\pm} = \left(S - \dfrac{1}{2}\right)\dfrac{g'_{N}}{2S}l -2l - 1 \pm\sqrt{(1 + l)^{2} + (2S - 1 - l)\dfrac{g'_{N}}{2S}l 
+ \dfrac{1}{4}l^{2}\left(\dfrac{g'_{N}}{2S}\right)^{2}} \, .
\end{equation}
Spin~1 is a special case where $\lambda_{-}$ is always negative, while $\lambda_{+} > $1/4 if $g'_{N}/2  >  g'_{c}$, where:
\begin{equation}
g_{c} = \dfrac{3}{4l}\dfrac{(3 + 16l + 16l^{2})}{9 + 4l}
\end{equation}
For  higher spins $S > 1$, both $\lambda_{\pm}$ can exceed $1/4$, if $\lambda_{+} > 1/4$ for $g'_{N}/2 > g_{c-}$ or $\lambda_{-} > 1/4$ for $g'_{N}/2 > g_{c+}$. For $S = 5/2$:
\begin{equation}
(g_{c})\mp = \dfrac{36 + 28l \mp \sqrt{1161 + 1296l + 64l^{2}}}{12l} \, .
\end{equation} 
Thus, $^{27}_{13}$Al will bind in either of these states, or the lowest angular momentum state,  since $g'_{N}/2 = 7.56$ and $1.374 < g_{c-} < 1.67$ and $1.67 < g_{c+}  <  4.216$.  
The preceding considerations indicate that  aluminium - a nucleus with large and positive anomalous magnetic moment - is a good choice for the MoEDAL Monopole Trapping Detector volumes.

\subsection{Binding of the monopole-nucleus state to the material matrix}

A practical question for a monopole trapping detector is how well the monopole-nucleus state is bound to the material lattice. The decay rate of a such a state has been estimated~\cite{GAMBERG} to be:
\begin{equation}
\Gamma \sim n^{-1/2}10^{23}s^{-1}  \exp{\left[-\dfrac{8\sqrt{2}}{3 \times 137}\left(\dfrac{-E}{m_{e}}\right)^{3/2} 
\dfrac{B_{0}}{nB} A^{1/2}\left(\dfrac{m_{p}}{m_{e}}\right)^{1/2}\right]} \, ,
\end{equation}
where the characteristic field, defined by $eB_{0}$ = m$_{e}^{2}$, is 4 $\times$ 10$^{9}$T. If we assume $B = 1.5~{\rm T}$ and $A = 27$, $-E = 2.6$~MeV (for  $^{27}_{13}$Al), to get a lifetime of at least a 10~years, the binding energy would need to be greater than around 1~eV. According to this calculation any state that is bound at the keV level or more will be stable for around $10,000$~years. 

This calculation suggests that a major disruption of the lattice would be required to dislodge the monopole-nucleus state. If monopoles bind strongly to nuclei they will not be extracted by the fields available in the LHC
experiments~\cite{GAMBERG}. However, another estimate~\cite{GOTO} suggests that magnetic fields of the order of 10~kG would be sufficient to release trapped monopoles. MoEDAL is deployed in a ``zero'' field region of the LHCb, with only a small fringe field from the LHCb main dipole magnet, much smaller than the field required to free a trapped monopole, for both cases considered above.

\section{Magnetic Monopoles and Dyons}

As we have seen from the above discussion the MoEDAL detector is designed to exploit the energy loss mechanisms of 
magnetically charged particles in order  to optimize its   potential to discover such revolutionary messengers of new physics.
 We will now discuss the various theoretical  scenarios in which magnetic charge would be produced  at the LHC. There are several more extensive reviews of magnetic monopoles available in the literature~\cite{Milton:2006cp,Rajantie:2012xh,ShnirBook,WeinbergBook}.

In classical electrodynamics, as described by Maxwell's equations, the existence of magnetic charge might be expected because it would make the theory more symmetric. In vacuum, the Maxwell equations are symmetric under a duality transformation that mixes the electric and magnetic fields,
\begin{equation}
 \vec{E}+i\vec{B}\rightarrow e^{i\phi}(\vec{E}+i\vec{B}),
\end{equation}
but this duality symmetry is broken if magnetic charges do not exist in Nature. This is because the duality would transform the electric charge density $\rho_{\rm E}$ to magnetic charge density $\rho_{\rm B}$,
\begin{equation}
 \rho_{\rm E}+i\rho_{\rm M}\rightarrow e^{i\phi}(\rho_{\rm E}+i\rho_{\rm M}).
\end{equation}
The symmetry of the equations therefore motivates the expectation that both electric and magnetic charges should exist in Nature.

Magnetic charges can be incorporated into classical electrodynamics trivially by modifying the magnetic Gauss's law to become
\begin{equation}
\vec{\nabla}\cdot\vec{B}=\rho_{\rm M}.
\end{equation}
An isolated magnetic point charge $g$ will then have a magnetic Coulomb field
\begin{equation}
\label{equ:Bfield}
\vec{B}(\vec{r})=\dfrac{g}{4\pi}\dfrac{\vec{r}}{r^3}.
\end{equation}
The description of the electromagnetic fields requires the introduction of the vector potential $\vec{A}$, which is related to the magnetic field $\vec{B}$ by the relation
\begin{equation}
 \vec{B}=\vec{\nabla}\times\vec{A}.
\end{equation}
For any smooth vector potential $\vec{A}$, the magnetic field is then automatically sourceless, $\vec{\nabla}\cdot\vec{B}=0$, and this appears to rule out the existence of magnetic charges. However, Dirac showed in 1931 that this conclusion is premature~\cite{Dirac1931kp}. Because the vector potential $\vec{A}$ is not an observable quantity, it does not have to be smooth. Dirac found that the magnetic field (\ref{equ:Bfield}) of a an isolated magnetic charge $g$ can be represented by the vector potential
\begin{equation}
\label{equ:Diracpot}
 \vec{A}(\vec{r})=\dfrac{g}{4\pi r}\dfrac{\vec{r}\times\vec{n}}{r-\vec{r}\cdot\vec{n}}
\end{equation}
everywhere except along a line in the direction of the unit vector $\vec{n}$. Along this line, which is known as the Dirac string, the magnetic field is singular. However, if one tries to probe the Dirac string with an electrically charged particle with electric charge $q$, one finds that it is unobservable if the magnetic charge $g$ satisfies the Dirac quantization condition,
\begin{equation} \label{equ:Diraccondition}
 qg\in2\pi\mathbb{Z}.
\end{equation}
where $\mathbb{Z}$ is the integer number set. If the electric charges of all particles satisfy this condition, meaning that they are quantized in units of $2\pi/g$, then the Dirac string is completely unobservable. In that case the vector potential (\ref{equ:Diracpot}) describes physically a localized magnetic charge at the origin, surrounded by the magnetic Coulomb field (\ref{equ:Bfield}), with the only singularity at the origin. The condition (\ref{equ:Diraccondition}) implies that the electric charge has to be quantized, i.e., an integer multiple of the elementary charge $q_{0}=2\pi/g$. This means that not only are magnetic charges allowed by quantum mechanics, but their existence also explains the observed quantization of electric charge. 

Obviously, the quantization condition (\ref{equ:Diraccondition}) also implies that the magnetic charge is quantized in units of $g_0=2\pi/q_0$, where $q_0$ is the quantum of electric charge, which in the Standard  Model is the charge of the positron, $q_0=e$. The quark electric charges are quantized in units of $e/3$, but because they are confined, they are not relevant for the quantization argument~\cite{Preskill1984gd}. Note that the minimum magnetic charge $g_0$ is very large. The strength of the magnetic Coulomb force between two charges $g_0$ is $g_0^2/q_0^2=4\pi^2/q_0^4\approx 4700$ times stronger than the electric Coulomb force between two elementary charges $q_0$. The magnetic analogue of the fine structure constant $\alpha=q_0^2/4\pi\approx 1/137$ is $\alpha_{\rm M}=g_0^2/4\pi=\pi/\alpha\approx 430$.

Instead of the description (\ref{equ:Diracpot}) with a singular but unobservable Dirac string,
it is also possible to describe the Dirac monopole by defining the fields on two overlapping
charts which cover the space around the monopole, for example the northern and southern hemispheres~\cite{Wu:1975es}. The fields can then be smooth on both charts provided that they are related by a topologically non-trivial gauge transformation along the equator where they overlap. Because the overlapping region has the topology of a circle, this argument shows that the Dirac string is an unphysical coordinate singularity if the gauge group $G$ is not simply connected, i.e., $\pi_1(G)\neq 1$. This is directly related to the Dirac quantisation condition (\ref{equ:Diraccondition}), because electromagnetism can be described by a compact $U(1)$ gauge group only if electric charges are quantised. Otherwise the gauge group is the non-compact covering group ${\mathbb R}$ of real numbers.

Schwinger~\cite{Schwinger1969ib} generalized the quantization condition to dyons, particles that carry both electric and magnetic charge. In this case, the charges of all particles have to satisfy the condition
\begin{equation}
\label{equ:quantcond}
q_1g_2-q_2g_1\in 4\pi{\mathbb Z}.
\end{equation}
His argument also implies that the minimum charge for a magnetic monopole should be
$g=4\pi/q_0$, twice the Dirac charge.

\subsection{GUT Monopoles}

The discussion in the previous section focussed on static monopoles. Formulating a quantum field theoretic description of dynamical magnetic monopoles is challenging,
but there examples
 of weakly-coupled quantum field theories in which magnetic monopoles appear as non-perturbative solutions. The most important case is the 't~Hooft-Polyakov monopole solution~\cite{tHooft1974qc,Polyakov1974ek} in the Georgi-Glashow model, which is an $SU(2)$ gauge field theory with an Higgs field $\Phi$ in the adjoint representation. When the Higgs field has a non-zero vacuum expectation value, $\langle\Phi\rangle=v>0$, the $SU(2)$ gauge symmetry breaks spontaneously to $U(1)$, thereby giving rise to electrodynamics. In this broken phase the theory has a smooth, spherically symmetric ``hedgehog'' solution with magnetic charge $g=4\pi/q$, where $q$ is the electric charge of the $W^+$ boson. The monopole has a non-zero core size $r_0\sim 1/qv$ and a finite mass 
$M\sim v/q^2$.
Therefore it appears as a particle state in the spectrum of the theory. Lattice field theory simulations~\cite{Rajantie2005hi} have confirmed the validity of the classical mass estimate also in the quantum theory at weak coupling.

The theory also has dyon solutions with both electric and magnetic charges~\cite{Julia1975ff}. At the semiclassical level they can have arbitrary electric charge, but quantum effects force the charges to be quantized according to Eq.~(\ref{equ:quantcond}). In the presence of CP violation the electric charges are non-integer valued~\cite{Witten1979ey}. 
Dyons are generally heavier than electrically neutral monopoles and can therefore decay into monopoles and electrons or positrons.

Although the Georgi-Glashow model by itself is not a realistic theory of elementary particles, the same solutions also exist in all Grand Unified Theories (GUTs), in which the strong and electroweak forces are unified into a simple gauge group. In the simplest $SU(5)$ GUT~\cite{Georgi1974sy}, with the symmetry breaking pattern
\begin{equation}
SU(5)\rightarrow SU(3)\times SU(2)\times U(1)
\end{equation}
the mass of these monopoles is determined by the GUT scale: $M\sim \Lambda_{\rm GUT}/\alpha\sim 10^{17}~{\rm GeV}$, and therefore well beyond the reach of the LHC. The lightest monopoles in the theory have a single Dirac charge $g=2\pi/e$, but they can form doubly-charged bound states.

In some GUTs, it is possible to find 't~Hooft-Polyakov monopole solutions with lower mass. For example, the Pati-Salam model~\cite{Pati1974yy} with the symmetry breaking pattern 
\begin{equation}
SO(10)\rightarrow SU(4)\times SU(2)\times SU(2)\rightarrow SU(3)\times SU(2)\times U(1) \, ,
\end{equation}
has monopole solutions with mass $M\sim 10^{15}~{\rm GeV}$~\cite{Lazarides1980cc} and charge $g=4\pi/e$, twice the Dirac charge. Family unification models with the symmetry group $SU(4)\times SU(3)\times SU(3)$ can have multiply-charged monopole solutions with masses as low as $M\sim 10^7~{\rm GeV}$~\cite{Kephart2006zd}.

Monopole solutions exist also in theories with compactified extra dimensions~\cite{Sorkin1983ns,Gross1983hb}. In this case the $U(1)$ gauge group of electrodynamics corresponds to the compactified dimension. At the monopole core the size of the extra dimension shrinks to a point in a smooth way. The natural mass values for these Kaluza-Klein monopoles are above the Planck scale, $M\sim M_{\rm Pl}/\alpha\sim 10^{20}~{\rm GeV}$ but, again, there are models in which they are considerably lighter.

These examples show that magnetic monopoles can exist consistently in quantum field theories, and also make it possible to investigate their properties and behaviour using normal quantum field theory methods. They are also well-motivated physical theories, and therefore they indicate that it is likely that superheavy monopoles of mass $M\gtrsim 10^{17}~{\rm GeV}$ exist. Of course it would be impossible to produce such monopoles in any conceivable experiment. 
However, it is perfectly possible that there is some new unexpected physics between the electroweak and GUT scales, and therefore there can well be lighter magnetic monopoles that are not related to grand unification or compactified extra dimensions.

\subsection{Monopole-like structures in the electroweak theory}

The discovery of a Higgs-like boson in July 2012 by the ATLAS and CMS~\cite{Higgs-like, Higgs-like1} experiments at the Large Hadron Collider (LHC) at CERN has led to the possibility that the puzzle of the Standard Model spectrum may be  completed. More recently, this new particle looks increasingly like a Standard Model (SM) Higgs boson~\cite{SMHiggs,SMHiggs1,SMHiggs2,SMHiggs3}, reinforcing the  electroweak theory of Glashow, Salam and Weinberg as a successful theory. The experimental verification of a monopole-like solution within the framework of the SM would explain from first principles the quantization of the electric charge, something that the present SM cannot do.

\subsubsection{Electroweak Monopole}




There has been some discussion in the literature of the possible  existence of monopoles within the Standard Electroweak Theory.
The electroweak symmetry breaking pattern is $SU(2)\times U(1)_Y \to U(1)_{\rm EM}$.  
A topological monopole requires nontrivial  topology of the quotient group $ SU(2)\times U(1)_Y/U(1)_{EM}$ which, because of the
residual $U(1)$ symmetry would be putatively nontrivial and of the sort which supports monopoles.   
It is therefore possible to repeat Dirac's argument and write down the potential (\ref{equ:Diracpot}) describing a static monopole with a non-zero magnetic hypercharge. At long distances, the Dirac quantisation condition (\ref{equ:Diraccondition}) must be satisfied by the electromagnetic charges, and therefore the monopole will also have a non-trivial $SU(2)$ gauge field configuration. 

However, unlike the 
model which is used in the discussion of the 't Hooft-Polyakov monopole   where the scalar field was an $SU(2)$ triplet, 
the standard Higgs field is an $SU(2)$ doublet 
and it is not possible to make the usual spherically symmetric hedgehog monopole configuration from a doublet.  
To see this in more detail, consider the Lagrangian of the Weinberg-Salam model,
\begin{align}\label{weinbergsalammodellagrangian}
{\mathcal L}&=  (D_\mu\phi)^\dagger D^\mu\phi-\frac{\lambda}{2}\left( \phi^\dagger\phi - \frac{\mu^2}{\lambda}\right)^2 -\frac{1}{4}F_{\mu\nu}F^{\mu\nu}
-\frac{1}{4}G_{\mu\nu}G^{\mu\nu} \\
D_\mu\phi &= \left(\partial_\mu  -i\frac{g}{2}\vec\tau\cdot\vec A_\mu -i\frac{g'}{2}B_\mu\right)\phi \\
F_{\mu\nu} &= \partial_\mu A_\nu^a-\partial_\nu A_\mu^a+\frac{g}{2}\epsilon^{abc}A_\mu^b A_\nu^c
~,~
G_{\mu\nu} = \partial_\mu B_\nu-\partial_\nu B_\mu 
\end{align}
where we use the $(1,-1,-1,-1)$ signature of the Minkowski space metric.
The ground state of the standard model has a nonzero value for the Higgs field, $\phi^\dagger \phi=\mu^2/\lambda$,  
and the rest of the fields equal to zero. The usual fields, $W^\pm,Z^0$, the
Higgs boson and the photon are quanta of fluctuations about this ground state.   

It has been pointed out, initially by Cho and Maison \cite{Cho1996qd}, that there is a singular topological monopole solution of this model.  
Their ansatz for a time-independent, spherically symmetric solution  is 
\begin{align}
&\phi (r,\theta,\varphi)= \rho(r) \left[ \begin{matrix} \frac{i}{\sqrt{2}}e^{-i\varphi}\sin\frac{\theta}{2} \cr -\frac{i}{\sqrt{2}}\cos\frac{\theta}{2}\cr\end{matrix}\right]
\\
&A_0^a =\frac{ \hat r^a}{g} A(r)~~,~~A_i^a(r,\theta,\varphi) = \frac{1}{g}(f(r)-1)\epsilon_{aij}\frac{r_j}{r^2}
\\
&B_0 = \frac{1}{g'}B(r) ~~,~~  B_i(r,\theta,\varphi) = \frac{1}{g'}(1-\cos\theta)\nabla_i\varphi
\end{align}
The angular dependence of the fields in this anzatz are determined by the requirement of rotational covariance, that is,   
 the requirement that the solution leads to a spherically
symmetric monopole and, moreover,  that the solution has a magnetic monopole charge for the degrees of freedom
which will become the electromagnetic fields.  The field $\vec B(r,\theta,\phi)$ has the usual Dirac string   at $\theta=\pi$ which is invisible
to particles with appropriately quantized charges.  

To see that this solution must be a singular
solution of the Weinberg-Salam theory (\ref{weinbergsalammodellagrangian}), we substitute the ansatz into the energy functional to get
the energy of the classical field configuration,
\begin{align}
{\mathcal E}= \frac{2\pi}{g^2}\int _0^\infty \frac{dr}{r^2} \left[\frac{g^2}{{g'}^2}+\dot f^2+(1-f^2)^2+ \frac{g^2}{2}(r\dot\rho)^2
+\frac{g^2}{4}f^2\rho^2  +\frac{g^2r^2}{8}(A-B)^2\rho^2  \right. \nonumber \\  \left.  +
\frac{\lambda g^2r^2}{4}\left(\rho^2-\frac{\mu^2}{\lambda}\right)^2 
+\frac{g^2}{2{g'}^2}(r\dot B)^2 +\frac{1}{2}(r\dot A)^2+f^2A^2 \right]
\end{align}
Here, first of all, we see the energy must diverge.  The integral of the first term in the integrand, 
which arises from the magnetic term $\int (\vec\nabla\times\vec B)^2$,
 is divergent, and all of the other terms are
positive semi-definite.    The first, divergent term therefore provides a lower found on the energy. .  
The minimum energy configuration is found by the simplest ansatz there all of the other terms
in the energy vanish, that is, 
where $A=B=0$, $f=1$ and $\rho=\mu/\sqrt{\lambda}$ and where the only nonzero field is
\begin{align}\label{minimalchomaisonmonopole}
 B_i(r) = \frac{1}{g'}(1-\cos\theta)\nabla_i\varphi
\end{align}
This classical field is very similar to the  Dirac monopole.  
However, as Cho and Maison pointed out \cite{Cho1996qd},  due to the way that the electromagnetic charge arises from an SU(2) weight, 
it must have an even number of 
units of Dirac monopole charge in general and two units in this specific case.  

The divergence of the energy is due to the point-like nature of the monopole.    It should be resolved by an ultraviolet regularization.
One would naively expect that this would make the first term in the energy
proportional to the ultraviolet cutoff, which  would obtain its scale from physics beyond the standard model, and making the 
monopole mass very large.  However,
there are some suggestions for resolution of this singularity which would result in lighter monopoles, with masses in the 
range of multi-TeV. 

This divergence may be regularised by new physics beyond the Standard Model, which would lead to an effective ultraviolet cutoff $\Lambda\gtrsim 1~{\rm TeV}$.
In that case, the classical energy would be
\begin{equation}
\label{equ:Mestimate}
E\sim \frac{\Lambda}{\alpha},
\end{equation}
where $\alpha$ is the fine structure constant.
Because $\alpha\ll 1$, this would appear to imply that the mass of a magnetic monopole would have to be 
significantly higher than the energy scale $\Lambda$ of new physics. 
This would set a lower limit of several ${\rm TeV}$ for the 
monopole mass. One would also generally expect that the same physics that cuts off the 
divergence also predicts other new particles with masses $m\sim \Lambda$ which have not been seen at the LHC yet.

However, comparison with the renormalisation of other particle masses in quantum field theory suggests that the 
estimate (\ref{equ:Mestimate}) may not be relevant. The same argument, when applied to the electric field of
the electron, would suggest that its mass should have an ultraviolet divergence $\sim e^2\Lambda$, but a proper 
 quantum field theory calculation shows that because of chiral symmetry the actual divergence is only logarithmic 
 $\sim e^2m\log \Lambda/m$. Furthermore, the ultraviolet divergence is actually cancelled by the bare mass (or the mass counterterm) 
of the particle. As a result, the fermion masses in the Standard Model are free parameters, not determined by the theory.
If the same applies to magnetic monopoles, then it means that the physical monopole mass may be related 
to the scale of new physics $\Lambda$ in a different way, or not at all, and could be significantly lower.
In particular, monopoles may then be light enough to be produced at the LHC.

A complete quantum field theory calculation of the monopole mass requires a
formulation with dynamical magnetic monopoles, instead of the above treatment
in which the monopole appears as a static, classical object. The Standard Model itself does not contain magnetic monopoles, but 
it may be possible to add them as additional elementary degrees of freedom.
However, this has turned out to be very difficult to do in practice, even in the simpler theory of
quantum electrodynamics~\cite{Schwinger1966nj},\cite{Zwanziger1970hk},\cite{Blagojevic1985sh}. The formulation requires 
two vector potentials and is not manifestly Lorentz invariant, although physical observables are Lorentz invariant 
if the charges satisfy the quantisation condition (\ref{equ:quantcond}). 
Even if these problems with the formulation of the theory are overcome, any quantum field theory in which magnetic monopoles appear as
dynamical particles will also suffer from the practical problem that their coupling to the electromagnetic field is 
strong. The loop expansion would give a power series in $\alpha_M$, and would therefore not converge, making
perturbation theory inapplicable. The full calculation will therefore require non-perturbative techniques such as lattice field theory simulations.

\subsubsection{The Cho-Maison Mass Estimate}

There is no 't Hooft-Polyakov monopole within the SM,
because of the trivial topology of the quotient  group $SU(2) \times U(1)_{Y} / U(1)_{EM}$ 
after the spontaneous symmetry breaking of  $SU(2) \times U(1)_{Y}  \to U_{EM}(1)$. 
Such a quotient  group space  does not possess the non-trivial second homotopy 
required for the existence of 't~Hooft-Polyakov-like monopole solutions. 
 
However, 
Cho and Maison~\cite{Cho1996qd} suggested that the SM could be viewed as a gauged $CP^{1}$ model, 
with the Higgs doublet field interpreted as the $CP^{1}$ field. This would be an
extension of our view of the SM. It would bypass the previous argument of the trivial group-space  topology, 
since the second homotopy group of the gauged $CP^{1}$ model
is the same as that  of  the Georgi-Glashow model that contains the 't Hooft-Polyakov monopole.

The Cho-Maison monopole can therefore be seen as a hybrid between  the 't Hooft-Polyakov and Dirac monopoles.
Unlike the Dirac monopole, it would carry magnetic charge $(4\pi)/e$ because
the $U(1)_{EM}$ in the SM has a period of $4\pi$, not $2\pi$, as it comes from the $U(1)$ subgroup of  $SU(2)$. 
The $SU(2)$ part of the Cho-Maison monopole is finite, like the `t Hooft- Polyakov monopole,
whereas its $U(1)$ part (corresponding to the $U(1)_{EM}$ gauge field) has a real Dirac monopole-like
singularity at the origin. This singularity is responsible for the fact that its mass is indeterminate,
whereas the 't~Hooft-Polyakov monopole has a finite  total energy,
thanks to the EW gauge group being embedded in a simple unified gauge group.

Recently, an estimate of the mass of the Electroweak  monopole has been made~\cite{CHO-KIM-YOON}, 
based on the assumption that  the monopole is a topological soliton. 
Such solitons are topologically non-trivial field configurations that are characterized by finite total energy. 
When representing the Cho-Maison monopole as a  topological soliton, it is necessary
to regularize the solution by means of a cut-off representing physics beyond the SM that
eliminates the short distance singularity due to the Abelian gauge field configuration at the origin. 
The procedure adopted was to  create an effective theory by embedding the electroweak theory in 
another microscopic model. The authors asserted that such a finite-total-energy soliton must be stable
under a rescaling of its field configuration, according to Derrick's theorem~\cite{derrick}. 
This implies that one could consider a simple scale transformation of the coordinates, 
$\vec x \to \lambda^\prime \vec x$, under which the stable monopole configuration should
satisfy certain relationships among its constituent quantities pertaining to spatial integrals of the 
various field configurations that enter the solution. 
  
However there are some subtle requirements that the cut-off theory has to satisfy in order for
such mass estimates to be correct, which we outline now. To this end, 
we first remark that, in the notation of~\cite{CHO-KIM-YOON},
the total SM contribution to the energy (and hence mass) of the monopole is the sum of four contributions 
\begin{equation} \label{energy}
E=K_A+K_B+K_\phi+V_\phi~,
\end{equation}
where $K_A,K_B$ and $K_\phi$ stand for the kinetic energies of the non-Abelian gauge field $\bf{A_{\mu}}$,
the Abelian gauge field $B_{\mu}$ and the Higgs field $\phi$ respectively, while $V_\phi$ is the Higgs potential energy:
\begin{eqnarray}
&&K_A=\dfrac{1}{4}\int d^3x ~\vec F_{ij}^2~~,~~K_B=\dfrac{1}{4}\int d^3x~ G_{ij}^2\\
&&K_\phi=\int d^3x~|D_i\phi|^2~~,~~V_\phi=\int d^3x~V(\phi)\nonumber~.
\end{eqnarray}
For the spherically-symmetric Cho-Maison monopole configuration,
$K_A,K_\phi$ and $V_\phi$ are finite, and $K_B$ is expressed as an integral
which, as mentioned previously, has a short distance ultraviolet (UV) singularity 
at $r=0$, 
which is regularized by introducing a short-distance cut off $\epsilon$.
The scaling properties of the cut-off are crucial in ensuring that the energy satisfies 
Derrick's theorem and is independent of the coordinate scale factor $\lambda^\prime$. 
To see this, one needs to  calculate explicitly the contribution $K_B$ for the Cho-Maison monopole:
one obtains 
\begin{equation} \label{kb}
K_B=\dfrac{1}{4}\int r^2\, {\rm \sin}\theta \,dr\, d\theta \, d\varphi~G_{ij}^2=\dfrac{\pi}{g'^2}\int_\epsilon^\infty \dfrac{dr}{r^2}
=\dfrac{\pi}{g'^2\epsilon}~.
\end{equation}
We next introduce the coordinate rescaling $\vec r\to\lambda^\prime \vec r$, under which scalar, vector and tensor quantities scale as 
\begin{equation}
\phi(\vec r)\to\phi(\lambda^\prime \vec r)~~,~~
B_i(\vec r)\to\lambda^\prime B_i(\lambda^\prime \vec r)~~,~~
G_{ij}(\vec r)\to{\lambda^\prime}^2G_{ij}(\lambda^\prime \vec r)~, 
\end{equation}
We then impose the requirement that the domain of space integration be scale-invariant
(which, in view of Derrick's theorem~\cite{derrick} would also imply that the energy density, 
not only the energy, would be invariant in the vicinity of $\lambda^\prime \simeq 1$). 
This is guaranteed if and only if the short-distance cut-off scales as:
\begin{equation} \label{cutoffrescaled}
\epsilon\to\epsilon/\lambda^\prime~.
\end{equation}
In this case, 
\begin{equation} \label{KBlambda}
K_B\to \tilde K_B=\frac{\lambda^{\prime 4}}{4}\int_{\epsilon/\lambda^\prime}^\infty r^2dr \int \sin\theta \, d\theta \, d\phi ~G_{ij}^2(\lambda^\prime \vec r)~,
\end{equation}
which, by means of a change of variable, leads to 
\begin{equation}
\tilde K_B=\frac{\lambda^\prime}{4}\int_\epsilon^\infty r^2dr \int \sin\theta \, d\theta \, d\phi ~G_{ij}^2(\vec r) 
=\lambda^\prime K_B~.
\end{equation}
The other three contributions $K_A,K_\phi,V_\phi$ are finite and thus cut-off independent, 
and therefore are trivially rescaled as
\begin{equation}
K_A\to\tilde K_A=\lambda^\prime K_A~~,~~K_\phi\to\tilde K_\phi=({\lambda^\prime})^{-1}K_\phi~~,~~V_\phi\to\tilde V_\phi=({\lambda^\prime})^{-3}V_\phi~,
\end{equation}
so that the energy (\ref{energy}) is rescaled as
\begin{equation}
E\to\tilde E=\lambda^\prime K_A+\lambda^\prime K_B+({\lambda^\prime})^{-1}K_\phi+({\lambda^\prime})^{-3}V_\phi~.
\end{equation}
Derrick's scaling argument~\cite{derrick} consists then in imposing that the energy
$\tilde E$ be locally invariant in the vicinity of $\lambda^\prime =1$:
\begin{equation}
\left.\frac{\partial \tilde E}{\partial\lambda^\prime}\right|_{\lambda^\prime=1}=0~,
\end{equation}
which leads to the relation found in \cite{CHO-KIM-YOON} 
\begin{equation}\label{condition}
K_A+K_B=K_\phi+3V_\phi~.
\end{equation}
This relation allows for the quantity $K_B$ to be expressed in terms of the finite quantities
$K_A,K_\phi,V_\phi$, in a manner independent of its  regularization, 
and leads to an estimate of a few TeV for the monopole mass, 
following the analysis in Ref.~\cite{CHO-KIM-YOON}. 
After regularization the  initially divergent quantity $K_B$ becomes dependent on the cutoff energy
itself. Thus, since the $K_B$ is expressed in terms of three other finite quantities which are estimated 
to be of order TeV, then this implies that the cutoff must also be of order TeV.
How this regularization can be achieved microscopically is being actively investigated.

If one accepts the above scaling arguments,
the total energy of the Cho-Maison soliton depends on quantities that can be calculated using SM  data,
such as the weak mixing  angle, the $W$-boson mass and the mass of the Higgs-like boson, $m_H \simeq 125$~GeV, 
discovered at the LHC in 2012~\cite{Higgs-like, Higgs-like1}.
Using these and other well-established parameters, the total energy of the EW monopole
was estimated in~\cite{CHO-KIM-YOON} to be $E \sim 0.35 \cdot M_W/\alpha_{em} \simeq 3.85$~TeV,
where $\alpha_{em}$ is the electromagnetic fine structure constant. 
Thus,  the mass of the monopole depends on the mechanism of spontaneous symmetry
breaking that gives masses to the weak gauge bosons. 
The non-perturbative nature of the monopole is clearly identified by the inverse relationship between its mass 
and  the electromagnetic fine structure constant (coupling). 

We note, however, that this estimate does not include quantum corrections, which
lead to renormalization of the couplings of the SM, requiring counterterms in the Lagrangian
to take into account such corrections. The SM is a perturbatively  renormalizable theory and this can be done,
but only order by order in the couplings. However, monopoles are non-perturbative solutions of the theory,
where all quantum corrections must be resummed. Since the complete set of  quantum corrections is
not known, their effect on the energy of the EW monopole is also unknown.
  

Other theoretical arguments have been used~\cite{CHO-KIM-YOON,Cho2014} to provide
alternative estimates of the EW monopole mass.  These involve  adding higher dimensional operators to the 
SM Lagrangian.  Additional operators in the Lagrangian result in additional terms
 in the field equations and there are a few scenarios where the new equations have regular solutions 
 with masses in the 1-10 TeV range.  Whether the additional operators that are needed for this are allowed by 
 SM phenomenology,  particularly the constraints imposed by precision electroweak measurements, is a question 
 that is yet to be studied in detail.

\subsubsection{Singularity resolution within string/brane  theory} 
 
 There is another possibility for singularity resolution which occurs in bottom up string theory  constructions of SM-like gauge
 theories \cite{Buican:2006sn}.  These are found by engineering configurations of D-branes whose low energy degrees of freedom
  are those of a gauge field theory resembling the standard model.    Verlinde has argued that, in the context of these constructions,  
string compactifications with D-branes may exhibit regular magnetic monopole solutions \cite{Verlinde:2006bc}.  These solutions 
have the novel aspect that their  presence does not rely on broken non-abelian gauge symmetry.   Moreover, these stringy monopoles 
exist on interesting  metastable brane configurations, such as anti-D3 branes inside a flux compactification or D5-branes wrapping
 2-cycles that are locally stable but  globally trivial. Further to this, Verlinde finds that, in brane realizations of SM-like gauge theories, 
 the monopoles carry one unit of magnetic  hypercharge. He argues that their mass can range from the string scale down to the 
 multi-TeV regime and give some arguments about how they can be light even in the 1-10 TeV range.


Independent of the above detailed arguments on the mass of the EW monopole,
there is a simple qualitative argument that is consistent with the above estimate. Roughly speaking, the mass of an
EW monopole should receive a contribution from the same mechanism that generates the mass of the weak bosons,
except that the coupling is given by the monopole charge. This means that the monopole mass should be of the
order of $M_W/\alpha_{em}\simeq 10$~TeV, where $\alpha_{em}$ is the electromagnetic fine structure constant.

Thus, the LHC could be the first collider to produce EW monopoles. However, the 
monopole-antimonopole pair production rate at LHC is quite uncertain, and currently the subject of
further study. If the production rate were to exceed that
for $WW$ production above the threshold energy, the
MoEDAL experiment should easily be able to detect the EW monopole, if it is kinematically accessible.

\subsubsection{Electroweak strings} 

In addition to the electroweak monopoles discussed in the previous Section, there are other defect solutions commonly known as electroweak strings. We recall that a topological defect is classified by the vacuum manifold, denoted by $M$, and homotopy groups. Topological monopoles can exist only if the homotopy group $\pi_{2}\left(M\right)\neq Identity$. It has more recently been emphasized that, for certain parameter ranges, a simply-connected $M$ does not necessarily imply the absence of defects that are stable when the model has both global and gauge symmetries~\cite{AV}. Within the context of the SM, Nambu~\cite{N} realized this a long time ago and made the interesting suggestion that such electroweak strings should, for energetic reasons, terminate in a (Nambu) monopole and anti-monopole at either end. Certainly, the monopole and antimonopole would tend to annihilate, whilst rotation
would obstruct longitudinal collapse." An example is the $Z$-string that carries the flux of the $Z$ boson.

Furthermore, Nambu estimated that such dumbbell-like configurations could have masses in the TeV range. Given
the energy range of the LHC this early suggestion can take on new interest. Independently, this suggestion has resurfaced at a theoretical level through primarily the work of Achucarro and Vachaspati~\cite{AV} who coined the phrase ``semilocal strings'' for such non-topological strings. The defects they consider are stable, not for topological reasons, but due to the interactions of their constituents, the scalar and gauge fields. We consider such defects if they are of finite energy.

The Nambu monopole is estimated to have a mass, $M_{N}$, whose order of magnitude
is given by~\cite{N} 
\begin{equation}\label{mN}
M_N \simeq \dfrac{4\pi}{3e}\sin^{5/2}\theta_{W}\sqrt{\dfrac{m_{H}}{m_{W}}}\mu ,
\end{equation}
where $\mu=m_{W}/g$, $g$ is the $SU\left(2\right)$ gauge coupling, $m_{W}$ is the $W$ boson mass, and $m_{H}$ is the Higgs mass. From the recent experimental results on the Higgs boson mass, we obtain $M_{N}\simeq 689$~GeV. Just like isolated monopoles, the Nambu monopole satisfies the Dirac quantization condition. The dumbbell structures can rotate and so emit electromagnetic radiation. The lifetime of such dumbbells is greater than the non-spinning variety, and might be long enough to be observed at energies greater than $7$~TeV, but instabilities of the connecting $Z-$string may decrease the lifetime~\cite{JPV}.

\subsection{Vacuum decay and light 't Hooft-Polyakov monopoles}

When 't Hooft~\cite{tHooft1974qc} and Polyakov~\cite{Polyakov1974ek} independently discovered that the $SO(3)$ Georgi-Glashow model~\cite{Georgi1972cj} inevitably contains monopole solutions, they further realized that any model of unification with an electromagnetic $U(1)$ subgroup embedded into a semi-simple gauge group that is spontaneously broken by the Higgs mechanism possesses monopole solutions. This mechanism  leads, however, to a mass proportional to the vector meson mass arising from the spontaneous broken symmetry: $\sim M_W/\alpha_{em}$, $\alpha_{em}$ being the fine structure constant at the breaking scale, i.e., the GUT scale.

\begin{figure}[htb]
\begin{center}
\includegraphics[width=0.6\textwidth]{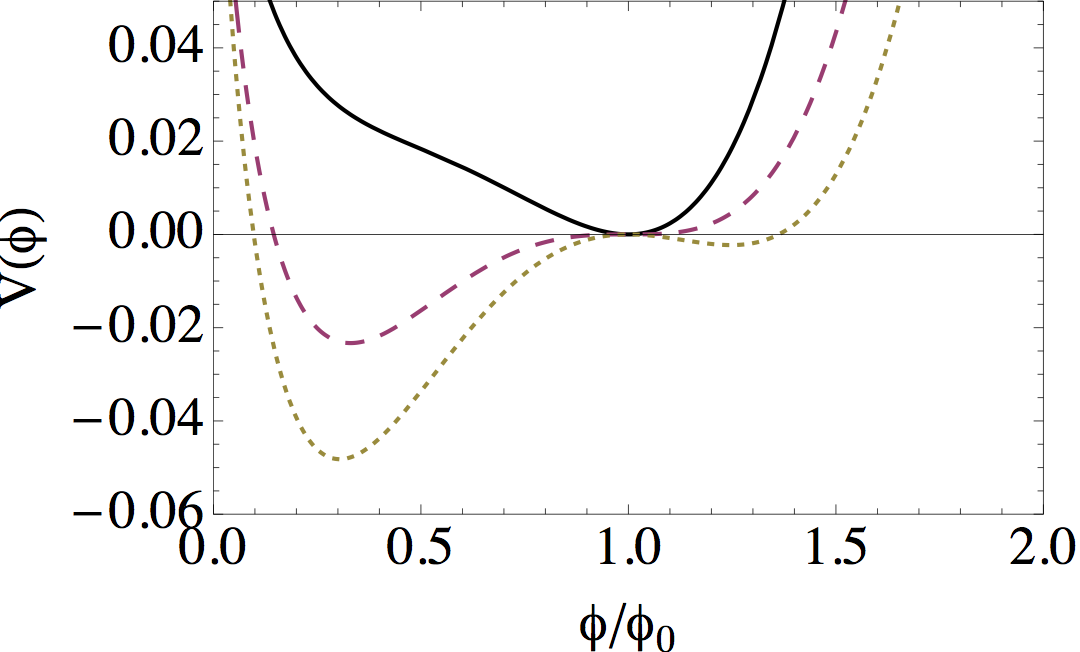}
\caption{The curves show the effective potential for $\epsilon=1$ and $\mu= - 0.3$ (solid), $- 0.5 $(dashed),$- 0.7$ (dotted).}
\label{potential} 
\end{center}
\end{figure}

Vacuum metastabilities may exist at GUT scales, and appear for particular choices of the effective  potential. The Higgs minimum remains, but a  second lower minimum appears which allows a vacuum decay. This type of modification, see Fig.~\ref{potential}, can lead to a smaller monopole mass. 


%

The energy density is defined as 
\begin{equation}
E(\epsilon,\mu)= \dfrac{M_W}{\alpha_{em}}  f(\epsilon, \mu),
\end{equation}
where the function $f(\epsilon,\mu)$ characterizes the dynamics. For convenience  we will fix $\epsilon =1$  and study the variation of $f$ with $\mu$ for fixed $\epsilon$ \cite{Vento2013jua}. The qualitative features of our analysis extend to any $\epsilon$.

\begin{figure}[htb]
\begin{center}
\includegraphics[width=0.6\textwidth]{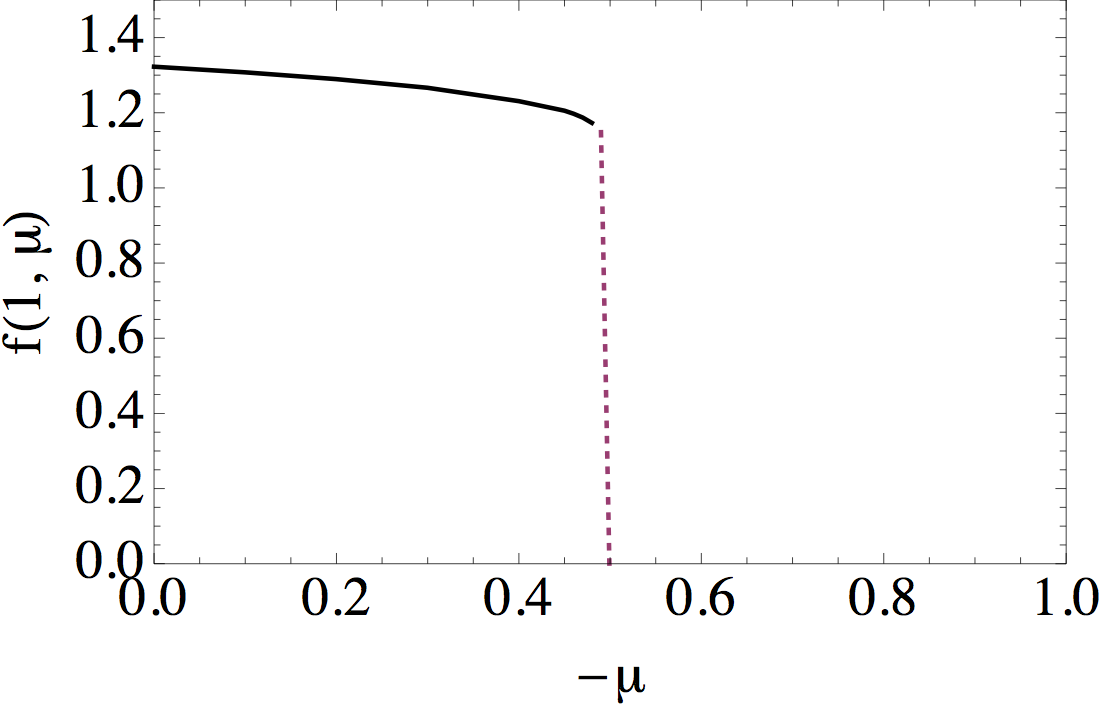}
\caption{The function $f(1,\mu$) for the potential in \protect Fig. \ref{potential} as a function of $\mu$. The function has a singularity at $\mu = -0.5$. For this value  the minimum transforms into an inflection point.}
\label{massmu}
\end{center}
\end{figure}

The result of the calculation for the potential density in Fig. \ref{potential} with different values of $\mu$ is shown in Fig.~\ref{massmu}. It is apparent that the region of interest is around $\mu \sim -0.5$, and we see that the potential in Fig. \ref{potential} has an inflection point at $ \phi/\phi_0 = 1$ for $\mu = -0.5$. For  values of $ -0.5 < \mu < 0$  the potential has a minimum at $\phi/\phi_0 = 1$ and is bounded for large fields. For $\mu < -0.5$ there exist two minima, one for  $\phi/\phi_0 <1$ and one for  $\phi/\phi_0 > 1$.  The solution then is ill-defined, i.e.\ the monopole solution is absent~\cite{Vento2013jua}. 


\begin{figure}[htb]
\begin{center}
\includegraphics[width=0.6\textwidth]{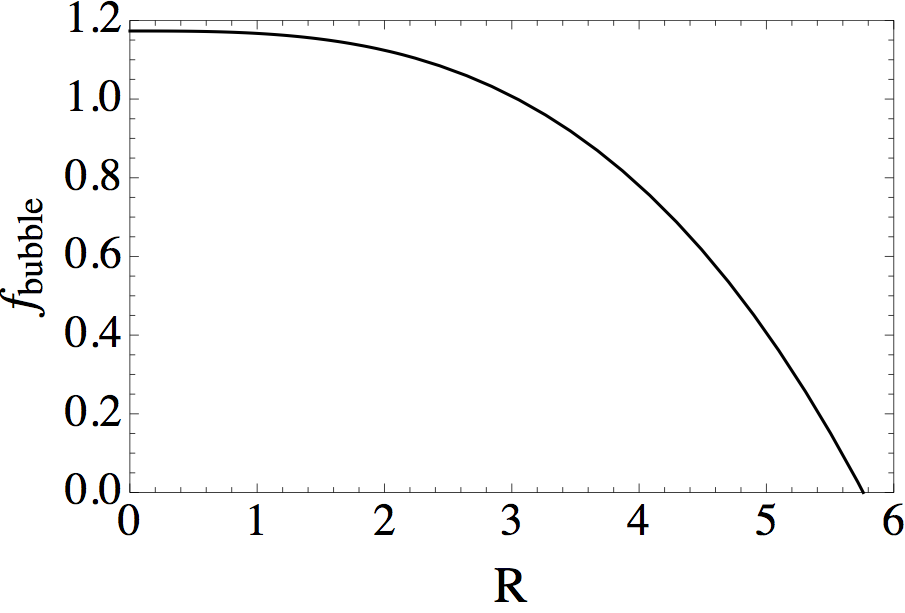} 
\caption{We plot the monopole mass function $f$ as a function of bubble radius $R$ for $\epsilon = 1$ and $\mu =0.48$.}
\label{fbubble}
\end{center}
\end{figure}

The region of interest is therefore $\mu > -0.5$, where the potential modification leads to a reduction of the energy density producing a slightly smaller mass, but most importantly, the Higgs minimum remains, and a second lower minimum appears which allows a quantum vacuum decay. The original (false) Higgs vacuum decays into a new (true) vacuum of lower energy. This decay is by bubble formation and we envisage a scenario in which small bubbles of true vacuum containing a monopole surrounded by larger ones of false vacuum, represent a decaying monopole with effective masses ranging from its GUT mass to zero (see Fig.~\ref{fbubble}). A new scale enters the description, the size of the bubble, which can easily compensate for the GUT scale. 


\subsection{Monopolium}

In the absence of uncontroversial evidence for the existence of  magnetic monopoles, most assume that, if magnetic monopoles do exist, their mass is too great and/or their abundance is too small for them to be detected in cosmic rays or at  existing accelerators. However, Dirac proposed an alternate explanation why monopoles have not been conclusively observed so far~\cite{Diracs_idea,khlopov}. His idea was that monopoles are not seen freely because they are confined by their strong magnetic forces in monopole - anti-monopole bound states called monopolium~\cite{Monopolium, Monopolium1}.


Some researchers have proposed that monopolium, due to its bound-state structure, might be easier to detect than free monopoles~\cite{Epele,Epelea}, and the possibility that the LHC might be able to discover monopolium has been advocated in~\cite{Epele,Epelea,Epele1}. Monopolium is a neutral state, and is therefore difficult to detect directly in a collider detector, although its decay into two photons would give a very nice signal for the ATLAS and CMS detectors \cite{Epele1} that is not visible in the MoEDAL detector. We discuss here possible scenarios in which monopolium could be seen by the MoEDAL experiment.
 
\begin{figure}[htb]
\begin{center}
\includegraphics[width=12pc]{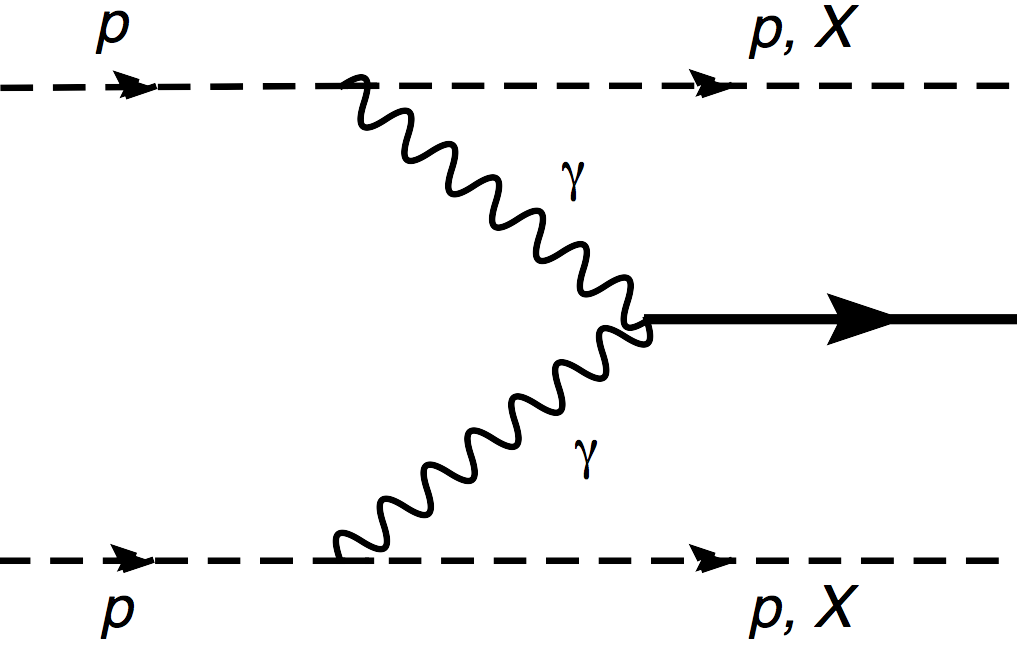}
\caption{Diagrammatic description of monopolium production at the LHC}
\label{Fig:Monopolium-prod}
\end{center}
\end{figure}

The production of monopolium at LHC, namely:
\begin{equation}
p + p \rightarrow p(X) + p(X)  + M
\end{equation}
would be expected to occur predominantly via photon fusion, as shown in Fig.~\ref{Fig:Monopolium-prod}, where $p$ represents the proton, $\gamma$  the photon,  $X$ an unknown final state and $M$ the monopolium. This diagram  summarizes the three possible processes:
\begin{itemize}
\item [i)] inelastic $p+ p \rightarrow X+X +(\gamma \gamma)
\rightarrow X + X + M$
\item [ii)] semi-elastic $ p + p \rightarrow p + X + (\gamma \gamma)
\rightarrow p + X + M$
\item [iii)] elastic $p + p \rightarrow p + p + (\gamma \gamma)
\rightarrow p + p + M$.
\end{itemize}

%
In inelastic scattering, both intermediate photons are radiated from partons (quarks or antiquarks) in the colliding protons. In the semi-elastic scattering case one intermediate photon is radiated by a quark (or antiquark), as in the inelastic process, while the second photon is radiated coherently from the other proton, coupling to the total proton charge and leaving a final-state proton intact. In the elastic scattering case, both intermediate photons are radiated from the interacting protons leaving both protons intact in the final state. The full $\gamma\gamma$ calculation includes contributions from these three individual regimes.

\begin{figure}[htb]
\begin{center}
\includegraphics[width=12pc]{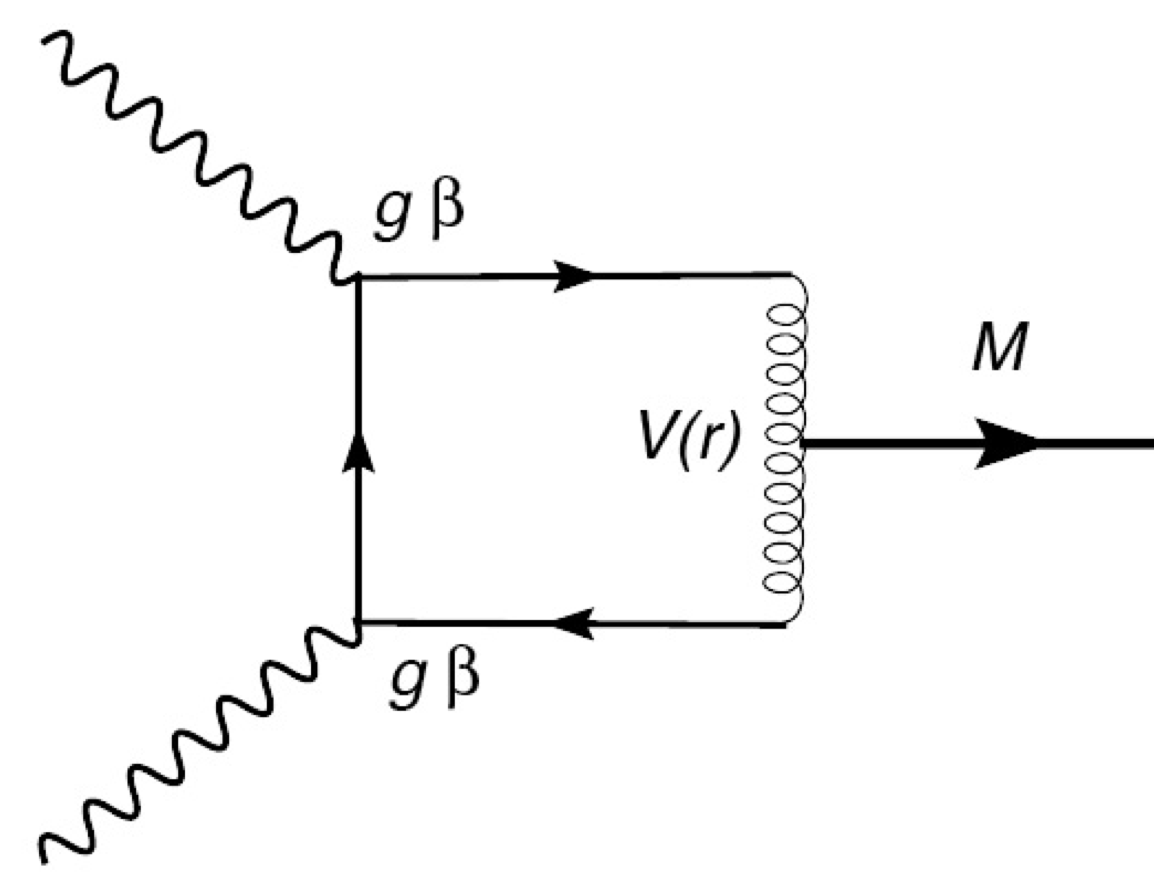}
\caption{Diagrammatic description of the elementary subprocess for monopolium production from photon
fusion, where $V(r)$ represents the interaction binding the monopole-antimonopole pair to form monopolium.}
\label{Fig:Monopolium}
\end{center}
\end{figure}

 
For the case of monopole interactions at energies higher than their mass there is no universally accepted effective field theory~\cite{Schwinger1966nj,Zwanziger,Gamberg}. 
We will employ a minimal model of monopole interaction which assumes an effective monopole-photon coupling that is proportional to
$g\beta$ for a monopole moving with velocity $\beta$~\cite{Mulhearn,Epele1,Dougall,Kalbfleisch}. The elementary subprocess calculated is shown in Fig.~\ref{Fig:Monopolium}. Since the Dirac quantization condition does not specify the spin of the monopoles, we choose here monopoles of spin $1/2$ coupled into monopolium of spin $0$, in order to have an $s$-wave radial structure with  minimal energy.

The standard expression for the cross section of the elementary subprocess for producing monopolium of mass $M$  and width $\Gamma_{M}$ is given by:
\begin{equation}
\sigma (2 \gamma \rightarrow M) = \dfrac{4\pi}{E^2}  \dfrac{M ^2
\,\Gamma (E) \, \Gamma_M}{\left(E^2 - M^2\right)^2 +
M^2\,\Gamma_M^2}, \label{ppM}
\end{equation}
where $\Gamma(E)$,  with $E$ off mass shell, describes the production cross section.
In the small binding limit the width $\Gamma(E)$ is proportional to $\beta^{4}$ and therefore monopolium can be very long lived
close to threshold \cite{Georgi1972cj} \cite{Vento2013jua} since
 $\Gamma_{M}$ arises from the softening of the delta function, $\delta(E^{2} - M^{2})$ and is therefore, in principle, independent of the production rate $\Gamma(E)$  and can be attributed to the beam width~\cite{Jauch,Peskin}.

\begin{figure}[htb]
\begin{center}
\includegraphics[width=0.6\textwidth]{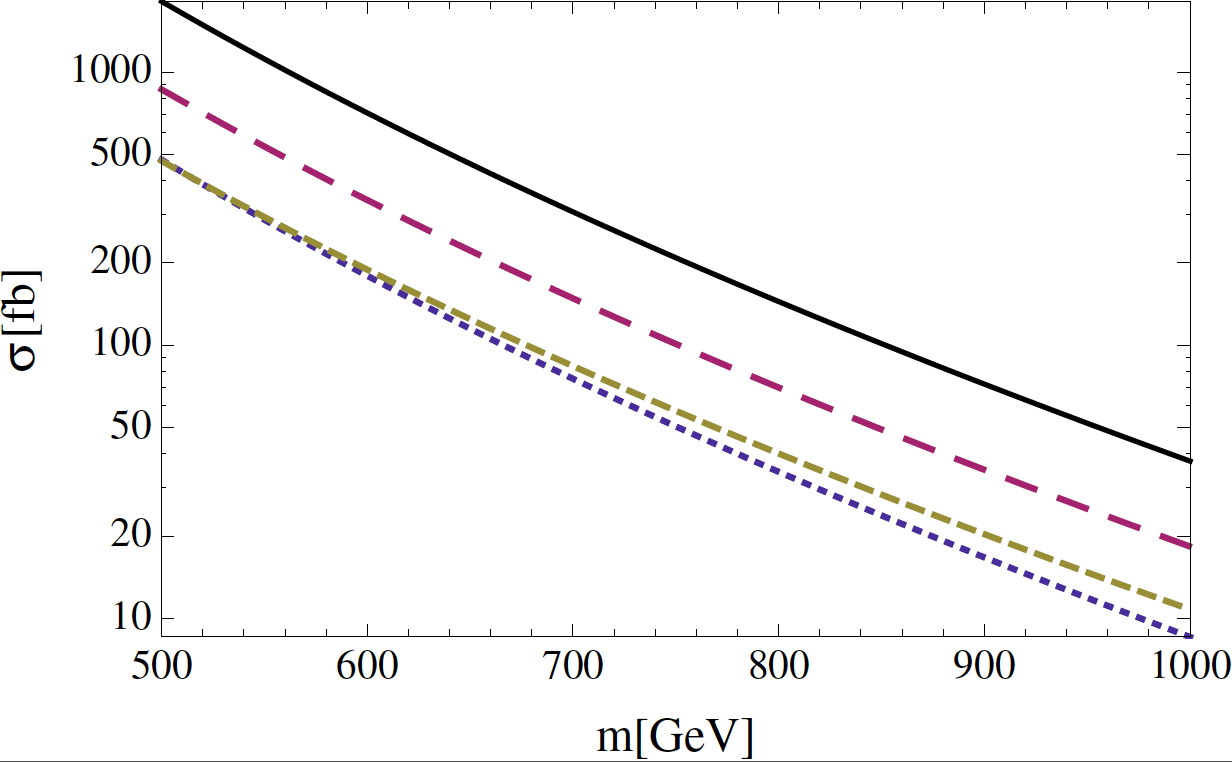}
\caption{Total cross section for monopolium production at LHC with $7$~TeV beams for monopole masses ranging from $500$ to $1000$~GeV (full curve). The broken curves represent the different contributions to the total cross section as described in the text: semielastic (dashed), elastic (shorst dashed) and inelastic (dotted). We have chosen a binding energy $\sim 2\;m/15$ and  $\Gamma_M = 10$ GeV.}
\label{Fig:Monopolium-production}
\end{center}
\end{figure}

In Fig.~\ref{Fig:Monopolium-production} we show the total cross section for monopolium production from photon fusion under future LHC running conditions, i.e.\ a center-of-mass energy of 14~TeV, for a monopole mass ($m$)  ranging from $500$ to $1000$~GeV. In the figure the binding energy is fixed for each mass ($2\;m/15$), chosen so that for our case study, $m=750$ GeV, the binding energy is $100$ GeV and thus $M=1400$ GeV. With this choice the monopolium mass ($M$) ranges from $933$ to $1866$ GeV. We note that detection with present integrated luminosities is possible even for binding energies below $10\%$ of its mass.

Monopolium in its ground state is a very heavy neutral object, and thus the only property suitable to be detected is its heavy mass via collisions with the large molecules of the detector. This mechanism is not very effective. However, the monopolium ground state has  a very large magnetic polarizability $ d \sim  r_M^3 B  \sim (\alpha E_{binding})^{-3}  B$, where $r_M$ is the monopolium size. The presence of large magnetic fields might provide the monopolium ground state with sufficient magnetic strength to be able to ionize the detector medium. However, in the vicinity of the MoEDAL detector there are only  very weak stray fields from the LHCb dipole magnet. 

\begin{figure}[htb]
\centering
\subfigure[$M$ in excited state]{ 
\includegraphics[width=0.3\textwidth]{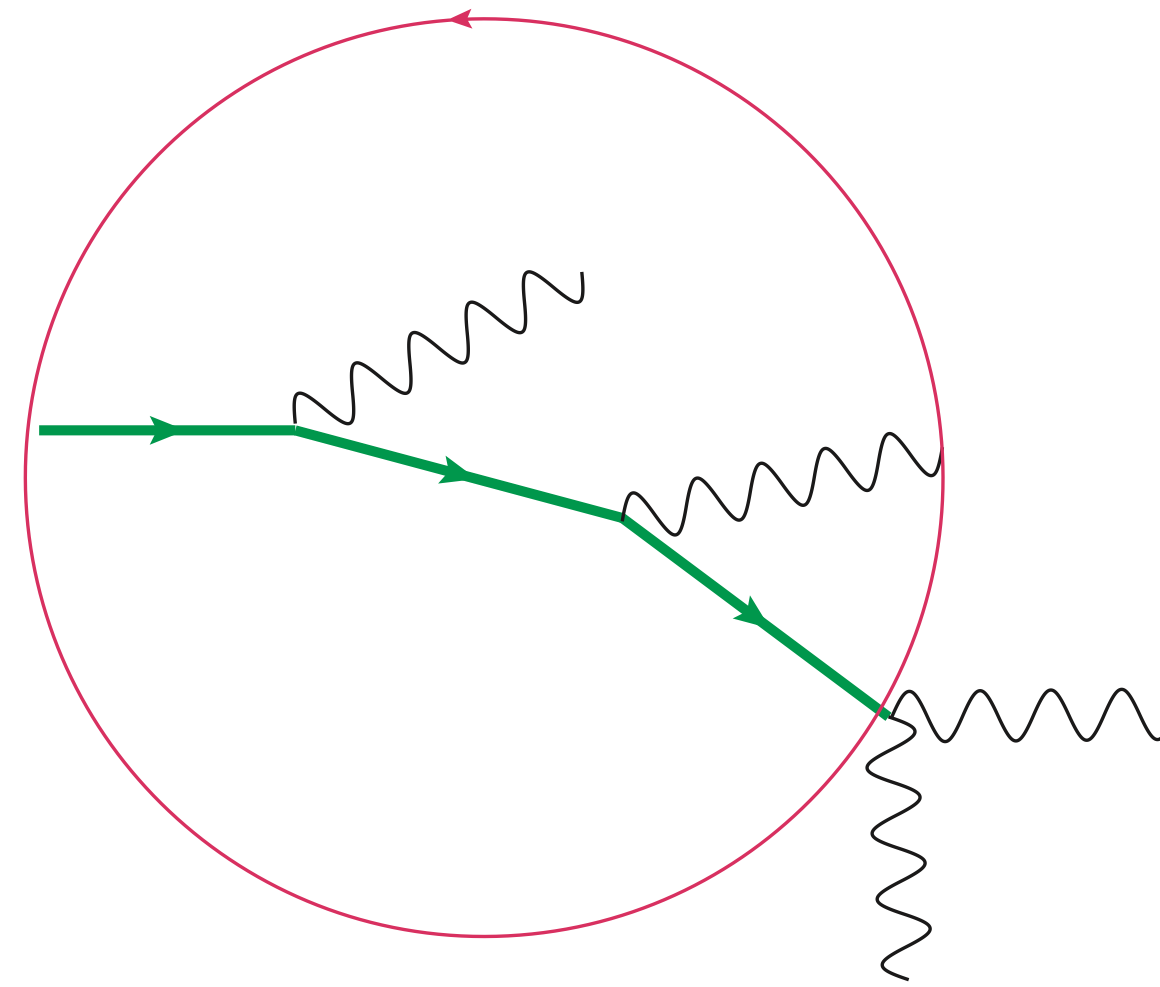}
\label{sfig:excited}
}
\quad
\subfigure[Dyon production]{
\includegraphics[width=0.35\textwidth]{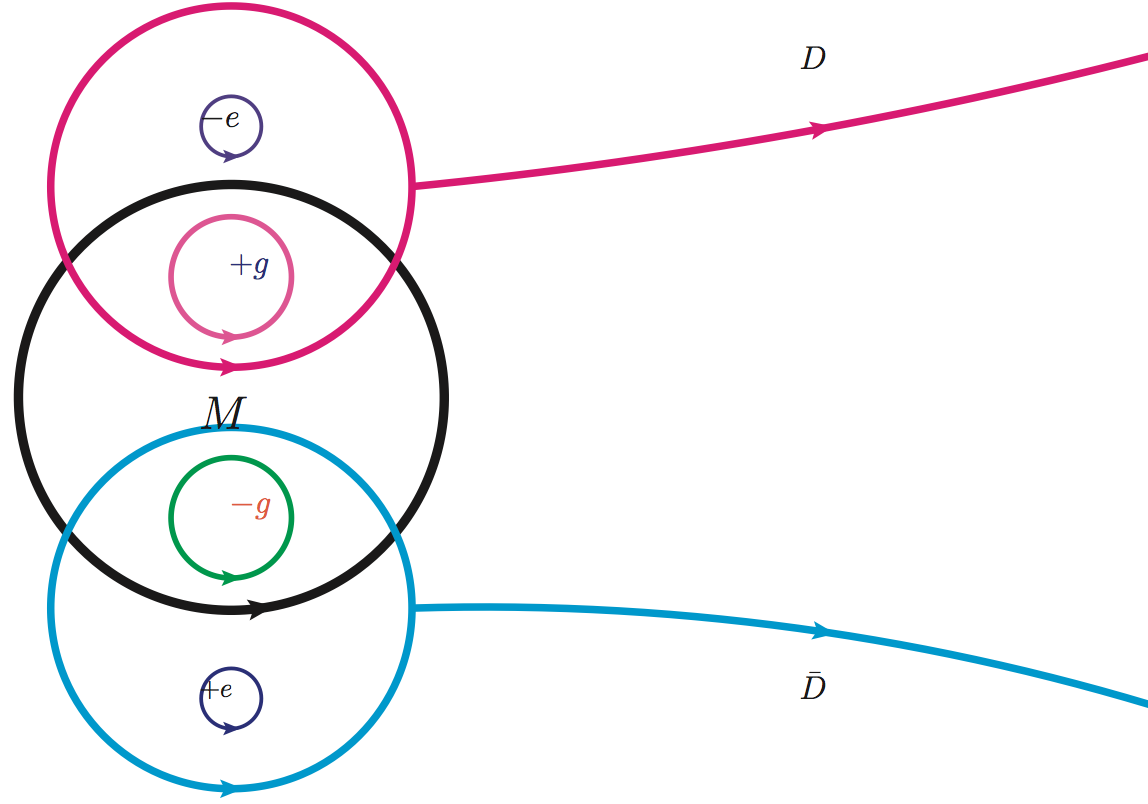}
\label{sfig:dyon-prod}
}
\quad
\subfigure[Di-dyon signature]{
\includegraphics[width=0.22\textwidth]{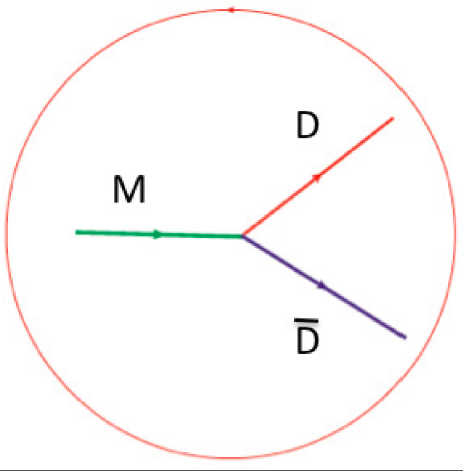}
\label{sfig:dyon-signal}
}
\caption{Monopolium may be produced in excited state that might be a magnetic multipole and thus will be highly ionizing. Its cascade decay via (undetected) photon emission will lead to a peculiar trajectory in the medium. (a) The circle encloses the expected observation consisting of a ---probably curved--- polyhedric trajectory. In another case study, a monopolium in the medium might break up into highly-ionizing dyons (b) producing a very clear di-dyon signal (c).}
\label{Fig:monopolium-dyon}
\end{figure}

On the other hand, the binding of monopole-antimonopole pairs might occur in excited states. If those states have angular momentum they will be magnetic multipoles that will be strongly-ionizing and thus directly detectable. Moreover, their decay into lower-lying multipole states will show a peculiar structure of the trajectory in the detector medium (see Fig.~\ref{sfig:excited}) which would be easy to isolate from other background trajectories.
If the lifetime is governed by the the $\beta$-scheme the lifetime of these states will be long enough to be detectable by MoEDAL.

Finally, if monopolium is weakly bound or produced in a highly-excited bound state, the presence of electrons and protons in the medium might allow for the formations of dyons (see Fig.~\ref{sfig:dyon-prod}). Dyons are highly-ionizing particles and therefore MoEDAL will detect a very clear signal (see Fig.~\ref{sfig:dyon-signal}).

\subsection{Summary of accelerator experiments}

If the monopole mass $M$ is less than half of the centre-of-mass energy in a particle collision, it is kinematically possible to produce a monopole-antimonopole pair. In particular, production of Cho-Maison monopoles with mass $4-7~{\rm TeV}$ should therefore be possible at the LHC~\cite{chopin}.

Searches for magnetic monopoles have been carried out with all major accelerators. Because there is significant uncertainty about the monopole pair-production cross-section, the results are generally expressed as upper bounds on the cross-section $\sigma$, rather than as lower bounds on the mass. Even these bounds  depend on assumptions about the kinematic distributions of the produced particles, which is usually assumed to be the same as for the Drell-Yan pair-production process~\cite{Kalbfleisch2003yt}. 

During the last twenty years collider searches for magnetic monopoles particles have been made with  dedicated  MODAL \cite{MODAL} and  OPAL monopole detector \cite{OPALDEDICATED}  at LEP \cite{LEP}.  The best LEP limits for magnetic monopole pair production, where the monopoles have single Dirac charge, are as follows. The OPAL experiment at LEP gave an upper bound $\sigma<0.05~{\rm pb}$ for the mass range $45~{\rm GeV}<M<102~{\rm GeV}$~\cite{Abbiendi2007ab} in electron-positron collisions. The CDF experiment  at the Tevatron \cite{CDF} yielded the limits  $\sigma<0.2~{\rm pb}$ for $200~{\rm GeV}<M<700~{\rm GeV}$ in proton-antiproton collisions.  Rather than identifying the highly ionizing nature of the monopole the H1 experiment at HERA sought monopoles trapped in the HERA beam pipe. Upper limits on the monopole pair production cross section have been set for monopoles with magnetic charges from 1 to 6g$_{D}$ or more and up to a mass of 140 GeV  \cite{monopole-range}.  Most recently, ATLAS placed an upper bound $\sigma <1 6$ - 145~${\rm fb}$ for masses $200~{\rm GeV}<M<1200~{\rm GeV}$ in proton-proton collisions~\cite{Aad2012qi}.

Turning these results into mass constraints is difficult because the large  magnetic charge makes high-order loop contributions increasingly large and perturbation theory therefore inapplicable. 
However, for 't~Hooft-Polyakov monopoles one can calculate the monopole-pair creation cross section semiclassically, at least in principle. 
 
Because the monopoles are large non-perturbative objects which carry a large number $n\sim 1/\alpha$ of quanta, it has been argued~\cite{Drukier1981fq,Dobbins1993vc} that their production cross section would be suppressed by an exponential factor $\exp(-2/\alpha)$, and numerical calculations for kinks in $(1+1)$-dimensional scalar theory support this finding~{\cite{Demidov2011dk,Demidov2011eu}. This would mean that monopoles would practically never be produced in particle collisions, even if they are kinematically allowed. However, these semiclassical calculations are only valid for monopoles at weak coupling and do not apply to other types, such as TeV-scale Cho-Maison monopoles.

\section{Electrically-Charged Massive (Meta-)Stable Particles in Supersymmetric Scenarios}

As has been discussed above,  massive slowly moving ($\beta \lesssim 5$)  electrically charged particles are  potential  highly ionizing avatars of new physics. If they  are sufficiently long-lived to  travel  a distance of at least $O$(1)m  before decaying and their $\Z/\beta  \gtrsim  0.5$,  then they will be detected in the MoEDAL NTDs.  Supersymmetric scenarios provide a number of candidate particles  that satisfy these criteria. 

In supersymmetric theories, all the known particles are accompanied by supersymmetric partners (spartners) with spins that differ by half a unit, e.g., leptons and sleptons $\tilde \ell$, quarks and squarks $\tilde q$, the photon and photino $\gamma$, gluons and gluinos $\tilde g$~\cite{SUSY}. No highly-charged particles are expected in such a theory, but there are several scenarios in which supersymmetry may yield massive, long-lived particles that could have electric charges $\pm 1$, potentially detectable in MoEDAL if they are produced with low velocities.

In the minimal supersymmetric extension of the Standard Model (MSSM)~\cite{MSSM}, the gauge couplings $g$ are the same as in the Standard Model, and there are Yukawa interactions $\lambda$ related to quark and lepton masses. The sensitivities to the production of different sparticle species vary as functions of the LHC center-of-mass energy in the manner shown in  Fig.~\ref{fig:susy1}. We see, for example, that the sensitivity to direct production of a pair of stop squarks $\tilde t$ extends to 800~GeV with 100~fb$^{-1}$ ($= 10^5$~pb$^{-1}$) of data at 14~TeV~\cite{Raklev:2009mg}.

\begin{figure}[h!]
\centering
\includegraphics[scale=0.4]{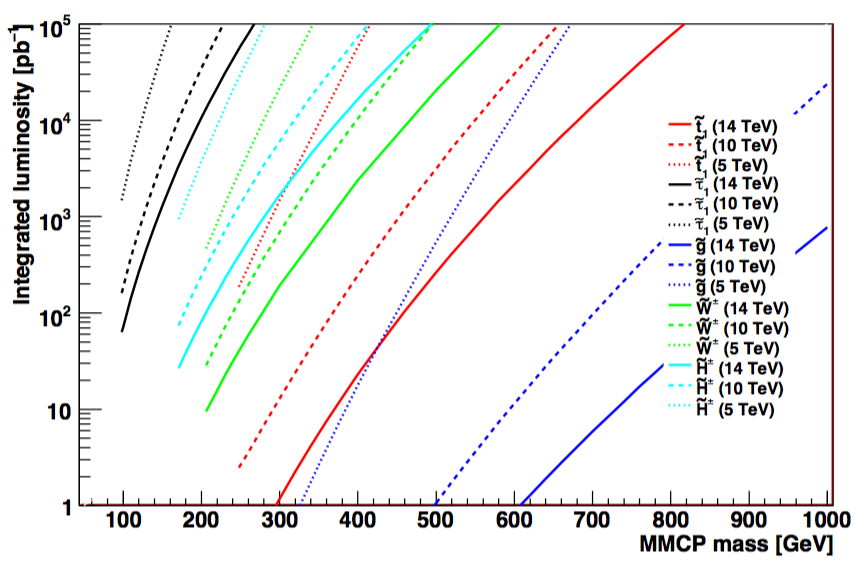}
\caption{The sensitivity to the direct production of different sparticle species at the LHC at various center-of-mass energies up to 14~TeV~\protect\cite{Raklev:2009mg}}
\label{fig:susy1}
\end{figure}

One complication is that in supersymmetric models there must be two Higgs doublets, one responsible for the masses of charge $2/3$ quarks and the other for the masses of charge $-1/3$ quarks and charged leptons. There is a supersymmetric coupling between the two Higgs doublets, called $\mu$, and unknown ratio of their vacuum expectation values, called $\tan\beta$. In addition, there are a multitude of unknown supersymmetry-breaking parameters, including scalar masses $m_0$, gaugino masses $m_{1/2}$, trilinear soft couplings $A_\lambda$, and a bilinear soft coupling $B_\mu$. For simplicity, and motivated by the agreement of rare and flavor-changing processes with Standard Model predictions, it is often assumed that these parameters are universal, i.e. there is a single $m_0$, a single $m_{1/2}$, and a single $A_\lambda$ at some high input renormalization scale. This scenario is called the constrained MSSM (CMSSM)~\cite{CMSSM, CMSSM1,CMSSM2,CMSSM3,CMSSM4,CMSSM5,CMSSM6,CMSSM7}. Later we consider variants in which the soft supersymmetry-breaking contributions to the Higgs masses are allowed to be non-universal (the NUHM)~\cite{NUHM}.

The lightest supersymmetric particle (LSP) is stable in many models because of conservation of $R$~parity, where$R \equiv (-1)^{2S - L+3B}$, where $S$ is spin, $L$ is lepton number, and $B$ is baryon number~\cite{Fayet}. It is easy to check that particles have $R = +1$ and sparticles would have $R = -1$. Hence, sparticles would be produced in pairs, heavier sparticles would decay into lighter sparticles, and the LSP would be stable because it has no allowed decay mode, and hence present in the Universe today as a relic from the Big Bang~\cite{EHNOS}. The LSP should have no strong or electromagnetic interactions, for otherwise it would bind to conventional matter and be detectable in anomalous heavy nuclei~\cite{EHNOS}. Possible weakly-interacting neutral scandidates in the MSSM include the sneutrino, which has been excluded by LEP and direct searches, the lightest neutralino $\chi$ (a mixture of spartners of the $Z, H$ and $\gamma$) and the gravitino, which would be a nightmare for astrophysical detection, but not in conflict with any experimental limits.

On the basis of the above discussion, we can identify several scenarios featuring metastable charged sparticles that might be detectable in MoEDAL. One such scenario is that $R$~parity {\it may not be exact}~\cite{RV}, in which case the LSP would be an unstable sparticle, and might be charged and/or coloured. In the former case, it might be detectable directly at the LHC as a massive slowly-moving charged particle. In the latter case, the LSP would bind with light quarks and/or gluons to make colour-singlet states, and any charged state could again be detectable as a massive slowly-moving charged particle. 

However, even if $R$~parity {\it is} exact, the next-to-lightest sparticle (NLSP) may be long-lived. This would occur, for example, if the LSP is the gravitino, or if the mass difference between the NLSP and the neutralino LSP is small, offering more scenarios for long-lived charged sparticles. The experimental signatures of $R$-violating scenarios are generically similar to those in the $R$-parity conserving scenarios that we now discuss.

In {\it neutralino dark matter} scenarios based on the CMSSM the most natural candidate for the NLSP is the lighter stau slepton ${\tilde \tau_1}$~\cite{stauNLSP}, which could be long-lived if $m_{\tilde \tau_1} - m_\chi$ is small. In {\it gravitino dark matter} scenarios with more general options for the pattern of supersymmetry breaking, other options appear quite naturally, including the lighter selectron or smuon, or a sneutrino~\cite{sleptonNLSP}, or the lighter stop squark ${\tilde t_1}$~\cite{stopNLSP}. Another possibility in models with split supersymmetry would be the gluino, whose lightest bound state might be charged. In all these cases, the NLSP would be naturally very long-lived, with a decay interaction of near-gravitational strength.

In subsequent Sections we discuss each of these scenarios in more detail. Before doing so, we make one general comment: it is a general feature of these scenarios that there is a range of sparticle lifetimes, typically ${\cal O} (10^3)$~s, where the bound-state dynamics and decays of metastable charged sparticles may serve a useful cosmological purpose, in improving the agreement of Big-Bang Nucleosynthesis calculations with measurements of the cosmological $^7$Li abundance without upsetting agreement with the measured abundances of the other light elements~\cite{SusyBBN,SusyBBN1,SusyBBN2,SusyBBN3,SusyBBN4,SusyBBN5,SusyBBN6,SusyBBN7,SusyBBN8,SusyBBN9,SusyBBN10}. This may provide additional motivation for studying in more detail scenarios for metastable charged sparticles.

\subsection{Metastable lepton NLSP in the CMSSM with a neutralino LSP}

We first consider the most constrained supersymmetric scenario, namely a stau NLSP in the CMSSM with a neutralino LSP, which is rather a natural possibility. We recall that there are several regions of the CMSSM parameter space that are compatible with the constraints imposed by unsuccessful searches for sparticles at the LHC, as well as the discovery of a Higgs boson weighing $\sim 126$~GeV~\cite{Higgs-like,Higgs-like1}. These include a strip in the focus-point region where the relic density of the LSP is brought down into the range allowed by astrophysics and cosmology because of its relatively large Higgsino component, a region where the relic density is controlled by rapid annihilation through direct-channel heavy Higgs resonances, and a strip where the relic LSP density is reduced by coannihilations with near-degenerate staus and other sleptons. It was found in a global analysis that the two latter possibilities are favored~\cite{MC8, MC8a}.

In the coannihilation region of the CMSSM, the lighter ${\tilde \tau_1}$ is expected to be the lightest slepton~\cite{stauNLSP}, and the $\tilde\tau_1-\tilde{\chi}_1^0$ mass difference may well be smaller than $m_\tau$: indeed, this is required at large LSP masses. In this case, the dominant stau decays for $m_{\tilde \tau_1} - m_{\tilde{\chi}_1^0} > 160$~MeV are expected to be into three particles: $\tilde{\chi}_1^0 \nu \pi$ or $\tilde{\chi}_1^0 \nu \rho$. If $m_{\tilde \tau_1} - m_{\tilde{\chi}_1^0} < 1.2$~GeV, the ${\tilde \tau_1}$ lifetime is calculated to be so long, in excess of $\sim 100$~ns, that it is likely to escape the detector before decaying, and hence would be detectable as a massive, slowly-moving charged particle~\cite{Sato}. The stau lifetime as a function of $m_{\tilde \tau_1} - m_{\tilde{\chi}_1^0}$ for typical supersymmetric model parameters is shown in the left panel of Fig.~\ref{fig:susy2}, and the right panel displays a typical pattern of decay branching ratios~\cite{CELMOV}. Three-body decays such as ${\tilde \tau_1} \to \tilde{\chi}_1^0 + \nu + \pi$ are important if $m_{\tilde \tau_1} - m_{\tilde{\chi}_1^0} > m_\pi$, whereas four-body decays ${\tilde \tau_1} \to \tilde{\chi}_1^0 + \nu + {\bar \nu} + e/\mu$ dominate if $m_{\tilde \tau_1} - m_{{\tilde{\chi}_1^0}} < m_\pi$.

\begin{figure}[ht]
\centering
\vskip 0.5in
\vspace*{-0.75in}
\begin{minipage}{8in}
\epsfig{file=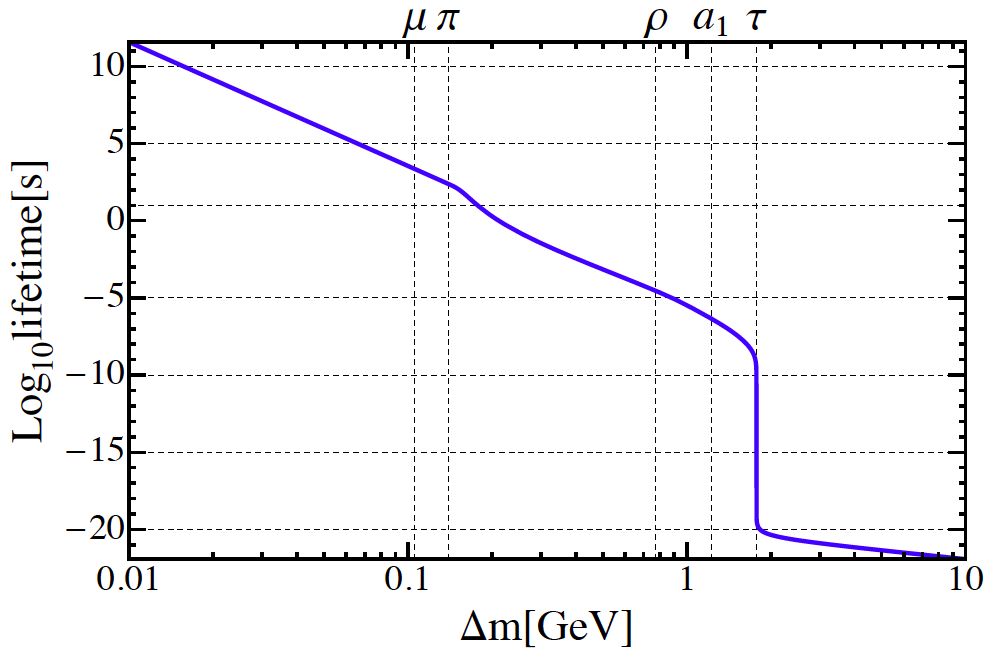,height=1.65in}
\epsfig{file=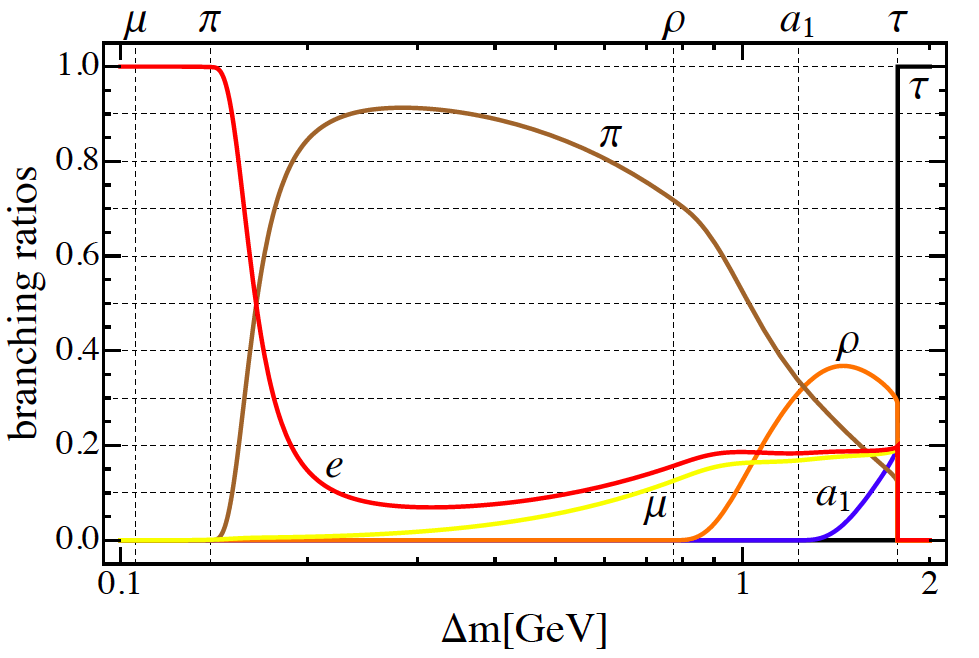,height=1.65in}
\hfill
\end{minipage}
\caption{
Left panel: The ${\tilde \tau_1}$ lifetime calculated for $m_{\tilde \tau_1} = 300$~GeV and a ${\tilde \tau_L} - {\tilde \tau_R}$ mixing angle $\theta_\tau = \pi/3$, as a function of $\Delta m \equiv m_{\tilde \tau_1} - m_{\tilde{\chi}_1^0}$ over the range $10~{\rm MeV} < \Delta m < 10~{\rm GeV}$, where the lifetime is between $\sim 10^{12}$ and $\sim 10^{-22}$~s~\protect\cite{CELMOV}.
Right panel: The principal ${\tilde \tau_1}$ branching ratios calculated for the same model parameters, as functions of $\Delta m \equiv m_{\tilde \tau_1} - m_{\tilde{\chi}_1^0}$ for $100~{\rm MeV} < \Delta m < 2~{\rm GeV}$~\protect\cite{CELMOV}. The black, blue, orange, brown, yellow, and red lines are for the final states with $\tau$, $a_1(1260)$, $\rho(770)$, $\pi$, $\mu$, and $e$, respectively, indicated by the labels adjacent to the corresponding curves. The vertical dashed lines correspond to the $\tau$, $a_1$, $\rho$, $\pi$ and $\mu$ masses, as indicated by the labels on the top of the panels.} 
\label{fig:susy2}
\end{figure}

\subsection{Metastable sleptons in gravitino LSP scenarios}

The above discussion is for the case of a neutralino LSP. If the gravitino ${\tilde G}$ is the LSP, the decay rate of a slepton NLSP is given by 
\begin{equation}
\Gamma ( {\tilde \ell} \to {\tilde G} \ell) = \dfrac{1}{48 \pi {\tilde M}^2} \dfrac{m_{\tilde \ell}^5}{M_{\tilde G}^2}
\left[ 1 - \dfrac{M_{\tilde G}^2}{m_{\tilde \ell}^2} \right]^{4},
\label{telldecay}
\end{equation}
where ${\tilde M}$ is the Planck scale. Since ${\tilde M}$ is much larger than the electroweak scale, the NLSP lifetime is naturally very long, as seen in Fig.~\ref{fig:susy3}~\cite{CAPTURE3}.

\begin{figure}[htb]
\centering
\includegraphics[scale=0.5]{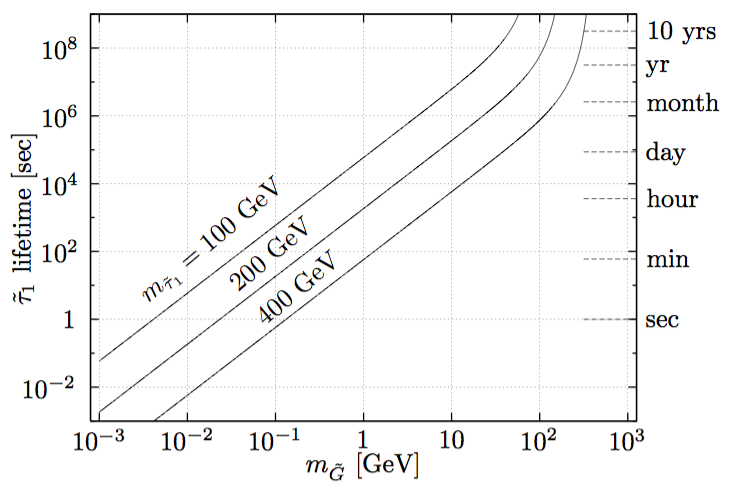}
\caption{The stau lifetime in a gravitino LSP scenario, for different values of the stau mass $m_{\tilde \tau_1}$ and the gravitino mass $m_{\tilde G}$~\cite{CAPTURE3}.}
\label{fig:susy3}
\end{figure}

There are many possibilities for the NLSP in scenarios with a gravitino LSP, some of which are illustrated in Fig.~\ref{fig:susy4}~\cite{sleptonNLSP}. This displays a $(\mu, m_A)$ plane in a variant of the MSSM with universal input squark and slepton masses $m_0 = 100$~GeV and gaugino masses $m_{1/2} = 500$~GeV, $\tan \beta = 10$, $A_0 = 0$ and non-universal soft supersymmetry-breaking contributions to the Higgs boson masses (the NUHM). We see different regions of the plane where the NLSP would be the lighter stau (coloured orange), the lighter selectron (yellow), the tau sneutrino (lighter blue) or the electron sneutrino (darker blue).

\begin{figure}[htb]
\centering
\includegraphics[scale=0.4]{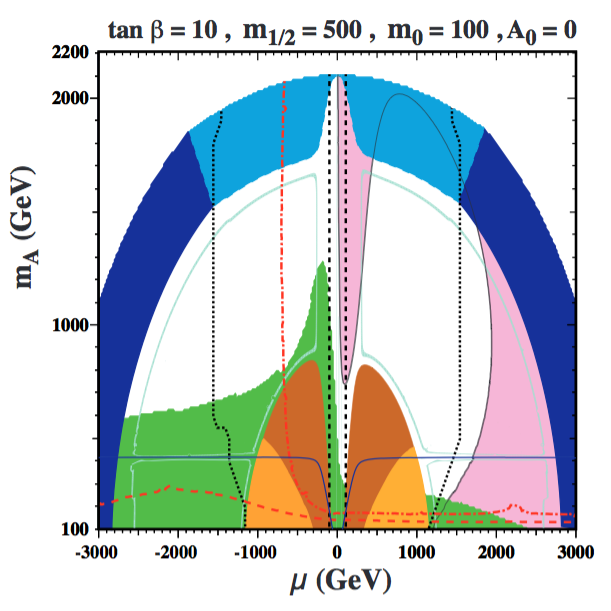}
\caption{A sample $(\mu, m_A)$ plane in the NUHM with a gravitino LSP and universal input squark and slepton masses $m_0 = 100$~GeV, gaugino masses $m_{1/2} = 500$~GeV, and $\tan \beta = 10$, $A_0 = 0$, illustrating different possibilities for the metastable NLSP: the lighter stau (coloured orange), the lighter selectron (yellow), the tau sneutrino (lighter blue) or the electron sneutrino (darker blue)~\protect\cite{sleptonNLSP}. Regions favored by the supersymmetric interpretation of the $g_\mu - 2$ measurement are shaded pink, and regions excluded by measurements of $b \to s \gamma$ are shaded green.}
\label{fig:susy4}
\end{figure}

\subsection{Metastable stop squark scenarios}

There are also scenarios in which the NLSP is the lighter stop squark, ${\tilde t_1}$~~\cite{stopNLSP},
though these are more tightly constrained, in particular since the LHC has established
stronger lower limits on a metastable ${\tilde t_1}$ mass because of its larger
production cross section. Three scenarios can be envisaged for stop decays:
\begin{enumerate}

\item Case~1: $m_{\tilde t_1} - m_{\widetilde G} > m_t$, i.e., small $m_{\widetilde G} <
m_{\tilde t_1} - m_t$. In this case, the stop can decay directly into a top 
quark and a gravitino, and the rate for this dominant decay is
\begin{eqnarray}
\Gamma &=& \dfrac{1}{192 \pi} \dfrac{1}{M_{\rm Pl}^2 m_{\widetilde G}^2 m_{\tilde t_1}^3} 
\left[ 4\left( m_{\tilde t_1}^2 - m_{\widetilde G}^2 - m_t^2 \right) 
+ 20 \, \sin \theta_{\tilde t} \, \cos \theta_{\tilde t} \, m_t \, m_{\widetilde G} 
\right]  \nonumber \\
&& \times \left[ ( m_{\tilde t_1}^2 + m_{\widetilde G}^2 - m_t^2)^2 - 4 m_{\tilde t_1}^2 m_{\widetilde G}^2
\right] \left[ ( m_{\tilde t_1}^2 + m_t^2 - m_{\widetilde G}^2 )^2 - 4 m_{\tilde t_1}^2 m_t^2
\right]^{1/2}.
\label{2bodylife}
\end{eqnarray}
This decay rate is similar to that for stau decay into tau plus gravitino (\ref{telldecay}), but in this case $m_t$ cannot be neglected. Typical values of the ${\tilde t_1}$ lifetime in this case are displayed in the left panel of Fig.~\ref{fig:susy5}.

\begin{figure}[htb]
\centering
\includegraphics[scale=0.3]{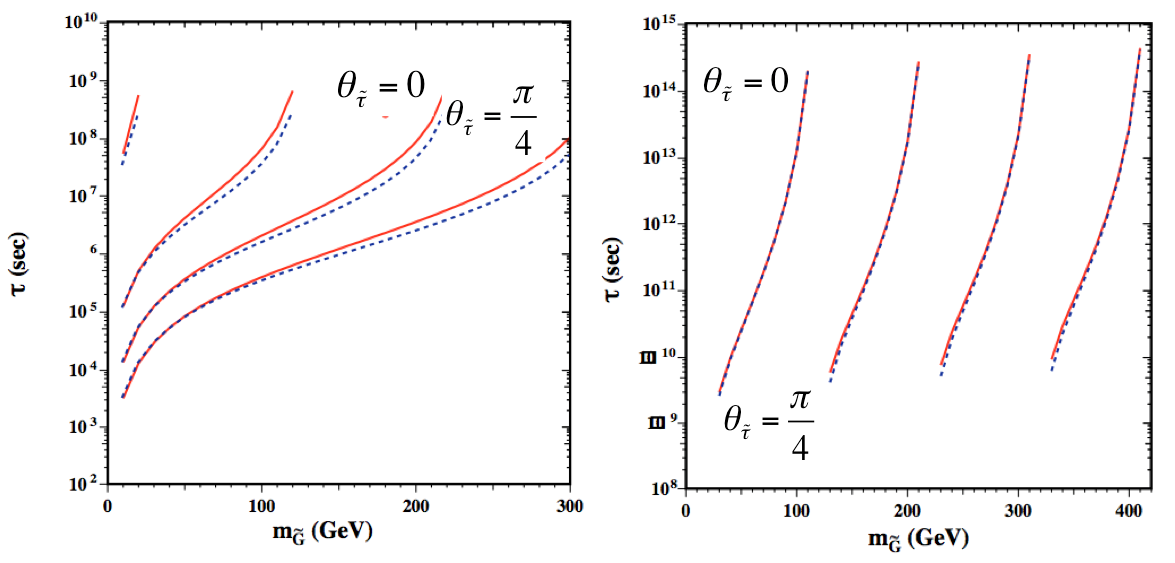}
\caption{The lifetime of the lighter stop LSP in gravitino LSP scenarios for two-body decays (left panel) and three-body decays (right panel) for different assumed values of the stop mixing angle: $\theta_{\tilde t_1} = 0$ (solid red line) and $\pi/4$ (dotted blue line)~\protect\cite{stopNLSP}.}
\label{fig:susy5}
\end{figure}

\item Case~2: $m_W + m_b < m_{\tilde t_1} - m_{\tilde G} < m_t$. In this case, the dominant decays are into the three-body final state ${\tilde t_1} \to \tilde{G} + W + b$. The formulae for the decay rate in this case are quite complicated, and can be found in Ref.~\cite{stopNLSP}. Typical values of the ${\tilde t_1}$ lifetime in this case are displayed in the right panel of Fig.~\ref{fig:susy5}, where we see that they are typically much longer than in Case~1.

\item Case~3:  $m_b + \Lambda_{QCD} < m_{\tilde t_1} - m_{\tilde G} < m_W + m_b$. In this case, the dominant decays are four-body: ${\tilde t_1} \to {\tilde G} + b + {\bar q}q$ or $\ell \nu$. The decay rate is further suppressed compared to Case~2, and likely to exceed $10^{12}$~s, in which case there would be important constraints from the CMB data, that have not been explored.

\end{enumerate}

The long-lived stop squark would hadronize immediately after production, forming a stop `mesino' ${\tilde t}-{\bar q}$ or a stop `baryino' ${\tilde t}-qq$, which might be either charged or neutral. As such stop hadrons pass through matter, they would in general change charge by nuclear interactions, complicating track-finding in a conventional LHC detector. There is no consensus on the charge of the lightest stop hadron, which might be charged, and hence detectable by MoEDAL.

\subsection{Long-lived gluinos in split supersymmetry}

The above discussion has been in the context of the CMSSM and similar scenarios where all the supersymmetric partners of Standard Model particles have masses in the TeV range. Another scenario, suggested following the non-discovery of supersymmetric particles at LEP, is `split supersymmetry', in which the supersymmetric partners of quarks and leptons are very heavy whilst the supersymmetric partners of Standard Model bosons are relatively light~\cite{splitSUSY, splitSUSY1}. In such a case, the gluino could have a mass in the TeV range and hence be accessible to the LHC, but would have a very long lifetime:
\begin{equation}
\tau \approx 8 \left( \dfrac{m_s}{10^9~{\rm GeV}} \right)^4 \left( \dfrac{1~{\rm TeV}}{m_{\tilde{g}}} \right)^5~{\rm s}.
\label{gluinotau}
\end{equation}
Long-lived gluinos would form long-lived gluino hadrons, including gluino-gluon (gluinoball) combinations, gluino-${\bar q}q$ (mesino) combinations and gluino-$qqq$ (baryino) combinations. The heavier gluino hadrons would be expected to decay into the lightest species, which would be metastable, with a lifetime given by (\ref{gluinotau}), and it is possible that this metastable gluino hadron could be charged.

In the same way as stop hadrons, gluino hadrons may flip charge through conventional strong interactions as they pass through matter, and it is possible that one may pass through most of a conventional LHC tracking detector undetected in a neutral state before converting into a metastable charged state that could be detected by MoEDAL. 

\subsection{Supersymmetric scenarios with $R$-parity violation}

The supersymmetric scenarios discussed in the previous Sections are all in frameworks where $R = (-1)^{2S - L + 3B}$ is conserved. However, this may not be true in general: there is no exact local symmetry associated with either $L$ or $B$, and hence no fundamental reason why they should be conserved. Indeed, they are violated in simple models for neutrino masses and grand unified theories, although in these cases they are often violated in such a way that the particular combination appearing in $R$ is conserved. In general, however, one could consider various ways in which $L$ and/or $B$ could be violated in such a way that $R$ is violated, as represented by the following superpotential terms~\cite{RV}:
\begin{equation}
W_{RV} \; = \; \lambda^{\prime \prime}_{ijk} {\bar U}_i {\bar D}_j {\bar D}_k
+  \lambda^{\prime}_{ijk} {L}_i {Q}_j {\bar D}_k
+ \lambda_{ijk} {L}_i {L}_j {\bar E}_k
+ \mu_i L_i H,
\label{Rviolation}
\end{equation}
where ${Q}_i, {\bar U}_i, {\bar D}_i, L_i$ and ${\bar E}_i$ denote chiral superfields corresponding to quark doublets, antiquarks, lepton doublets and antileptons, respectively, with $i, j, k$ generation indices. The simultaneous presence of terms of the first and third type in (\ref{Rviolation}), namely $\lambda$ and $\lambda^{\prime \prime}$, is severely restricted by lower limits on the proton lifetime, but other combinations are less restricted. The trilinear couplings in (\ref{Rviolation}) generate sparticle decays such as ${\tilde q} \to {\bar q} {\bar q}$ or $q \ell$, or ${\tilde \ell} \to \ell \ell$, whereas the bilinear couplings in (\ref{Rviolation}) generate Higgs-slepton mixing and thereby also ${\tilde q} \to q \ell$ and ${\tilde \ell} \to \ell \ell$ decays. For a nominal sparticle mass $\sim 1$~TeV, the lifetime for such decays would exceed a few nanoseconds for $\lambda,  \lambda^{\prime}, \lambda^{\prime \prime} < 10^{-8}$. As already mentioned, there is no strong reason why any of these couplings should vanish, but equally there is no strong reason to expect any non-zero couplings within this range.

If $\lambda_{ijk} \ne 0$, the prospective experimental signature would be similar to the stau NLSP case that was discussed earlier. On the other hand, if $\lambda^{\prime}$ or $\lambda^{\prime \prime} \ne 0$, the prospective experimental signature would be similar to the stop NLSP case that was also discussed earlier, yielding the possibility of charge-changing interactions while passing through matter. This could yield  a metastable charged particle, created whilst passing through the material surrounding the intersection point,  that would be detected by MoEDAL.

\subsection{Heavy sleptons from Gauge Mediated Supersymmetry Breaking scenarios}

In the Gauge Mediated Supersymmetry Breaking  (GMSB) scenario~\cite{GMSB, GMSB1, GMSBFISCHLER1, GMSBFISCHLER2,
 GMSBFISCHLER3, GMSBFISCHLER4} supersymmetry is broken at a low scale, within a few orders of magnitude of the weak scale 
 and the Standard Model gauge interactions serve as `messengers' of supersymmetry breaking, giving rise to a high degree of 
 degeneracy among squarks and sleptons. GMSB offers the the possibility of solving the flavor problem. Moreover, since the 
 relevant dynamics occur at an energy scale much smaller than the Planck mass, GMSB models require no input from quantum gravity.
    
In gauge-mediated theories the supersymmetric mass spectrum is determined in terms of relatively few parameters.  The most important parameter is $\Lambda  = F/M$, where $M$ is the mass scale of the messenger fields,  $F$ the order of the mass-squared splittings inside the messenger supermultiplets and $\Lambda$ sets the scale of supersymmetry breaking in the observable sector. The supersymmetric particle masses are typically a one-loop factor smaller than $\Lambda$. Supersymmetric particle masses depend only logarithmically on $M$.  This scale can vary roughly between several tens of TeV and $10^{15}$~GeV. The lower bound on $M$ is determined by experimental results; the upper bound from the argument that  contributions should be small enough not to reintroduce the flavor problem.
    
In GMSB models the NLSP can be the lightest neutralino $\chi_{1}^{0}$, the lightest stau $\tilde{\tau}$ or, in a very small corner of parameter space, the lightest sneutrino. There is  also the possibility of having co-NLSPs. This occurs when the mass difference between NLSP and co-NLSP is small enough to suppress the ordinary supersymmetric decay and when $F$ is adequately low to allow for a sizeable decay rate into gravitinos. Candidates for co-NLSP include $\chi_{1}^{0}$ (with $s\tilde{\tau}$ NLSP), $\tilde{\tau}$ (with $\chi_{1}^{0}$ NLSP). 

From the NLSP decay rate~\cite{GMSBREVIEW} the average distance travelled by an NLSP with mass $m$ and produced 
with energy $E$ is:
\begin{equation}
L = \dfrac{1}{\kappa_{\gamma}}\left(\dfrac{\rm 100 GeV}{m}\right)^{5}\left(\dfrac{\sqrt{F/k}}{\rm 100 TeV}\right)^{4}
\sqrt{\dfrac{E^{2}}{m^{2}} - 1} \times 10^{-2} cm
\end{equation} 
where $\kappa_{\gamma}$ is 1 for the stau. Mainly depending on the unknown value of $\sqrt{F/k}$, the NLSP can either decay within microscopic distances or decay well outside the solar system.

For larger $\sqrt{F}$, for $\sqrt{F} \gtrapprox 10^{6}$~GeV, 
the NLSP lifetime is longer and the collider phenomenology can resemble the well-known missing-energy supersymmetric signatures (for a neutralino NLSP) or can lead to a long-lived heavy charged particle (for a stau NLSP). This signature is quite novel, with a stable charged massive traversing the detector, leaving an anomalous ionization track. If the particle is sufficiently slow it will be detected by MoEDAL.

\subsection{Metastable charginos}

In certain regions of parameter space, charginos --- the superpartners of $W$-bosons and/or charged Higgs bosons --- can have lifetimes of order centimeters to meters. As such they may leave tracks inside MoEDAL and the other LHC detectors. For instance, this occurs when the LSP is the neutral wino i.e.\ the superpartner of the neutral $W$-boson. Such LSPs arise, for example, in `anomaly mediated supersymmetry breaking' (AMSB)~\cite{AMSB1,AMSB2, AMSB2a} and the $G_2$-MSSM model which derives from string/$M$ theory~\cite{G2-MSSM1,G2-MSSM2}.

The neutral wino is part of an $SU(2)$ triplet whose other two members are two charginos: $\tilde{\chi}_{1}^{+}$  and $\tilde{\chi}_{1}^{-}$. {\it In the limit of exact $SU(2)$-symmetry all three winos are degenerate in mass}. This mass degeneracy can thus only be broken by the Higgs sector. Often, the mass splitting is suppressed by couplings and one-loop effects, which are relatively small, typically leading to splittings of order one or two pion masses (see Ref.~\cite{2loopAMSB} and references therein). 

The near degeneracy between $\tilde{\chi}_{1}^{+}$ and $\tilde{\chi}_{1}^{0}$ leads to a striking prediction: that $\tilde{\chi}_{1}^{+}$ can easily travel distances which range from centimetres to meters before decaying into $\tilde{\chi}_{1}^{0}$ plus additional, low mass and momentum states which will either be a positron and neutrino pair or a charged pion. If, in addition, R-parity is violated, other decay modes may be significant.



For $ \Delta M_{\tilde{\chi}_{1}} \lesssim M_{\pi}$, $\tilde{\chi}_{1}^{+}$ can easily travel a meter or more. If it does so the chargino can be distinguished as a heavy ionizing track  (e.g., if  $\beta\gamma < 0.85$  the track will cause an ionization at least twice that of a minimum ionizing particle (MIP)). This type of long, heavily ionizing track signal will be detected by MoEDAL as long as its ionization is at least five times that of a MIP.

We now discuss different production mechanisms for these charginos at the LHC. One such mechanism is for charged winos to arise in the decay products of gluinos. In many models with wino LSP, gluinos typically have masses which range from a few to ten times that of the wino due to renormalization group effects from high scales to the LHC scale. When the gluino mass is of order a few TeV or less, they can be pair produced at the LHC and their subsequent decay products can often include $\tilde{\chi}_{1}^{+}$ and/or $\tilde{\chi}_{1}^{-}$~\cite{4top,G2Wino}. This is typically simple when squark and slepton masses are out of the LHC reach as might be suggested by the measured value of the Higgs boson mass.

Independently of the gluino channel, charged and neutral winos can be produced through electroweak processes in the supersymmetric analogue of $W$-boson pair production and $W$-$Z$-boson production at the LHC. Often, direct production of ($\tilde{\chi}_{1}^{+}$, $\tilde{\chi}_{1}^{-}$) pairs and/or ($\tilde{\chi}_{1}^{+}$, $\tilde{\chi}_{1}^{0}$) pairs can be the dominant production mode of supersymmetric particles with a larger cross section than gluino pair production. These channels are particularly interesting because MoEDAL has a distinct advantage over ATLAS and CMS. Consider, for instance, the production of ($\tilde{\chi}_{1}^{+}$, $\tilde{\chi}_{1}^{0}$) pairs. $\tilde{\chi}_{1}^{+}$ decays to $\tilde{\chi}_{1}^{0}$ plus, say, a very soft pion. The pion is so soft that it will never leave the magnetic field of the ATLAS or CMS inner detector leaving a track which will be impossible to distinguish from other products of $pp$ collision. Hence, effectively, the final state consists of two $\tilde{\chi}_{1}^{0}$'s which leave the detector without depositing any energy. Such events will not trigger  ATLAS or CMS and so the observation of such events requires associated production of additional high $p_{\rm T}$-jets and/or charged leptons, thereby reducing the sensitivity. In MoEDAL by contrast, the mere presence of the chargino with sufficient ionization is enough to provide a signal.

Another relevant class of models is Gaugino Anomaly Mediation~\cite{INOSMB}. This is a scenario of supersymmetry breaking suggested by the  phenomenology of flux compactified type~IIB string theory. The main element of this scenario is that the gaugino masses are of the anomaly-mediated SUSY breaking form, while scalar and trilinear soft SUSY breaking terms are highly suppressed. 
 
Renormalization group effects give rise to  an experimentally viable sparticle mass spectrum, while at the same time avoiding charged LSPs. SUSY-induced flavor and CP-violating processes are also suppressed since scalar and trilinear soft terms are highly suppressed. Under these premises the lightest SUSY particle is the neutral
wino, while the heaviest is the gluino. 
 
As far as LHC phenomenology is concerned in this scenario, there should be a strong multi-jet plus missing energy $E{\rm _T^{miss}}$ signal from squark pair production. Also, a double mass edge from the opposite-sign/same flavor dilepton invariant mass distribution should be visible. Importantly,  short - yet visible - highly ionizing tracks from quasi-stable charginos, which  should provide a smoking-gun signature for Gaugino Anomaly Mediation, that would be detectable by MoEDAL.
 
Charginos which are predominantly the superpartners of charged Higgs bosons can also have suitably long lifetimes and these have been studied in Ref.~\cite{Higgsino}.

\subsection{The Fat Higgs model}

The Fat Higgs~\cite{FATHIGGS} is a particular, interesting solution to the ``supersymmetric little hierarchy problem''~\cite{SUSYLHP}. It proposes an alternative to the standard MSSM picture of electroweak symmetry breaking (EWSB) and results in a heavier `light' CP-even Higgs than can be realized in that standard scenario, thus naturally evading the LEP-II bounds. A new variant of the Fat Higgs Model~\cite{NEWFATHIGGS} has been produced, where  the Higgs fields remain elementary, alleviating the supersymmetric fine-tuning problem while maintaining unification in a natural way. 
 
A latest incarnation of the supersymmetric Fat Higgs model has been introduced in which the MSSM Higgs bosons and the top quark are composite~\cite{FATHIGGSFATTOP}. The underlying theory is a confining $SU(3)$ gauge theory with the MSSM gauge groups realized as gauged sub-groups of the chiral flavor symmetries. This motivates the requirement for a large top mass and SM-like Higgs of mass greater than that of the $Z$-boson in a natural way as the residual of the strong dynamics responsible for the composite fields. This solves the fine-tuning problem  associated with these couplings present in the original Fat Higgs and `New Fat Higgs' models, respectively.

The model also has a number of additional chiral multiplets. The colored quark singlets $q_{1}$  and $q_{2}$  have masses of order of the compositeness scale ($\lambda$) whereas the color neutral particles are expected to have masses  of order 200~GeV. We expect the lightest of these to be the singlet $\chi$ fields, with the slightly heavier charged ($\pm$) fields ($\psi$) to be be slightly heavier.  The scalar components  are  expected to be slightly heavier than their fermionic partners due to SUSY-breaking contributions to the scalar masses.

The dynamically generated super-potential has a symmetry which has all of the exotic particles coupling in pairs. Since all of the exotic states must decay through $q_{1}$ whose mass is of order a PeV, the exotics are typically very long lived and have complicated multi-particle final states. In the case of $\psi$ this results in electrically charged fermions and their scalar partners which are collider stable appearing as massive charged objects. It is expected that the LHC will cover the entire parameter space~\cite{Ambrosanio}.

\subsection{XYons from 5D SUSY breaking}

In a recently proposed 5D~model~\cite{xyon} a general framework is presented for supersymmetric theories that do not suffer from fine tuning in electroweak symmetry breaking. Supersymmetry is dynamically broken at a scale $ \Lambda \sim (10-100)$~TeV, which is transmitted to the supersymmetric standard model sector through standard model gauge interactions. The dynamical supersymmetry breaking (DSB) sector possesses an approximate global $SU(5)$ symmetry, whose $SU(3)\times SU(2) \times U(1)$ subgroup is identified as the standard model gauge group. This  $SU(5)$ symmetry is dynamically broken at the scale $\Lambda$, leading to TeV scale exotic scalars with the quantum numbers of GUT XY bosons appear, the so-called XYons.

If a condition analogous to $R$~parity holds in the DSB, the Xyons, that have multi-TeV mass, are long lived.
 Their precise quantum numbers depend on the details of the DSB; in general they are both colored and charged. In the simplest $SU(5)$ case they lie in a colour triplet isospin doublet with electric charges  $Q = -1/3, -4/3$, but also $SO(10)$ assignments are possible, for example an additional doublet with $Q= 1/3, -2/3$, 
 could easily be possible. 

The xyonic fermionic mesons are: $\tilde{T}^{0} \equiv  \phi\uparrow\bar{d}$,  $\tilde{T}^{-} \equiv \phi\uparrow\bar{u}, \tilde{T}'^{-} \equiv  \phi\downarrow\bar(d)$ and $ \tilde{T}'^{--} \equiv \phi\downarrow\bar{u}$. In the case of $SU(5)$ XYons, the masses of these XYonic mesons (XYmesons) split due to the mass difference between $\phi\uparrow$ and $\phi\downarrow$ ---  XYmesons containing $\phi\downarrow$ are heavier than those containing $\phi\uparrow$ by about 600~MeV. This implies that $\tilde{T}'^{-} ( \tilde{T}'^{--})$ decays into $\tilde{T}^{0} ( \tilde{T}^{-})$ and a charged pion with the lifetime of about a picosecond. 
 
The mass splitting between $\tilde{T}^{0}$ and $\tilde{T}^{-}$  (and $\tilde{T}'^{-}$ and $\tilde{T}'^{--}$) is of order a few MeV, which comes from isospin breaking effects due to electromagnetic interactions and the $u-d$ mass difference. Because the two effects work in the opposite direction, it is not clear which of $\tilde{T}^{0}$ and $\tilde{T}^{-}$ is lighter. The decay of the  heavier into the lighter state is through weak interactions with lifetime is of order $10^{-1}$  to $10^{2}$  seconds, so that both $\tilde{T}^{0}$ and $\tilde{T}^{-}$ are essentially stable for collider purposes. The multi-TeV mass of the charged XYmeson means that it is likely that will it be slow moving a highly ionizing enough at the LHC to be detectable by MoEDAL.
 
There are also bosonic baryons formed by XYons and the standard model quarks. The lightest states of these xyonic baryons (xybaryons) will come either from a $(2,1)$ scalar multiplet,
\begin{equation} 
\tilde{U}^{0}_{S} \equiv \phi\uparrow[u,d], ~~~~~\tilde{U}^{-}_{S} \equiv \phi\downarrow[u,d],
\end{equation}
or a $(2,3)$ vector multiplet
\begin{gather}
\tilde{U}^{+}_{V} \equiv \phi\uparrow uu, ~~~~\tilde{U}^{0}_{V} \equiv \phi\uparrow\{ud\}, ~~~~\tilde{U}^{-}_{V} \equiv \phi\uparrow dd   \\
~~\tilde{U}'^{0}_{V} \equiv \phi\downarrow uu, ~~~~\tilde{U}'^{-}_{V} \equiv \phi\downarrow\{ud\}, ~~~~\tilde{U}'^{--}_{V} \equiv \phi\downarrow dd
\end{gather}
where $\{ \}$ and $ [ ] $ denote summarization and antisymmetrization, respectively.

In the model described here~\cite{xyon} only the lighter of baryons $\tilde{U}^{ 0}_{S}$ and $\tilde {U}^{+}_{V}$, $\tilde{U}^{0}_{V}$ and $\tilde{U}^{-}_{V}$ are collider stable particle(s) at colliders. The stability of these lighter XYbaryons is ensured by baryon number conservation. The charged baryons again with multi-TeV masses leave highly-ionizing tracks so that they can be detected relatively easily by MoEDAL.

There are also be other signals that can be used to detect XYons. When anti-XYmesons or XYbaryons traverse a detector they can exchange isospin and charge with the background material through hadronic interactions, and so  make transitions between neutral and charged states. This leaves a distinct signature of intermittent highly ionizing tracks which are detectable using MoEDAL.

\subsection{Current LHC limits on sparticles}

SUSY searches in collider experiments involving promptly-decaying sparticles typically focus on events with high transverse missing energy, arising from (weakly interacting) LSPs, in conjunction with no or few leptons ($e$, $\mu$), many jets and/or $b$-jets, $\tau$-leptons and photons. In the absence of deviations from SM predictions, lower bounds on sparticle masses have been set by ATLAS~\cite{atlas-det} and CMS~\cite{cms-det} in several SUSY scenarios using $pp$ collision data at $\sqrt{s}=7-8~{\rm TeV}$.

Strong production through $\tilde{g}\tilde{g}$, $\tilde{g}\tilde{q}$ and $\tilde{q}\tilde{q}$ is expected to be abundant at the LHC, followed by cascade decay into lighter sparticles that finally leads to the LSP. In the CMSSM case, the 95\% confidence limit (CL) on squark mass reaches 1750~GeV and on gluino mass is 1400~GeV if the results of various analyses are employed~\cite{atlas-susy-results}. 

Naturalness considerations suggest that the third-generation sfermions ($\tilde{t}_1$, $\tilde{b}_1$ and $\tilde{\tau}_1$) are the lightest colored sparticles. In gluino-mediated $\tilde{t}_1$ production, gluinos with masses 560--1300~GeV have been ruled out for a massless LSP by CMS~\cite{cms-susy-results}. For a heavy gluino, $\tilde{t}_1\tilde{t}_1$ production yields $(m_{\tilde{t}_1},\,m_{\tilde{\chi}_1^0})$ limits that depend of mass hierarchy and decay branching ratios. The stringent bounds reach the $~680$~GeV ($250$~GeV) in stop (neutralino) mass at 95\% CL as set by ATLAS~\cite{atlas-susy-results}. 

If all squarks and gluinos are above the TeV scale, weak gauginos and sleptons with masses of few hundred GeV may be the only sparticles accessible at the LHC. Charginos with masses up to 740~GeV are excluded by CMS for a massless LSP in the chargino-pair production with an intermediate slepton/sneutrino~\cite{cms-det}. 

Both ATLAS and CMS experiments have also probed prompt $R$-parity violating SUSY through various channels, either by exclusively searching for specific decay chains, or by inclusively searching for multilepton events.  Of more interest to the MoEDAL physics program are searches for unstable long-lived LSPs. Such an analysis looking for a multi-track displaced vertex (DV) that contains a high-momentum muon at a distance between millimeters and tens of centimeters from the $pp$ interaction point has been performed by ATLAS~\cite{atlas-dv}. The results are interpreted in the context of an $R$-parity breaking SUSY scenario, where such a final state occurs in the decay $\tilde{\chi}_1^0\rightarrow\mu q\bar{q}'$, allowed by the non-zero RPV coupling $\lambda'_{2ij}$. The limits are reported as a function of the neutralino lifetime and for a range of neutralino masses and velocities, which are the factors with greatest impact on the limit. Indicatively squark masses of up to 700~GeV have been ruled out for $\tilde{\chi}_1^0$ LSP decay lengths from 1~mm to 1~m if squark-pair production and 50\% branching fraction for the LSP decay is assumed.

ATLAS has searched for $\tilde{\chi}_1^{\pm}$ decays into a neutralino and a soft (undetectable) pion, experimentally being observable as `disappearing tracks'. Such particles would decay inside the inner detector volume and would be identified by well-reconstructed tracks in the inner-tracker layers, but with low numbers of hits in the outer-tracker layers. Constraints on the chargino mass, the mean lifetime and the mass splitting are set, which are valid for most scenarios in which the LSP is a nearly pure neutral wino. In the AMSB models, a chargino having a mass below 270~GeV is excluded at 95\% CL~\cite{atlas-kink}.

Both experiments have searched for stable leptons, which being charged and penetrating, are expected to interact as if they were heavy muons. A recent analysis by ATLAS~\cite{atlas-slepton} is based on a measurement of the mass of slepton candidates, as estimated from their velocity and momentum based on their interactions in the inner detector, the calorimeters and the muon spectrometer. The null results are interpreted in the context of GMSB models where the $\tilde{\tau}_1$ is the NLSP. Long-lived $\tilde{\tau}_1$ in the GMSB model considered are excluded at 95\% CL at masses below 402--347~GeV, for $\tan\beta=5Ð50$. Exclusion limits on the $\tilde{\tau}_1$ mass up to 342 (300)~GeV are set in the hypothesis that $\tilde{\tau}_1$ are produced directly or via other slepton ($\tilde{e}$, $\tilde{\mu}$) pair production, assuming a mass splitting between the light slepton and stau of 1 (90)~GeV. In decoupled scenarios, where $\tilde{\tau}_1$ pair production is the only SUSY signature, exclusion limits up to 267~GeV are set on the $\tilde{\tau}_1$ mass. 

A more general search for heavy stable charged particles has been carried out recently by CMS~\cite{cms-slepton} involving the momentum, energy deposition, and time-of-flight of signal candidates. Leptons with an electric charge between $e/3$ and $8e$, as well as bound states that can undergo charge exchange with the detector material, such as R-hadrons, have been considered utilizing long time-of-flight to the outer muon system and anomalously high (or low) energy deposition in the inner tracker. The lower limits on gluino masses range between 1233 and 1322~GeV depending on fraction of $\tilde{g}g$ production and the interaction scheme. For stop production, the corresponding limits range between 818 and 935~GeV. Drell-Yan like signals with $|Q| = e/3, 2e/3, 1e, 2e, 3e, 4e, 5e, 6e, 7e, 8e$ are excluded with masses below 200, 480, 574, 685, 752, 793, 796, 781, 757, and 715~GeV, respectively.

A fraction of the R-hadrons that may produced at the LHC could lose all of their momentum, mainly from ionization energy loss, and come to rest within the detector volume, only to decay to a $\tilde{\chi}_1^0$ and hadronic jets at some later time. In the latest analysis performed by ATLAS with data at $\sqrt{s}=7$ and 8~TeV, candidate decay events are triggered in selected empty LHC bunch crossings in order to remove $pp$ collision backgrounds~\cite{atlas-stopped}. In the absence of an excess of events, limits are set on gluino, stop, and sbottom masses for different decays, lifetimes, and neutralino masses. With a neutralino of mass 100~GeV, the analysis excludes gluinos with mass below 832~GeV, for a gluino lifetime between 10~s and 1000~s in the generic R-hadron model with equal branching ratios for decays to $qq\tilde{\chi}_1^0$ and $g\tilde{\chi}_1^0$. Under the same assumptions for the neutralino mass and squark lifetime, top squarks and bottom squarks in the Regge R-hadron model are excluded with masses below 379 and 344 GeV, respectively.

\section{ Scenarios with Extra Dimensions}

Over the past two decades, new models based on compactified extra spatial 
dimensions (ED) have been proposed, which could explain the large gap between 
the electroweak (EW) and the Planck scale of  $M_{EW}/M_{PL}  \approx  $10$^{-17}$. 
The four main LHC search scenarios discussed in this arena are the
Arkani-Hamed-Dimopoulos-Dvali (ADD) model of large extra dimensions~\cite{ADD,ADD1,ADD2}, the 
Randall-Sundrum (RS) model of warped extra dimensions~\cite{Randall}, 
TeV$^{-1}$-sized extra dimensions~\cite{TEV-1, TEV-1a,TEV-1b}, and the Universal Extra 
 Dimensions (UED) model~\cite{UED}.

\subsection{ Extra dimensions and microscopic black holes }

The ADD model~\cite{ADD,ADD1,ADD2}, 
foresees the existence of $n$ spatial flat extra dimensions, which are accessible 
only to gravity. The other fields, on the contrary, are 
localized on a three-dimensional brane. The extra dimensions are compactified on a torus 
whose size $R$ is related to the fundamental mass scale $M_D$ by the expression:
\begin{equation}
M_{PL}^{2} = M_{D}^{2+n} R^{n}
\end{equation}
needed to restore at large distances the correct value of the Newton gravitational 
constant. Using the above equation for $M_{D}$  of the order of 1 TeV,  $R$  can be, 
depending on $n$, as large as a tenth of a millimetre, motivating the name ``large extra 
dimensions'' for this framework. In this model the weakness of gravity is due to 
the fact it spreads into the large bulk  of the extra dimensions away from the brane 
and the hierarchy is removed.

The compactification of the the extra dimensions gives rise to a tower of massive 
Kaluza-Klein (KK) states with a mass gap $\Delta m \sim 1/R $. The KK gravitons can be 
produced, at sizeable rates, in high-energy particle collisions, and since they are 
weakly coupled to matter (the couplings are of gravitational strength) they escape 
the detector, resulting in missing energy. Even if the detailed aspect of the theory at energies above 
$M_{D}$  is not known, the graviton emission rates can be calculated in a effective theory 
which holds for energies below $M_D$~\cite{GUIDICE}.

Searches at the LHC for evidence in favour of this model can be performed by
 detecting an excess of events with a single energetic jet or a single energetic
  photon recoiling against the graviton (missing $E_{T} + jet$ or missing 
  $E_{T} + \gamma$) compared to the expected rate of these events from 
  Standard Model processes, mainly from $ jet + Z \rightarrow \nu\nu$ and 
  $jet + W \rightarrow l\nu$. The presence of ADD extra dimensions can be 
  detected at the LHC also by investigating the contribution of virtual KK graviton
   exchange to Drell-Yan processes, in particular di-lepton or di-photon production. 
   Virtual graviton exchange produces deviations from the Standard Model 
   expectations in the high invariant mass ($ m_{ll}  > $  2~TeV) region.  

The RS model~\cite{Randall} is another framework for extra dimensions. 
The simplest realization of this model posited
    the existence of one extra dimension, bounded by two four-dimensional branes,
    the Standard Model fields being localized on one of these branes while 
    gravity lives both on the branes and in the bulk. A distinctive feature of this 
    model is that the metric of the five-dimensional space is non-factorizable, with
     a warping factor of the four-dimensional Minkowski metric $\eta_{\nu\mu}$,
     which depends on the position along the extra dimension:
\begin{equation}
ds^{2} = e^{-k|y|}dx^{\mu}dy^{\nu}n_{\mu \nu} \, .
\end{equation}
This ``warp factor" reformulates the hierarchy problem. The sole fundamental scale 
     is of the order of  $M_{PL}$, and the TeV scale $\Lambda$ is generated on the 
     Standard Model brane by the warp factor:
\begin{equation}
\Lambda = M_{PL}e^{-k\pi r_{c}}
\end{equation}
where $k\pi r_{c} \sim 11 - 12$ and the value of $k$ is of the order of $M_{PL}$. 

In this model the gravitons are coupled to the Standard Model fields and they decay
  into fermions or bosons. Two parameters determine the model, which can be chosen
   as the mass $m_{G}$ of the first excited KK graviton and, the coupling $c = k/M_{PL}$
  that determines the width of the resonances. The clear signature of the RS  
  gravitons at the LHC is the presence of resonances in the invariant mass spectrum of 
  di-leptons, $m_{ll}$, or di-photons $m_{\gamma\gamma}$, with an angular
  distribution characteristic of spin 2. The contribution of the
  standard Drell-Yan processes to the $m_{ll}$  spectrum falls continuously and
   becomes very low at high $m_{ll}$ $\sim$ 2 TeV.

The existence of extra spatial dimensions \cite{ADD,Randall} and a
sufficiently small fundamental scale of gravity opens up
the possibility that microscopic black holes can be produced
and detected \cite{ED1, ED1a,ED1b,ED1c, ED1d, bhevaporation, fischler1, CHARYBDIS, ED2,ED3,ED3a}  at the
LHC. 

Once produced, the black holes will undergo an evaporation process.
The evaporation process can be categorized in three characteristic stages~\cite{bhevaporation, fischler1}. 
The first stage is called the ``balding phase'' when  
  the black hole radiates away its inherited  multi-pole moments  and settles down 
  into a hairless state, losing  a  fraction of the initial mass to  gravitational radiation. 
The second stage is the ``evaporation phase'' which starts  with a spin-down phase in
which  Hawking radiation~\cite{hawkingradiation,hawkingradiation1} carries away the angular momentum,
 after which it proceeds with the emission of thermally-distributed quanta until the black hole
reaches the Planck  mass.  The third and final stage is the Planck phase when the black hole mass in
near to the Planck mass. At this stage we are in the realm  of quantum gravity and predictions
 become increasingly difficult. It is generally assumed that the black hole will 
 decay completely to some last  few Standard Model particles. However, another intriguing possibility 
 is that the remaining energy is carried away by a stable remnant.

\subsubsection{Long-lived microscopic black holes}

In a regime of the RS model in which the brane cannot be neglected there are very few physical solutions
 to the higher-dimensional black hole problem \cite{RSBH,RSBH1}. 
A class of RS solutions have been found in which the solution is a 
Reissner-Nordstrom metric with the electric charge replaced by a tidal charge~\cite{tidalcharge}. If the effects of this
tidal-charge term are negligible, the solution becomes an effective four-dimensional solution
and the effects of low-scale gravity are unlikely to be observed. If the extra term is significant,
the effects of low-scale gravity may be observable if the fundamental scale of gravity is low
enough.

The decays of tidal-charged black holes have been studied in the context
of the microcanonical picture~\cite{ED2, microcanonical, microcanonical1}. The microcanonical corrections may be significant when the
object reaches the Planck size and the classical black hole description fails. 
The microcanonical corrections to the canonical decay treatment are larger in the RS
scenario for Planck-sized objects, and in certain cases they may live long enough
to be considered long-lived or even quasi-stable. This possibility is discussed in 
\cite{ED2, microcanonical,microcanonical1,gingrich-tidalcharge}.

At the LHC black holes are typically formed from the interaction of valence quarks as
those carry the largest available partonic momenta. Thus, the largest cross-section will
be for black holes with a charge of $\sim$4/3, as seen in Fig.~\ref{Fig:Qofremnants}.
which if the black hole is moving slowly ($\beta \lesssim$ 0.3) would often be sufficiently highly-ionizing to be detected
 by MoEDAL.

\subsubsection{Microscopic black hole remnants}

The final fate of a black hole is still an open question. The last stages
of the evaporation process are closely connected to the information-loss puzzle \cite{infoloss,infoloss1,infoloss2}. 
 When one tries to avoid the information-loss problem there are two possibilities. Either the
information is regained during the decay by some mechanism, or a stable black hole remnant is
formed that retains the information. An important argument against the
existence of remnants is that, since no evident quantum number prevents it,
black holes should radiate away completely.  On the other hand, it has been argued  that the 
generalized uncertainty principle~\cite{GUP,GUP1} may prevent the total evaporation of 
a black hole, not by symmetry but by dynamics,  as a minimum size and mass are  approached. 
Other arguments for black-hole remnants are given in~\cite{ARGSFORBHREMS, ARGSFORBHREMS1,ARGSFORBHREMS2,
ARGSFORBHREMS3,ARGSFORBHREMS4,ARGSFORBHREMS5, ARGSFORBHREMS6}.
The prospect of microscopic black hole production at the LHC has been
discussed within the framework of models with large extra dimensions 
by Arkani-Hamed, Dimopoulos and Dvali in~\cite{ADD,ADD1,ADD2}. 

\begin{figure}[htb]
\begin{center}
\includegraphics[width=20pc]{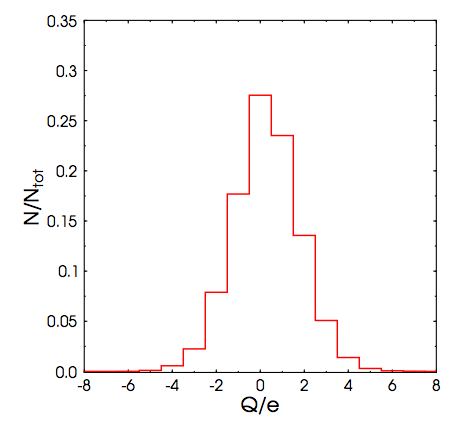}
\caption{The distribution of black-hole remnant charges in p-p interactions at $\sqrt{s}$ = 14 TeV calculated with the
{\tt PYTHIA} event generator and the {\tt CHARYBDIS} program.}
\label{Fig:Qofremnants}
\end{center}
\end{figure}

To compute the production details, the cross-section for black hole production may be approximated by the classical
geometric cross-section: $\sigma(BH) \approx \pi R^{2}_{H}$, where $R_{H}$ is the horizon radius of the black hole. 
This expression  contains only the fundamental Planck scale as a coupling constant. This cross-section
is a subject of ongoing research~\cite{NAIVEXS,NAIVEXS1}, but the classical limit may be used up to energies 
of at least $\sim$10M$_{f}$~\cite{NAIVEXSGOODTO,NAIVEXSGOODTO1,NAIVEXSGOODTO2,NAIVEXSGOODTO3}, 
where $M_{f}$ is  true fundamental  of gravity.
It has also been shown that the naive classical result remains valid in string theory~\cite{STRINGSOK}.


Black holes produced at the LHC are expected to decay with an average multiplicity of $\approx$ 10-25 into
Standard Model particles,  most of which will be charged, though the 
details of the multiplicity distribution depend on the number of extra
dimensions~\cite{BHMULT}. After the black holes have evaporated off enough energy to reach the remnant mass, some
will have accumulated a net electric charge. According to purely statistical considerations, the probability for being
left with highly-charged black hole remnants drops fast with the deviation from the average. The largest fraction of
the black holes should have charges $\pm$ 1 or zero, although a smaller but non-negligible fraction would be multiply-charged.

 The fraction of charged black-hole remnants has been estimated in~\cite{BHMULT} using the {\tt PYTHIA} event
generator and the {\tt CHARYBDIS} program~\cite{CHARYBDIS}. In this study  it was  assumed 
 that the effective temperature of the black hole drops towards zero for a finite remnant mass, $M_{R}$.
This mass of the remnant is a few time $M_{f}$ and a parameter of the model. Even though the temperature-mass
relation is not clear from the present status of theoretical studies, such a drop of the temperature can
be implemented into the simulation. The value of $M_{R}$ does not noticeably affect the  investigated charge distribution, 
as it results from the very general statistical distribution of the charge of the emitted particles.

Thus, independent of the underlying quantum-gravitational  assumption leading to the remnant formation, the authors found
that about 30\% of the remnants carry zero electric charge, whereas $\sim$ 40\% would be singly-charged
black holes, and the  remaining $\sim$30\% of remnants would be multiply-charged. (We note,
however, that another study finds a much smaller percentage of charged
black holes \cite{BELLAGAMBA}).  The distribution of the  remnant charges obtained  is shown 
in Fig.~\ref{Fig:Qofremnants}. The black hole remnants  considered here are heavy, with masses of a TeV or more. 
 A significant  fraction of the black-hole remnants produced would have a  Z/$\beta$ of greater than 5, sufficient 
 to register in the CR39 NTDs forming the LT-NTD  sub-detector of MoEDAL.

Another study  \cite{noncommutative} noted that, if the Planck scale is of the order of a TeV, non-commutative
geometry-inspired black holes could become accessible to experiments. One of
the main consequences of the model is the existence of a black-hole remnant whose mass
increases with a decrease in the mass scale associated with non-commutativity
and a decrease in the number of dimensions. The experimental signatures  differ 
from previous studies of black holes and remnants at the LHC in that  the mass of the remnant
could be well above the Planck scale, and in a few percent of the time the remnant is singly-charged. 
The large multi-TeV mass and charge of the non-commutative geometry-inspired black hole
make it also a candidate for detection by MoEDAL.

\subsection{Long-lived Kaluza-Klein particles from Universal Extra Dimensions}

Nowadays there are many models that contain extra dimensions, but the Universal Extra Dimensions 
(UED) model \cite{UED}  is the model closest to the original paradigm of Kaluza and Klein 
\cite{Kaluza1921tu,Klein1926tv}.  In this model, SM fields propagate into an extra 
dimension that is compactified, in the simplest case onto an $S^1/Z_2$ orbifolded cylinder.  This 
orbifolding endows the momentum modes in the higher dimensions with a property known as 
Kaluza-Klein (KK) parity, an additional quantum number means that only pair-production is possible.   

At tree level the only difference between the mass of the KK modes of the SM particles
 is their lower-dimensional mass, but radiative corrections spoil this and generally lead to a specific 
 spectrum \cite{Cheng2002iz}.  The $S^1/Z_2$ compactification mentioned above together with 
 minimal extra assumptions is referred to as minimal UED (mUED) scenario, and the entire mass spectrum 
 of the KK modes in this case depends only upon the Higgs mass, the compatification scale $R^{-1}$ 
 and the cut-off $\Lambda$ where the higher-dimensional (non-renormalizable) theory breaks down,
 requiring an ultraviolet completion.  
 The dependance of the KK mass spectrum upon the precise value of $\Lambda$ is weak, and it is 
 normally taken to be some multiple of $R^{-1}$ of order 20 or so.  

Now the Higgs mass has been observed to be $m_H\sim 125$ GeV.   
Unlike the dependence on the cut-off $\Lambda$, the mass spectrum depends sensitively on
 this value, so the parameter space is now quite tightly constrained and for all but the very largest
  compactifications the first KK mode of the hypercharge gauge boson is the lightest particle
   \cite{Cembranos2006gt,Cheng2002iz}.  In some sense this is a good thing, as it means 
   that the theory possesses a dark matter candidate which is not charged~\cite{Servant2002aq},
   but at the same time it means that in these very simple models there would be no charged particles observable at large 
   distances from the LHC interaction region for MoEDAL to detect.  
   Charged fermionic KK modes can of course be created at the LHC, but the mass splitting 
   between the lightest KK mode and the next-to-lightest KK mode would be such that the decay 
   rate of particles charged under the SM gauge group would be very rapid and occur 
   very close to the interaction point~\cite{Cembranos2006gt}.

If the lightest KK particle were charged, it would be a problem if we were to assume a normal 
cosmology with some relatively high reheat temperature, as the Universe would be full of stable 
charged particles, something we know is not the case because we have not detected any (for 
example in the same searches as those for anomalous $R$-hadrons in sea water \cite{SEAWATER,SEAWATER1}.

There is a slight complication here, due to the fact that there should also be a KK graviton 
associated with the compact space that does not suffer radiative corrections, and therefore 
has a mass of $R^{-1}$.  For compactification scales below 800 GeV, the KK graviton can 
be the lightest KK particle.  This in principle opens the door for the lightest gauge particle in 
the UED model to be charged, with decay to the graviton taking a very long time due to the
 the gravitational coupling~\cite{Cembranos2006gt}.  There would be the usual problems with 
 the non-observance of $\gamma$-rays by the Fermi satellite or the Compton Gamma-Ray Observatory, 
 but these in principle could be  circumvented by invoking a highly non-standard cosmology just before nucleosynthesis,
   with a low reheat temperature.  In this situation, charged KK particles could be observed by 
   MoEDAL. However, this would never take place in the mUED model, as in that case the lightest KK 
   excitation of a SM particle is not charged for a 125 GeV Higgs.
 
However, there are non-minimal versions of UED which contain additional boundary terms located at the
 fixed points of the orbifold.  In these models there are regions of the parameter space where the 
 charged component of the Higgs can be the lightest KK particle, and hence of interest to MoEDAL~\cite{Flacke2008ne}.

\subsection{D-matter} 

Modern versions of string theory incorporate higher-dimensional ``domain-wall''- like membrane (``brane'') 
 structures in space-time. Fundamental open strings - which represent elementary-particle excitations above the
 vacuum - have their ends attached to membranes embedded in higher-spatial dimensional ``bulk'' spaces. On
 the other hand, only gravity and closed string  modes (such as radions)  are free to propagate in the bulk space 
 between branes.  These brane structures are called D-branes because the attachment of the ends of the open 
 strings is described (in a world-sheet picture) by Dirichlet world-sheet boundary conditions.

However, once we accept the concept of higher-dimensional space-times with domain-world structures, it 
is also natural to consider cases where the bulk is ``punctured'' by lower-dimensional D-brane defects, 
which are either point-like or have their longitudinal dimensions compactified~\cite{westmuckett, westmuckett1,
westmuckett2, westmuckett3}. From a low-energy observer's 
point of view, living on a brane Universe with three spatial longitudinal,  uncompactified  dimensions, such 
structures would effectively appear to be point-like ``D-particles''. 

Unlike the  space-filling background D-brane worlds, the effectively point-like D-particles - obtained from
 Dp-branes with all their dimensions compactified -  have  dynamical degrees of freedom. Thus,  in contrast
  to the background D-branes,  D-particles can  be treated as quantum  excitations above the vacuum~\cite{westmuckett,shiu}
that are  collectively referred to as  {\it D-matter}. D-matter states are   non-perturbative stringy objects 
with masses of order $m_D \sim M_s/g_s$, where $g_s $ is the string coupling. However,  $g_s$ cannot be 
arbitrarily small since,  to reproduce the   observed gauge and gravitational couplings,  $g_s $ is typically of 
order one. Hence, the D-matter states could be  light enough  to be phenomenologically 
 relevant at the LHC.

\subsubsection {D-particles with magnetic charge}

The stability of a  D-brane is due to the charge it carries.  Depending on their type, D-branes could 
carry integral or torsion (\emph{discrete}) charges. The lightest D-particle (LDP) is stable, because it is the lightest
state carrying its particular charge. Therefore, just as in the case of the lightest supersymmetric particle (LSP)
the LDPs are possible candidates for cold dark matter~\cite{shiu}. D-particles, like all other D-branes, are solitonic
 non-perturbative objects in the string/brane theory. As discussed in the relevant literature~\cite{shiu}, 
there are similarities and differences between D-particles and magnetic monopoles, which are common in string 
 models, with non-trivial cosmological implications~\cite{Witten2002wb,westmuckett,mitsou}. 

\begin{table}[htb]
\tbl{Comparison between the
't Hooft-Polyakov monopole and the D-Matter discussed in this section. Here, $g_{\mbox{YM}}$ denotes  the Yang-Mills
 gauge coupling, $<\phi>$ denotes the vacuum expectation  value of the scalar field $\phi$ of the monopole
configuration,  $\lambda$ is its coupling constant, and $M_X$ is the symmetry breaking scale. The size of a D-matter 
particle depends on the probe, since D-branes can probe smaller distances than fundamental strings. 
From ref.~\protect \cite{shiu}.}
{\begin{tabular}{|c|c|c|}
\hline
& \quad {'t Hooft-Polyakov Monopole} \quad & D-Matter  \\
\hline
& & \\
Mass & $\dfrac{\langle \phi \rangle}{g_{\mbox{YM}}} \sim \dfrac{M_X}{g_{\mbox{YM}}^2}$ & 
$\dfrac{M_s}{g_s} \sim \dfrac{M_s}{g_{\mbox{YM}}^2} $ \\
& & \\
\hline && \\
(size)$^{-1}$ & $~~~~\lambda \langle \phi \rangle$ & $g_s^{\alpha} M_s$ \\
&  $g_{\mbox{YM}} \langle \phi \rangle$ & \\
& &
\qquad $\alpha = \left\{ \begin{array}{ll} -1/3 & \mbox{brane-probe} \\
0 & \mbox{string-probe}
\qquad
\end{array}
\right.$
\\
& & \\
\hline &&  \\
Interaction & $\propto \mu_m = \dfrac{n}{g_{\mbox{YM}}}$ where $n \in {\bf Z}$
& $\propto g_{\mbox{YM}}$ \\ 
& & \\
\hline
\end{tabular}  
\label{table:monopole}}
\end{table}

An important difference between the
D-matter states and other non-perturbative objects, such as magnetic monopoles,  is that they could have
{\it perturbative } couplings shown in  Table~\ref{table:monopole}. 
Magnetic monopoles are characterized by 
magnetic  charges, e.g.,  $\mu_n \sim n/{g_{\rm YM}}$ ($n \in $~\textbf{Z})  where $g_{\rm YM}$  
is the Yang-Mills (gauge)  coupling of the spontaneously-broken gauge theory that accommodates `t Hooft-Polyakov 
monopole states.  The  interactions of the D-matter are proportional  to  $g_{\rm YM} \sim \sqrt{g_s}$, where 
$g_s = O(1)$ is the string coupling and are  thus perturbative, with 
no magnetic charge in general.

Nevertheless, in the modern context of brane-inspired gauge theories, one can construct brane states that 
have properties  similar  to a  magnetic monopole. Thus, they can have  magnetic charges.  In this  case, they 
would manifest themselves in MoEDAL in a manner similar to  magnetically-charged monopoles. For instance, 
one may consider a D1 brane with  its ends attached to two coincident D3 branes. Such a state corresponds 
to a magnetic monopole  in the SU(2) gauge theory  that lives on the D3-brane world-volume. Such a 
construction is S-dual to the case of an open fundamental string  with its ends attached to the two 
D3 branes (corresponding to $g_s \to 1/g_s$). 

Perturbative, non magnetically-charged LDPs  are weakly-interacting, and  so they could be candidates for dark 
matter, depending on the string scale~\cite{shiu}. In general,  brane defects have spin 
structures, so D-matter states  could be bosonic or fermionic, corresponding to the bosonic or fermionic zero modes
 of D-branes, respectively.

\subsubsection{Electrically-charged D-particles}
 
 Non-magnetically-charged  D-matter  could be produced at colliders and also give  rise to interesting signatures 
 of direct relevance to the MoEDAL experiment. For instance, the excited states of D-matter (${\rm D}^\star$), which 
 formally can be obtained from the LDP by  attaching fundamental open strings, can be electrically-charged, 
 provided one end of the  open string is attached to the D3 brane Universe. Such charged states,
  are supermassive -  compared to  the other states of the SM, which are represented by open strings with
   their ends attached to the brane world with mass given by:
\begin{equation} \label{chargemass}
M^2_{{\rm D}^\star} = M_D^2 + n\, M_s^2  \,
\end{equation}
\noindent
where $n \in $~\textbf{Z}$^+$  is the resonance level, and $M_D \sim M_s/g_s$ is the LDP mass. For 
comparison, we give the masses of the towers corresponding to conventional string resonances,  Kaluza-Klein (KK) 
states and   winding modes~\cite{shiu}: 
\begin{center}
\begin{multline}
M_{\rm resonances} = g_s M_s^2 + n M_s^2, \,  M^2_{\rm KK} =
 g_s M_s^2 + n^2 M_C^2~,   \, \\
 M^2_{\rm winding} = g_s M_s^2 + n^2\, \frac{M_s^4}{M_c^2},
 \end{multline} 
 \end{center}
 where 
 $M_c = R^{-1}$ is the compactification scale.    We thus observe that the various towers of new particle excitations 
 have distinct patterns, and for low string scales $M_s$   of order of a TeV, they can all have distinct signatures at
  colliders. For typical string couplings of phenomenological  relevance, e.g. $g_s \sim 0.6$, and $M_s = 
  \mathcal{O}(1)$~TeV, we have that the first few massive levels may be accessible to the  LHC. 

Depending on the details of the microscopic model considered, and the way the SM is embedded,  such 
massive charged states can be relatively long-lived, thereby playing a role  analogous to the possible
long-lived charged particles in low-energy supersymmetric models, and could likewise be detectable in the MoEDAL detector. 

The coupling of D-matter to SM excitations can be understood as follows. As already mentioned, in the 
brane world set-up, the SM fields are open-string excitations with both endpoints of the open strings 
attached to the same D$p'$-brane. These open string states are  denoted by $C_{p'p'}$. There are also open strings
that stretch between the propagating D-matter state  and the background D$p'$-brane, and we denote these states
 by $C_{0 p'}$. 
 
 As illustrated in Fig.~\ref{fig:ddg},  the $C_{0 p'}$ states couple to the SM fields in the 
 $C_{p' p'}$ sector. Hence there is an effective coupling of D-matter to  Standard Model fields, e.g., gauge bosons,
 via the interactions between perturbative open-string states. The D-matter states are charged under 
 the gauge groups localized on the D$p'$-branes. Therefore, they  couple with matter fields on the branes through 
 gauge interactions.
 
\begin{figure}[ht]
\centering
\includegraphics[width=0.6\textwidth]{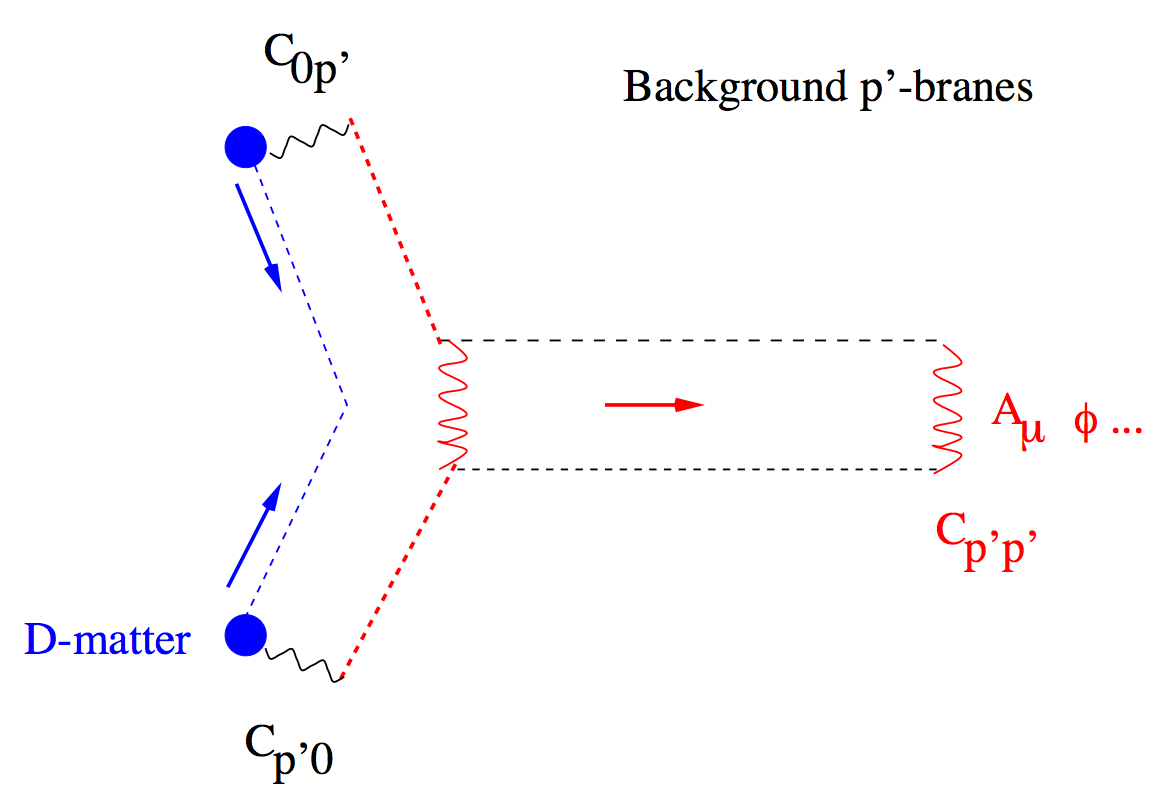}
\caption{ Interactions between D-matter via perturbative string states, which describe SM 
excitations. The enveloping rectangle denotes a D$p'$ brane with $p'$ uncompactified longitudinal dimensions. The $C_{0 p'}$ 
denotes open strings stretched between the D-particles (D0-branes) and the D$p'$ brane. Finally, the $C_{p' \, p'}$ denote open 
string excitations with their ends attached on the D$p'$ brane world (Picture taken from ref.~\protect \cite{shiu}).}\label{fig:ddg}
\end{figure}

Such trilinear couplings between $D-{\overline{D}}$ pairs and SM gauge bosons imply the production of 
D-matter in SM particle collisions, such as the quark-antiquark process indicated in Fig.~\ref{fig:dproduction1}, 
which is a typical process used in dark matter searches at the LHC.
\begin{figure}[ht]
\centering
\subfigure[Signal]{ 
\includegraphics[width=0.40\textwidth]{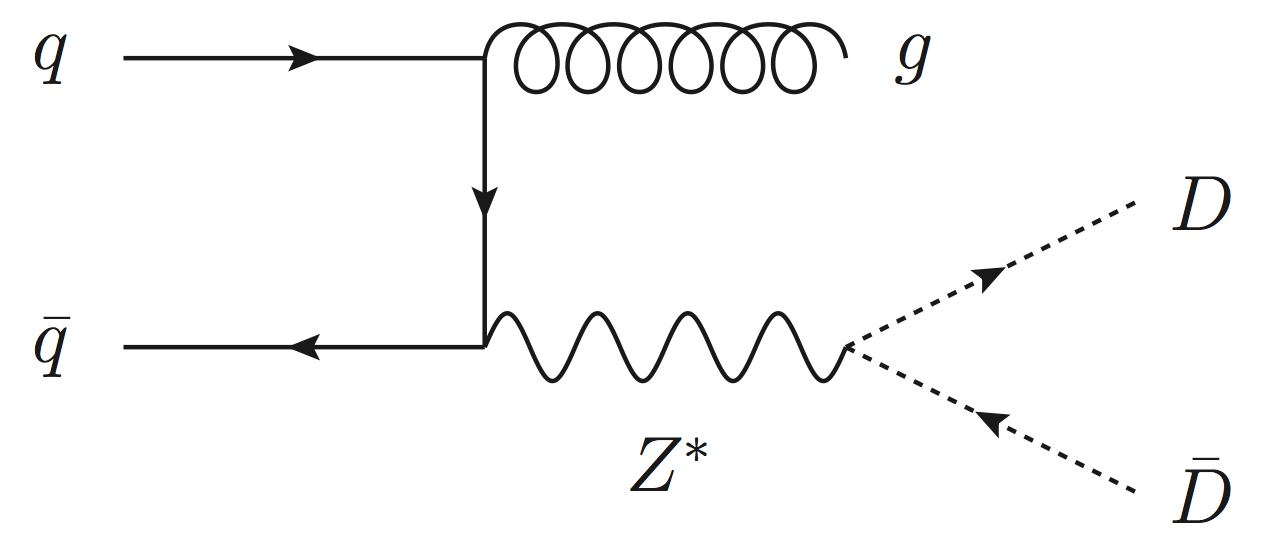}
\label{sfig:signal}
}
\quad
\subfigure[DM production]{
\includegraphics[width=0.40\textwidth]{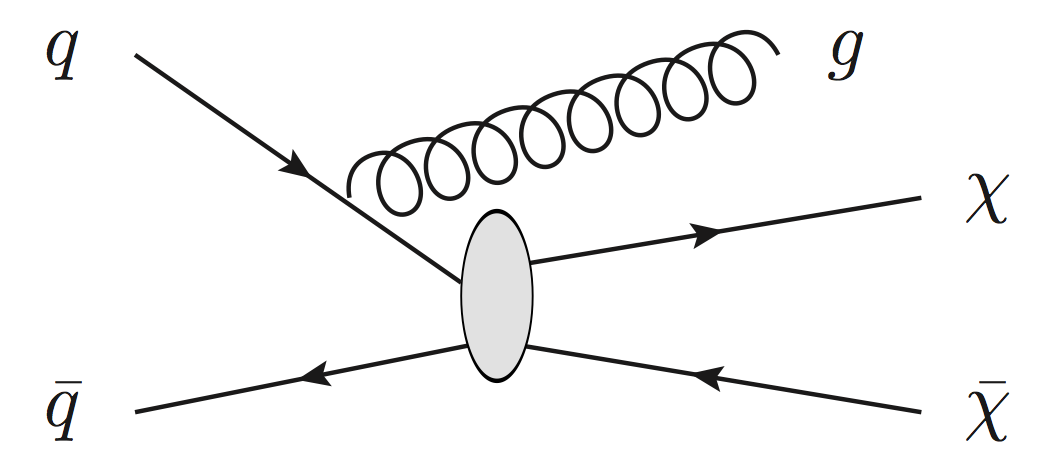}
\label{sfig:DM-prod}
}
~~~~
\subfigure[$Z\to\nu\nu$ background]{
\includegraphics[width=0.40\textwidth]{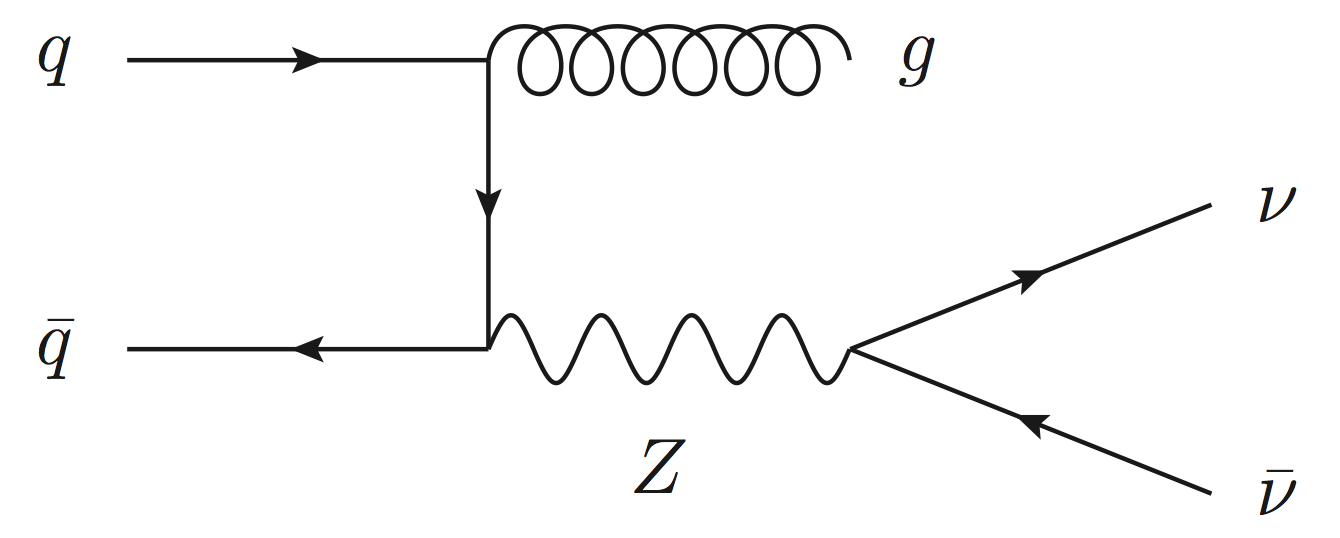}
\label{sfig:bkg1}
}
\quad
\subfigure[$W\to\ell_{\text inv}\nu$ background]{
\includegraphics[width=0.40\textwidth]{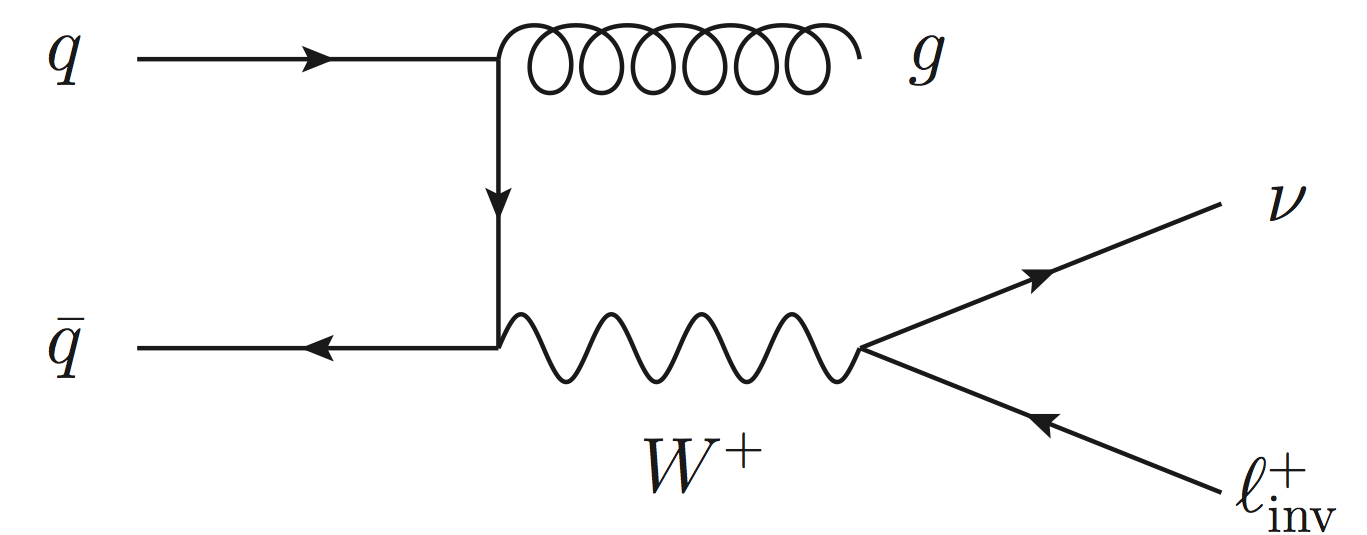}
\label{sfig:bkg2}
}
\caption{(a) Feynman diagrams at the parton level for the production of D-particles by, 
say, q$\overline{{\rm q}}$ collisions  in a generic D-matter low-energy model. This is only 
an example~\protect \cite{mitsou}. There are many other processes  for the production of D-matter from SM
boson decays or gauge boson fusion, which we do not consider in our qualitative   discussion
 here. (b) Production of conventional dark-matter particle-antiparticle ($\chi\bar{\chi}$)
  in effective field theories,   assuming that dark matter, which may co-exist with D-matter, 
  couples to quarks via higher-dimensional contact   interactions~\protect \cite{kolbmaverick,lhcmaverick}.  (c), (d)
   The dominant background processes within the 
 Standard Model framework.}
\label{fig:dproduction1}
\end{figure}
In a low-energy, string-inspired, effective field-theory action,  the leading 
interactions of the D-particles with Standard Model  matter are provided 
by terms with the generic structure (omitting Lorentz derivative or Dirac-matrix 
structures for brevity)~\cite{shiu}:
  \begin{equation} \label{eqn:waysD1}
   \propto g_{D} \, {\overline {\rm D}} \, {\rm D} \, {\rm (Gauge~Bosons}).
   \end{equation}
The symbol $\propto $ in front of each type of interaction is included to denote form factors that arise from tree-level
string amplitude calculations~\cite{shiu}.  As a result of (\ref{eqn:waysD1}), for instance, one may have the graphs of 
Fig.~\ref{sfig:signal}, arising from quark/antiquark scattering. 

D-matter/antimatter pairs can be produced~\cite{mitsou} by the decay of intermediate off-shell $Z$-bosons, which is in agreement with (\ref{eqn:waysD1}). The D-matter pairs produced in a hadron collider will traverse the detector and exit undetected, as they are only weakly interacting, giving rise to large transverse missing energy, $E{\rm _T^{miss}}$. Hence mono-$X$ analyses, targeting DM-pair production plus an initial-state-radiation jet, photon or gauge boson, such as the one shown in Fig.~\ref{sfig:DM-prod}, would be of highly relevant investigative tool. 
 
The dominant SM background in such searches involves the decay of a $Z$ to a neutrino pair and of $W^+$ to a lost (``invisible'' $\ell_{\rm inv}$) lepton and a neutrino, as depicted in  Figs.~\ref{sfig:bkg1} and~\ref{sfig:bkg2}, respectively. Such searches have been performed by the ATLAS and CMS 
 experiments of the LHC at centre-of-mass energies of 7 and 8~TeV giving null results. For example,
the null results found using, e.g., the ATLAS detector in searches for missing energy + jets, impose, in the context of our model, a 
bound on the dark matter mass $m_\chi$ and the D-particle coupling $g_{37}$. With the LHC data at $\sqrt{s}=8$~TeV 
and 20~fb$^{-1}$ integrated luminosity, the current lower bounds on the DM mass set by ATLAS using a mono-$W/Z$ analysis is ${\cal O}(1$~TeV)~\cite{atlas-dm2014}, improving significantly earlier bounds~\cite{lhcmaverick}. 
The full LHC potential ($\sim 300~{\rm fb}^{-1}$ at $\sqrt{s}=14$~TeV) 
will strengthen greatly the constraints on the parameters of TeV-scale D-particles. On the other hand,
 as mentioned previously, excited states of such a LDP, of mass $M_{\rm D}^\star$, involving stretched 
 strings between the D-particle and the brane world, can be charged and thus highly-ionising. Thus, they  are of relevance 
 to the MoEDAL detector searches, provided the string scale is sufficiently low.

\section{Highly-Ionizing Particles in Other Scenarios}

We now turn to a brief overview of exotic possibilities for (meta-)stable massive particle (SMP) states in
scenarios for physics beyond the SM (BSM) other than those that arise in supersymmetric or extra-dimensional scenarios. 
 We cannot be complete -  the model space is far too large - nor do we give a detailed 
 discussion of each scenario, but we hope to illustrate the spectrum of ideas that are relevant
 to the MoEDAL experiment, and   where possible point the reader to relevant literature 
 where further information can be found. 

\subsection{Long-lived heavy quarks}

A number of   models that predict  new heavy particles beyond the top quark are still consistent with
current experimental measurements. For example,  vector-like quarks -  those 
whose different chiral  components  transform identically under the electroweak 
gauge group -  are a common feature of BSM scenarios \cite{frampton}, including
extra-dimensional models, grand unified models and  little Higgs models.  However, those 
models involving a fourth sequential family of  quarks are now disfavoured by the recent 
results on the  the Higgs boson \cite{eberhardt}. 
 
 Non-chiral quarks can also  decouple in the heavy-mass limit, leading to SM-like signals. In models 
 where the mixings   of these states with the light SM fermions are suppressed, the Higgs production rates
  would not be easily distinguishable from the  SM expectations \cite{dawson}. These new quarks could
   be long-lived enough to be effectively stable as far as collider detectors are concerned.
 The compact nature of the MoEDAL detector would allow particles, with  lifetimes of the order of 
 nanoseconds, to be detected as effectively  ``stable''  highly ionizing particles.

It is important to recall that particles with nanosecond lifetimes or more would  hadronize, allowing for a  rich spectrum 
 of new heavy and exotic bound states. Such pseudo-stable, or stable,  massive particles are 
 hypothesized  in many new physics models, either due to  the fact that  the decays are suppressed 
 by kinematics or small couplings \cite{Fairbairn07} or because of  new conserved quantum 
number such as  $R$-parity in supersymmetric models.

\begin{table}[htdp]
\tbl{Vector-like multiplets allowed to mix with the SM quarks through Yukawa couplings. The electric charge is the
sum of the third component of isospin $T_3$ and of the hyper- charge $Y$.}{\begin{tabular}{|c|ccccccc|} \hline
$Q_{q}$   &     $T_{2/3}$  & $B_{-1/3}$  
& 
$
\begin{pmatrix}
X_{5/3} \\
T_{2/3} \\
\end{pmatrix}
$
&
$
 \begin{pmatrix} 
  T_{2/3} \\
   B_{-1/3} \\
\end{pmatrix}
$
&   
$
\begin{pmatrix}
B_{-1/3}\\
Y_{-4/3} \\
\end{pmatrix}
$
&
$
\begin{pmatrix}
X_{5/3} \\
T_{2/3} \\
B_{-1/3} \\
\end{pmatrix}
$
  &   
  $
  \begin{pmatrix}
  T_{2/3} \\
  B_{-1/3} \\
  Y_{-1/3} \\
  \end{pmatrix}
  $
  
         \\ \hline

$T_{3}$   &   0    &    0    &   1/2    &   1/2     &   1/2     &   1    &     1       \\ \hline
$Y$          & 2/3   & -1/3  &   7/6    &  1/6      &  -5/6     &  2/3 &  -1/3    \\ \hline
\end{tabular}
\label{tab:quarks}}
\end{table}

Such new states  can only mix with the  SM quarks through a limited number of gauge-invariant couplings. 
Classifying them into SU(2)$_{L}$ multiplets, their Yukawa terms only allow for seven distinct possibilities, i.e., the two singlets, 
the three doublets and the two triplets displayed in Table~\ref{tab:quarks},
whose notation is adapted from Ref.~\cite{aguilla}.

 The production of  new heavy quarks, either chiral or vector-like, is usually assumed to proceed
at the LHC dominantly through gluon fusion, $gg  \rightarrow QQ$. But electroweak single production can also provide 
an alternative mechanism, as it is not a affected by the large phase-space suppression of pair-production. 
For example, new heavy quarks can be produced singly in favour-changing processes via the electroweak 
interaction through, $q_{i} ^{(}\overline{q}_{j}^{)} \rightarrow V^{*} \rightarrow q_{k}Q $, where  $V = W, Z$ 
\cite{campbelletal,campbelletal1, buchkremer}}. Single electroweak production is also a promising discovery channel for new heavy quark
searches with $m_{Q}  \sim$ 1 TeV/c$^{2}$.

 The  partial decay width for a new sequential heavy quark $Q$ decaying on-shell to a light quark $q$
 through a charged current, assuming
$m_{Q}  \gg m_{q} $,   can be written as:
\begin{equation}
\Gamma(Q \rightarrow qW)  \approx 170.5  |\kappa_{Qg}|^{2} \dfrac{m^{3}_{Q}}{m^{3}_{W}}
\end{equation}
where $\kappa_{Qq}$ signifies the generic Q-g quark coupling, equal to the Cabibbo-Kobayashi-Maskawa
(CKM) matrix element for a new 
sequential family of quarks.

The classification given in Table~\ref{Tab:parameters}  summarizes the long-lived quark possibilities 
corresponding to the small mixing scenarios considered here. Scenario (a)  defines the short-lived scenario  
with no experimental differences compared to the direct searches carried out in the general purpose LHC detectors. 
The second scenario (b) arises for intermediate decay  lengths ranging between a few microns and centimetre distances. 
Possible exceptions are as follows. In  the case of  new  heavy multiplets with sizeable mass splittings,  
as allowed in extra generation models and possible extensions \cite{buchkremer, buchkremeretal, buchkremeretal1, buchkremeretal2},    if  
$m_{Q_{1}}  \lesssim  m_{Q_{2}} + m_{V}$, the heaviest quark  $Q_{1}$ can be short-lived and decay semi-weakly  
to $Q_{2}V^{(*)}$, while the lightest partner is  likely stable if all its decay modes suffer severe suppression. 
On the other hand If   $m_{Q_{1}} \simeq  m_{Q_{2}}$, all   heavy-to-heavy transitions are suppressed 
and both quarks   could be long-lived.

\begin{table}[htdp]
\tbl{Possible decay signatures for new long-lived quarks with masses $m_Q \sim 1$~TeV, as discussed in the text.}{\begin{tabular}{|c|ccc|} \hline
      &    (a)     &  (b)    &    (c)      \\
$|\kappa_{Qq}|$ & $\gtrsim 10^{-7}$   & 
$
\begin{array}{cc}
\gtrsim 10^{-7} \\
\gtrsim 10^{-9} \\
\end{array}
$
&  $\gtrsim 10^{-9}$  \\ \hline

$\Gamma(GeV)|$ & $\gtrsim 10^{-12}$   & 
$
\begin{array}{cc}
\gtrsim 10^{-12} \\
\gtrsim 10^{-16} \\
\end{array}
$
&  $\gtrsim 10^{-16}$  \\ \hline

$\tau(s) $ & $\lesssim 10^{-13}$   & 
$
\begin{array}{cc}
\gtrsim 10^{-13} \\
\lesssim 10^{-9} \\
\end{array}
$
&  $\gtrsim 10^{-9}$  \\ \hline
\end{tabular}
\label{Tab:parameters}}
\end{table}%

Scenario (c) of Table~\ref{Tab:parameters} describes long-lived  particles with decay lengths larger than 
standard LHC detector  dimensions.  If all heavy quark couplings with SM fermions are below the 10$^{-9}$ level, the 
stable case becomes a relevant scenario, perhaps  in conjunction with events with displaced vertices. if all
$Q - q$  quark couplings to the SM families are less  than around 10$^{-2}$, such new heavy 
fermions could hadronize. As a result, annihilation decays and hadronic transitions between 
the formed bound states would dominate. 

While $Q\overline{Q}$ quarkonium resonances would be impossible to observe in MoEDAL,
 the possibilities for  ``open-favour'' hadrons $Q\overline{q}$, ($\overline{Q}q$) and $Qqq$ ($\overline{Qqq}$)
with  $q$ being a light SM quark, provide a wide  spectrum of new heavy mesons and
baryons that could be sought at the LHC.  Such states can form by capturing a light quark
partner and transforming  as they pass through the detector into various slow-moving
heavy states. Table~\ref{Tab:newstates} lists these states assuming that they hadronize with $u$, $d$
and $s$ quarks. 

\begin{table}[htdp]
\tbl{Possible mesons and baryons  involving $Q = X_{5/3}, T_{2/3}, B_{-1/3}$ and $Y_{-4/3}$ vector-like quarks. The
states in bold font are hadrons whose yields are expected to be substantial at the LHC, as predicted in \protect 
\cite{mackeprang} for penetration depths between 0 and 3 meters. Only the neutral and positively-charged hadrons
 are displayed.}{\begin{tabular}{|c|c|c|} \hline
Chrg  &                                     Mesons                                                                        &                               Baryons     \\ 
Q=0    & $ \mathbf{T\bar{u}, \bar{T}u, B\bar{d},  \bar{B}d}, B\bar{s}, \bar{B}s$    & $Tdd, Tds, Tss, \mathbf{Bud}, \overline{Bud},
Bus, \overline{Bus}, Yuu, \overline{Yuu}$ \\
Q=1  & $ \mathbf{X\bar{u}, T\bar{d}},  T\bar{s}, \mathbf{ \bar{B}u, \bar{Y}d}, \bar{Y}s$   & $Xdd, Xds, Xss, \mathbf{Tud}, Tus, Buu, \overline{Bdd},
\overline{Bds}, \overline{Bss}, \overline{Yud}, \overline{Yus}$ \\ 
Q=2 &  $ \mathbf{X\bar{d}}, X\bar{s}, \bold{\bar{Y}u}$  &  $\mathbf{Xud}, Xus, Tuu, \overline{Ydd}, \overline{Yds}, \overline{Yss}$ \\
Q=3 &    -   &  $Xuu $   \\ \hline
\end{tabular}
\label{Tab:newstates}}
\end{table}%

Interestingly, the interactions of such open-flavour mesons with the material would be similar to those of
$R$-hadrons. As they move through and interact with  the detector material, most of the new states 
convert into baryons, allowing for $Quu, Qud$ and $Qdd$ states, as new heavy mesons are kinematically 
favoured to increase their baryon number by emitting one or more pions \cite{mackeprang}.

Interestingly, new mesons that convert at the beginning of the scattering chain could generate intermittent tracks,
disappearing and reappearing,  signalling possible baryonic or electric charge exchange. 
Indeed, $ Q\bar{q}$ and $\bar{Q}q$ bound states traversing a medium composed of light quarks likely 
 flip their electric charge,  frequently interchanging their parton constituents with those of the material nuclei. 
 
The pseudo-stable hadrons described here would lead to observable tracks due to the high ionization energy losses, allowing
for signatures similar to slow-moving muons with high transverse momentum. Searches for pseudo-stable charged particles,
with a highly-ionizing signature therefore provide a promising strategy to rule out or confirm the possibilities for novel exotic quarks
with long lifetimes. Such signatures are accessible to the MoEDAL detector.  

Limits derived from ATLAS and CMS searches \cite{buchkremer1}, can be placed on 
searches for heavy and long-lived quarks. For example,  ATLAS results \cite{ATLASHQ}  for 
$tW$  final states rules out cross-sections of $\sigma BR(Q \rightarrow  tW) >  $ 0.05 pb for lifetimes
of 10$^{11}  < \tau < $   3 $\times$ 10$^{-10}$s (3 mm$<  c\tau  <  $ 10 cm),
corresponding to an excluded heavy quark mass of $m_{Q}  <   650$~GeV.

The detection of an SMP in the multipurpose LHC detectors requires
the heavy state to propagate through the full tracking detector. The reinterpretation \cite{buchkremer}
of the CMS search \cite{CMSHQ}, assuming similar behaviour as for scalar long-lived top quarks, 
excludes cross-sections of $\sigma > $ 3 fb for very long lifetimes  greater than 10$^{-7}$ s ( $c\tau > $ 30 m). 
The mass limit $m_{Q}  <  $ 800 GeV/c$^{2}$  is still valid for lifetimes $\tau >$ 10$^{-8}$s ($c\tau > $ 3 m).

An important caveat in this discussion is that  these limits are based on the assumption  that  the new heavy quarks are
pair-produced, with kinematics independent of the lifetime of the particle \cite{buchkremer1}. But, as 
discussed in \cite{johansen}, the squark pair-production from light quarks can dominate the gluon 
fusion mechanism for large masses, owing  to the diminishing  probability for the gluons to carry
 a  large fraction of the proton momentum. Hence the production channels then become more model-dependent
 for increasing masses, requiring  dedicated MC simulations.

\subsection{Massive (pseudo-)stable particles from vector-like confinement}

A plausible extension of of the SM is obtained by adding new fermions in vector-like 
representations of the SM gauge groups, with a mass scale within the reach of the LHC. One can easily imagine
such fermions as remnants of  physics at a higher energy scale. The advantage of, say,  a new vector-like 
fermion is that it can have mass without coupling to electroweak symmetry breaking and affecting the
precision electroweak observables.  Such new fermions may also interact via new gauge forces, which 
may be weakly or strongly coupled.

In  the case where these  new fermions also feel a new strong gauge
 force, in addition to SM  gauge forces,  that confines at TeV energies, the  
 phenomenology is drastically different from that of the Standard Model.  
 Vector-like confinement augments the SM at the TeV scale in the same manner that
  QCD augments QED at the GeV scale. A new confining gauge interaction (hypercolour) 
  and new vector-like fermions (hyperquarks) are added to
the SM. 
  
  Hyperquarks are assumed to be light compared to the hypercolor confinement scale,
in a way that is analogous to the $u, d$ and $s$ quarks, which are light compared to the QCD confinement
scale. The hyperquarks, once pair-produced, rapidly form bound states due to hypercolor
confinement, allowing for both resonant and pair-production of  hypermesons  at the LHC, in 
a  way that is analogous to the resonant and pair-production of mesons and pions at a low-energy 
 $ e^{+}$ - $e^{-}$  collider.
 
 There are many possible models of vector-like confinement. However, the presence of  a spin-1
 bound state $\tilde{\rho}$ - analogous to the QCD $\rho$ meson -  and a pseudoscalar bound state $\tilde{\pi}$ - analogous
to the QCD pion - are completely general.  The scenario of vectorlike confinement
is discussed in detail elsewhere \cite{vectorlike}. The spin-1 resonances
decay predominantly into a pair of pseudoscalar bound states, just as the $\rho \rightarrow \pi^{+}\pi^{-}$.
 The resonant  $\tilde{\rho}$ production followed by $\tilde{\rho} \rightarrow \tilde{\pi}^{+}\tilde{\pi}^{-}$
is  the signature process of vector-like confinement. 
  
\begin{figure}[ht]
\centering
\includegraphics[width=0.8\textwidth]{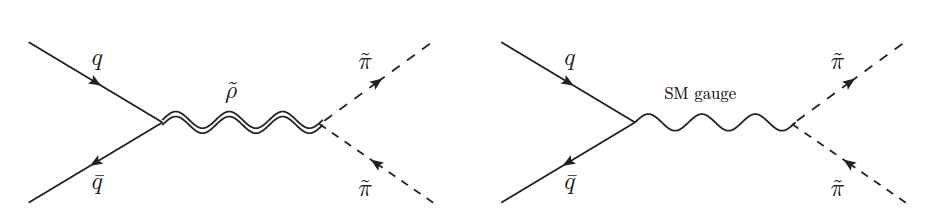}
\caption{On the left we show the diagram for the production of a $\tilde{\pi}$ pair via a 
$\tilde{\rho}$ resonance. On the right we show the diagram for Drell-Yan production of a $\tilde{\pi}$ pair. }
\label{fig:vectorlike}
\end{figure}    

A generic  phenomenological feature  in vector-like confinement models is
the existence of charged and/or colored massive pseudoscalars, or hyperpions, 
 that are stable on collider time
scales  \cite{KILIC}. If coloured, such hyperpions will hadronize with quarks and gluons,  thereby forming a massive stable
hadron, like an $R$-hadron,  which will carry a net electric charge some fraction of the time. 
Charged long-lived colour-neutral pseudoscalars and ``R-hadrons''  can be pair  produced via
a Drell-Yan process as well as the decay of a spin-1 resonance, as shown in Fig.~\ref{fig:vectorlike} (left),  
or  through an $s$-channel gluon, as in Fig.~\ref{fig:vectorlike} (right) \cite{KILIC}. 
In both cases the produced particles can be stable on collider distance and time scales. 
In many  vector-like confinement scenarios these massive stable particles are 
sufficiently slow-moving  ($\beta \lesssim$ 0.2)  to be detected by  MoEDAL.

  
 
 \subsection{Fourth-generation fermions}

An intriguing way to address the big hierarchy problem of the Standard Model is to introduce a
a new fourth family of leptons. Although  models that involve a fourth sequential quark are now
disfavoured by recent results on the Higgs boson \cite{eberhardt}.   A natural method to 
accomplish such a scenario is to have the Higgs
itself be a composite of these new fermions. This  setup was  investigated \cite{SANINNO} 
using as a template minimal walking technicolor \cite{MWT} with  a general heavy neutrino mass structure with
and without mixing with the SM families. By imposing an exactly-conserved $\varepsilon$-lepton number
the mixing between the fourth-generation $\varepsilon$ neutrino and the three light neutrinos can be forbidden. On the
other hand, the possibility that the new heavy leptons mix with the Standard Model leptons can also be allowed.

It is certainly possible that the neutrino will be the heavier than the charged leptons.
 In this case, the charged lepton can only decay through mixing with lighter generations, 
 and might thus be extremely long-lived. Collider signatures of long-lived charged leptons could
either be displaced vertices or, if the charged lepton decays outside the detector, a muonlike signal.
In the event that the heavy lepton is slow-moving it will be highly-ionizing and thus can be
detected by the MoEDAL experiment.

\subsection{`Terafermions'  from a  `sinister'  extension of the Standard Model}

In a model introduced by Glashow and Cohen \cite{SINISTER} based on the gauge group 
SU(3) $\times $ SU(2) $\times $ SU(2)$'$ $\times $ U(1), the  quarks and leptons of the Standard Model 
are accompanied by an equal number of ``terafermions''.
 An unconventional CP operation called CP$'$,  maps ordinary fermions into the conventional CP 
 conjugates of their tera-equivalents, and vice versa. This `sinister' (i.e., `left-left symmetric') 
 model is akin to certain `left-right symmetric' models for  which an unconventional space-reflection
  operation P$'$, rather than CP$'$, links ordinary and exotic fermions. 
 
 The model also involves heavy versions of   the weak intermediaries: W$'$ and Z$' $ bosons. 
 Soft CP$'$ breaking within a simple Higgs sector  (comprised of a  SU(2) doublet and a SU(2)$'$ doublet)
  leads to large and experimentally allowed  masses for the terafermions, and for the W$'$ and Z$'$. This
  model resolves two of the problems of the SM: the mass hierarchy and the strong CP problem. 
  A natural seesaw mechanism allows the observed neutrinos to have very small Dirac masses.
  This extension of the SM predicts the existence of novel heavy stable quarks and leptons that,
   through the formation of electromagnetically-bound states (``terahelium''), yield candidates for dark matter.
  
  In this model the tera-electron is the least massive charged terafermion. The other charged teraleptons
  are unstable, decaying rapidly via W$'$ exchange. The lightest teraquark, $U$, is stable with an estimated
  mass of a few TeV. The heavier teraquark ,  $D$,  is unstable, decaying for example via
  $D \rightarrow U  + E^{-} + \nu'_{e}$. Although the electric  charge of the  $U$ quark is only 2/3, the
slow-moving massive $U$  particle could have sufficient ionization to be detectable by MoEDAL.

\subsection{A massive particle from a simple extension to the SM}

The SM is  based on the gauge group SU(3)$\times$ SU(2) $\times$ U(1), which has 
quarks that transform non-trivially under all the three factor groups SU(3), SU(2) and
U(1). On the other hand, leptons transform non-trivially under only  two of these groups, namely SU(2) and U(1).
However, it is  possible to have a fermion that transforms nontrivially under U(1) \cite{RAJASEKARAN}. It is 
an electrically-charged fermion that does not have weak decays or strong interactions. If
in addition it is stable, it will be of interest to MoEDAL. Absolute stability in the case of a singly-charged
 particle is not possible if the Standard Model  Higgs doublet exists, unless a discrete symmetry is imposed. 
  However, in the scenarios described in Ref. \cite{RAJASEKARAN}  there is no need of the discrete symmetry.
   Conservation of the U(1) hyper- charge itself forbids the Higgs coupling of   as well as the off-diagonal mass 
   terms connecting  with the right-handed charged leptons. Thus,  a  doubly charged particle  is automatically stable 
   and potentially detectable by MoEDAL.

\subsection{Fractionally charged massive particles}

 \begin{figure}[ht]
\centering
\includegraphics[width=0.7\textwidth]{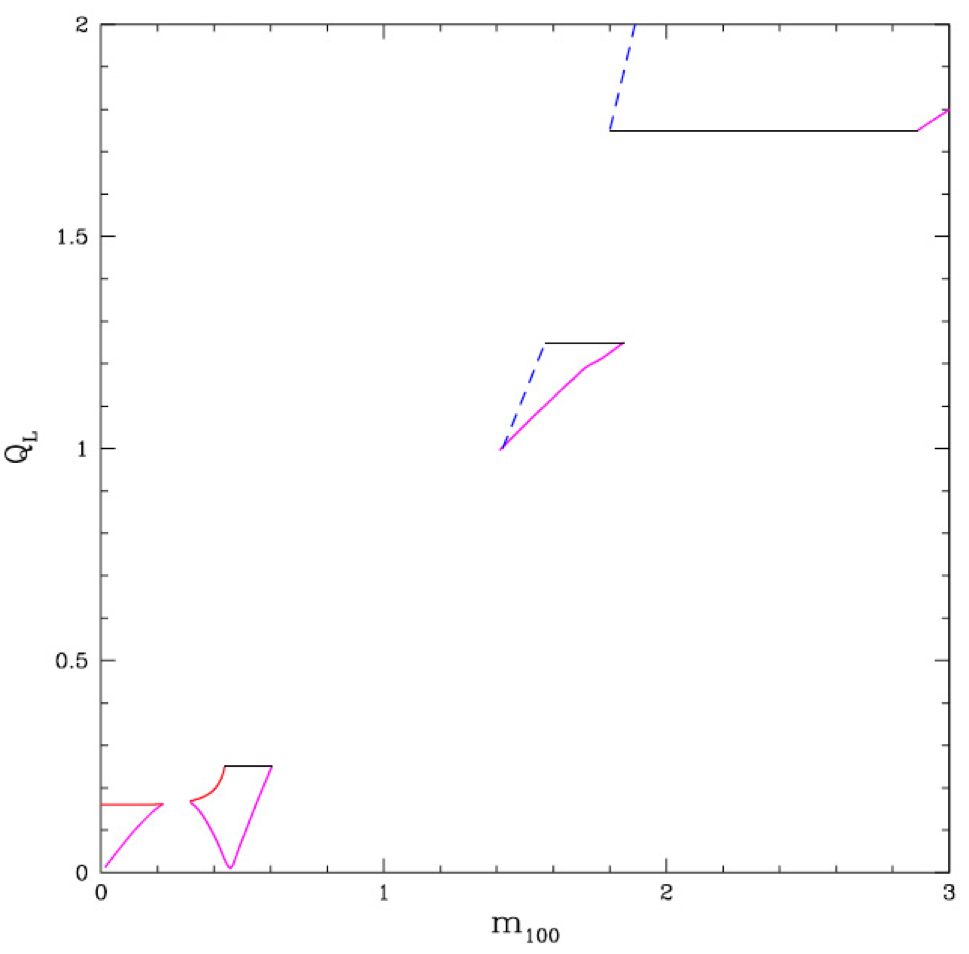}
\caption{Currently-allowed windows for FCHAMPs, consistent with the relic density, accelerator, and $Z$ width constraints, 
for which $|Q_{L} - n| \le 0.25$, $n = 0,1,2$, from Ref.~\protect \cite{fchamp}. 
The FCHAMP mass, $m_{L} \equiv$ 100m$_{100}$~GeV.}
\label{fig:fchamp}
\end{figure}    
A final topic we discuss briefly Fractionally Charged Massive Particles (FCHAMPS). Such particles 
exist in extensions of the SM that can be obtained in the low-energy limit of superstring theories.
 The lightest FCHAMP would be stable, and any of them produced during the Early Universe would be present today. 
 In \cite{fchamp}, the thermal production, annihilation and, survival of an FCHAMP, a lepton with electroweak, i.e., with non-trivial hypercharge $U(1)_Y$, but no strong interactions, of mass  $m$ and charge  $Q_L $ (in units of the electron charge)
 have been explored, taking into account standard cosmological constraints coming from primordial nucleosynthesis and 
 cosmic microwave background radiation. In addition, the  invisible width of the $Z$-boson of the Standard Model can be 
 used to provide constraints on the FCHAMP charge-mass relation. The surviving FCHAMP abundance on Earth 
 is orders of magnitude higher than the limits from terrestrial searches for fractionally-charged particles, 
 closing the window on FCHAMPs with $Q_{L} \ge 0.01$.  However, as $Q_L$ ~approaches an integer ( $|Q_{L} - n| \le 0.25$),
 these searches become increasingly insensitive, leaving some potentially unconstrained ``islands" in the charge-mass 
 plane (see Fig.~\ref{fig:fchamp}), which may be explored by searching for FCHAMPs in the cosmic rays and also at the LHC. 
 The fact that we have a large island with charge near $2e$ makes FCHAMPs candidates for detection by the MoEDAL detector.

\section{Scenarios with Doubly-Charged Massive Stable Particles}

Doubly-charged particles appear in many BSM scenarios. 
As examples, doubly-charged scalar states, often dubbed 
doubly-charged Higgs fields, appear in left-right symmetric models \cite{Pati1974yy, LRSM, LRSM1,LRSM2}  and
in see-saw models for neutrino masses with Higgs triplets \cite{seesaw,seesaw1,seesaw2,seesaw3,seesaw4,
seesaw5,seesaw6,seesaw7,seesaw8,seesaw9,seesaw10,seesaw11,seesaw12,seesaw13,seesaw14}. 
Doubly-charged fermions can appear in extra-dimensional models including
new physics models inspired by string 
 theories \cite{strings},
and as the supersymmetric partners of 
 the doubly-charged scalar fields in supersymmetric extensions of left-right symmetric
 models~\cite{ susylr,susylr1,susylr2,susylr3}.  Finally, models of new physics  with an extended
gauge group often include doubly-charged vector bosons \cite{extendedgg,extendedgg1,extendedgg2,extendedgg3,extendedgg4}.
It is also possible to consider vector states with double electric charge 
independently of any gauge-group structure, as in models with non-commutative
geometry or in composite or technicolor theories \cite{noncommtc, noncommtc1,
noncommtc2,noncommtc3,noncommtc4,noncommtc5,noncommtc6,WTCEF}. 

A general effective Lagrangian analysis of production and decay rates of doubly 
charged exotic particles (scalars, fermions and vectors) at the LHC indicates
 promising channels and distinct signatures and shows how to distinguish among 
 particles with different spins and $SU(2)_L$ representations \cite{Alloul:2013raa}.
We consider such new physics scenarios with potentially massive long-lived 
doubly-charged particles in more detail below. Such particles would give rise to
highly-ionizing particles that could be detected by the MoEDAL detector.

\subsection{XY gauginos and warped extra dimension models} 

Models that address supersymmetric  grand unification in warped extra dimensions
with ``GUT parity''~\cite{xygaugino1,xygaugino2,xygaugino2a,xygaugino3}  were
 introduced to alleviate the problems of the conventional supersymmetric desert picture. 
 In these models the combination of extra dimensions  and effective TeV-scale  
 supersymmetric grand unification results in KK towers, not  only of the SM gauge and Higgs
  fields, but also of their supersymmetric GUT partners, including XY bosons  of Grand Unification as 
  well as coloured Higgs multiplets. A parity can be
   chosen such that the MSSM particles are even and their GUT partners odd, 
hence the lightest ``GUT-odd'' particle (LGP) is stable or long-lived if this quantum number is 
conserved or approximately conserved, respectively.

 In  an early model   involving gauge coupling unification of the four-dimensional MSSM \cite{xygaugino1}
  we have a  scenario where the LGP is   a light isospin-up
(-down) colour-triplet XY gaugino, with electric charge -1/3 (-4/3).  Since
the XY gauginos are coloured, they hadronize by picking up an up or down quark, making
neutral or charged mesons $T^{0} \equiv  \tilde{X}\uparrow\bar{d}$,  $T^{-}  \equiv  \tilde{X}\uparrow\bar{u}$, 
$T'^{-} \equiv \tilde{X}\downarrow\bar{d}$ and $T^{ - - } \equiv  \tilde{X}\downarrow\bar{u}$, where $\tilde{X}\uparrow$
and $\tilde{X}\downarrow$ are the isospin up and down components of the XY gaugino doublets, respectively.
Among these mesons, the lightest one is either $T^{0}$ or $T^{-}$, and the heavier states can decay
through beta processes. However, the decay is slow enough that all the meson states are
effectively stable on collider distance and time scales, and the  singly- and doubly-charged  mesons, which have TeV-scale masses,   
will easily be seen because they leave highly-ionizing tracks capable of being detected by the MoEDAL detector. 
In more recent  five-dimensional models~\cite{xygaugino2,xygaugino2a,xygaugino3} many possibilities for 
highly-ionizing particle production are open.

\subsection{Doubly-charged leptons in the framework of walking technicolor models}

Technicolor models were originally rejected due to some serious shortcomings, for example 
their predictions of large Flavour-Changing Neutral Currents (FCNCs)~\cite{DE}. Also early  technocolour models 
led to a plethora of technimesons~\cite{hole},   for which there was no evidence. However, extended technicolor models
 where the technicolour sector has a  ``walking" behaviour - that is the slow running of the technicolour  gauge coupling
  over an extended range -  do not have a FCNC problem. Indeed, the latest walking technicolour models \cite{WTC,WTC1,WTC2,WTCEF} 
  provide a good description of particle physics phenomena.
 The discovery in 2012 by ATLAS and CMS of a Higgs-like boson with mass approximately 126~GeV~\cite{Higgs-like,Higgs-like1}
 was not generically predicted by walking technicolour models, but can be accommodated by them.

The minimal walking technicolor model \cite{WTC,WTC1,WTC2,WTCEF}  has two techniquarks, 
 U (up) and D (down), that transform under the adjoint representation of an
SU(2) technicolor gauge group. The global symmetry of the model is a SU(4) that breaks spontaneously
to an SO(4), and the chiral condensate of the techniquarks breaks the electroweak symmetry.
There are nine Goldstone bosons emerging from the symmetry breaking, three of which are eaten by the
$W$ and $Z$ bosons. The remaining six Goldstone bosons are $UU$, $UD$, $DD$ composites and their corresponding antiparticles. 
The effective theory of the minimal walking technicolor  model presented here has been  described in detail elsewhere \cite{WTCEF} 
 
The six Goldstone bosons carry technibaryon number since they are made of two techniquarks or two anti-techniquarks. 
If no processes violate technibaryon number, the lightest technibaryon will be stable.
The electric charges of the $UU$, $UD$, and $DD$ boson are given in general by $ n + 1, n $, and $ n - 1$
respectively, where $n$ is an arbitrary real number. 

The model requires in addition the existence of a fourth family of leptons,
\begin{equation}
{\nu'  \choose e'_{L}}  (\nu'_{R}, e'_{R}) ,
\end{equation}
 i.e., a new `neutrino' $\nu'$ and a new  `electron'  $e'$. 
Their hypercharges are  are $(-3y)/2$ and $ (-(3y-1)/2, -(3y+1)/2)$, respectively.
If we take $y = 1$,  the Goldstone bosons $UU$, $UD$, and $DD$ have electric charges 2, 1, and 0,
respectively. Using the convention $Q=T_{3} + Y$, the electric charges of the new lepton $\nu'$ and $e'$ are  -1
and -2, respectively.   

Thus two types of of stable doubly-charged particles can exist in the framework of Minimal
Walking Technicolour, the technibaryon $UU^{++}$ and the technilepton $e'^{++}$. The masses
of these particles are expected to exceed 100 GeV. The MoEDAL detector has a low enough threshold
to detect doubly-charged techniparticles with velocities smaller than around 0.4c.

\subsection{Doubly-charged Higgs bosons in the L-R symmetric model} 

The electroweak gauge symmetry of the SM is broken by the Higgs mechanism, 
which imparts masses to the $W$ and $Z$ bosons, the mediators of the weak forces. A number 
of models mentioned above include additional symmetries and extend the SM Higgs sector 
by introducing doubly-charged Higgs bosons. We consider
here in more detail one good example of such a model, the L-R Symmetric Model \cite{Pati1974yy, LRSM,LRSM1,LRSM2}.

One major puzzle of the SM is the fact that weak interaction couplings are strictly left-handed. In order 
to remedy this apparent arbitrariness of Nature, one can extend the gauge group of the SM 
to include a  right-handed sector. The simplest realization is a Left-Right Symmetric Model  
(LRSM) \cite{Pati1974yy, LRSM}  that postulates  a right-handed version of the weak interaction, whose gauge 
symmetry is spontaneously broken at high mass scale, leading to the parity-violating  SM. This model 
accommodates naturally recent data on neutrino oscillations~\cite{NUOSC} and 
the existence of small neutrino masses~\cite{SMALLNUMASS}.  The model generally requires 
Higgs triplets containing doubly-charged Higgs bosons ($H^{\pm \pm}$)  $\Delta_{R}^{++}$
and $\Delta_{L}^{++}$, which could be light in the minimal 
supersymmetric left-right model \cite{LRSUSY0,LRSUSY1,LRSUSY2}.

Single production of a doubly-charged Higgs boson  at the LHC is possible via vector boson
fusion, or via the fusion of a singly-charged Higgs boson with either a $W^\pm$ or another
singly-charged Higgs boson. The amplitudes of the  $W_{L} W_{L}$ and $W_{R} W_{R}$ vector
boson fusion processes are proportional to $v_{L,R}$, the vacuum expectation values 
of the neutral members of the scalar triplets of the  LRSM . 
 For the case of $\Delta_{R}^{++}$ production, the vector boson fusion process
dominates. For the production process $W^+ W^+ \rightarrow \Delta_{L}^{++}$ , the suppression 
due to the small value of the $v_{L}$  is somewhat compensated by the 
fact that the incoming quarks radiate a
lower-mass vector gauge boson.

Pair production of doubly-charged Higgs bosons is also possible via a Drell-Yan process,
with $\gamma$, Z or Z$_{R}$ exchanged in the s-channel, but at a high kinematic price since substantial
energy is required to produce two heavy particles. In the case of $\Delta_{L}^{++}$, 
double production may nevertheless be the only possibility if $v_{L}$ is very small or vanishing.

The decay of a doubly-charged Higgs boson can proceed via several channels. Dilepton decay
provides a clean signature, kinematically enhanced, but the branching ratios depend on
the unknown Yukawa couplings. Present bounds~\cite{DCHYUKAWACOUPS, DCHYUKAWACOUPS1} on the 
diagonal couplings h$_{ee,\mu\mu,\tau\tau}$ to charged leptons are consistent with values ${\cal O}(1)$ if 
the mass scale of the triplet is large. For the $\Delta_{L}^{++}$, this may be the dominant mode
if $v_{L}$ is very small. One would then have the golden signature $q\bar{q} \rightarrow
\gamma^{*}/Z^{*}/Z^{{\prime}*}  \rightarrow \Delta^{++}_{L}\Delta^{--}_{L} \rightarrow 4l$. 

In the case of very small Yukawa couplings $H_{ll} \lesssim $ 10$^{-8}$, the doubly-charged Higgs boson
could be quasi-stable. In this case  slowly moving  pseudo-stable Higgs bosons could be detected in 
the MoEDAL NTDs. For example with  CR39, one could
detect doubly-charged Higgs particles moving with speeds less than around $\beta \simeq 0.4$.

  \subsection{Doubly Charged Higgsinos in the L-R supersymmetric model}
  
  If the LR symmetric model is extended to include supersymmetry, the emerging model cures 
  some of the outstanding problems of MSSM. It disallows explicit R-parity violation 
  \cite{Kuchimanchi:1993jg}, provides a natural mechanism for generation neutrino 
  masses using Higgs triplet fields that transform as the adjoint representation of the 
  $SU(2)_R$ group. It also  provides a solution to the strong and electroweak  CP problem 
  in MSSM \cite{Mohapatra:1996vg, Mohapatra:1995xd}.
  
  The left-right supersymmetric models predict the existence of the fermionic partners of the 
  doubly charged Higgs bosons, the doubly charged higgsinos. If the scale for left-right symmetry 
  breaking is chosen so that the light neutrinos have the experimentally expected masses, the
   doubly charged higgsinos can be light. Such particles could be produced in abundance and 
   thus give definite signs of left-right symmetry at future colliders like the LHC  and at the linear collider. 
   These particles will be distinguished by certain characteristic signatures in regard to
their lepton and jet spectra in the final state. In particular, they give rise to a distinguishing 
   $4 \ell + \rm E^{miss}_{T}$ \cite{susylr,susylr1,susylr2,susylr3}.  As with doubly charged Higs bosons, doubly charged 
   Higgsinos can be long-lived enough to traverse the MoEDAL and be detected in its NTD system.

\subsection{Doubly-charged leptons in the framework of almost commutative geometry} 

The novel mathematical theory of almost-commutative (AC) geometry \cite{CONNES} 
has been invoked in an attempt to  to unify gauge models with gravity.
 The AC-model \cite{ACmodel,ACmodel1,ACmodel2}, which is based on almost-commutative geometry, 
 extends the fermion content of the SM by two heavy
particles with opposite electromagnetic and weak $Z$-boson charges.  Having no other SM gauge
charges, these particles (AC-fermions) behave as heavy stable leptons 
with charges  -2e and +2e, called here $A$ and $C$, respectively.

The AC-fermions are sterile relative to the SU(2) electroweak interaction, and
do not contribute to SM parameters. The masses
of AC-fermions  originate  from the non-commutative geometry of the
internal space, which is less than the Planck scale, and are not related
to the Higgs mechanism. The mass scale of the A and C fermions is fixed
on cosmological grounds. It was assumed in~\cite{ASSUME100,ASSUME100a}  that the masses of the
$A$ and $C$ leptons are greater than 100~GeV.

 In the absence of AC-fermion mixing with light fermions, AC-fermions can be
absolutely stable. Such absolute stability follows from a new U(1) interaction
of electromagnetic type and  strict conservation of the additional U(1) gauge charge, 
which is called Y-charge and is carried only by AC-leptons.
A heavy doubly-charged lepton with speed $\beta   \lsim$~0.4 would be detectable
by the MoEDAL detector.

\section{Highly-Ionizing  Multi-Particle Excitations}

An intriguing class of highly-ionizing electrically-charged particles is 
that  of multi-particle excitations. We discuss
three examples of such exotic  states -  Q-balls, strangelets and quirks.

\subsection{Q-balls}

 In theories where scalar fields carry a conserved global quantum number $Q$, 
 there can exist non-topological solitons that are stabilized by global charge 
 conservation. They act like homogenous balls of matter, with $Q$ playing the role 
 of the quantum number. Coleman called this type of matter Q-balls \cite{Coleman:1985ki}

The conditions for the existence of absolutely stable Q-balls may be satisfied
in supersymmetric theories with low-energy supersymmetry breaking~\cite{SUSYQballs}. 
The role of conserved quantum number is played in this case by the baryon
number, or by lepton number for sleptonic Q-balls.

These Q-balls can be considered as coherent states of squarks, 
sleptons and Higgs fields. Under certain assumptions about the internal 
self-interactions of these particles and fields, the Q-balls could be absolutely stable~\cite{StableQballs, StableQballs1}. 
Supersymmetric Q-balls fall into two classes: supersymmetric
electrically-neutral solitons (SENS) and supersymmetric
electrically-charged solitons (SECS). 
      
Low-charge, Q-balls  - called Q-beads - have been hypothesized ~\cite{kusenkosolitons}, which are 
also extended objects whose size is large in
comparison to their de Broglie wavelength.  These  could be produced at a collider,
although the  probability of producing them is probably  exponentially reduced
 by the size  of the Q-bead. But the question of the potential observability of Q-beads is
 by no means clear and needs further consideration. If  Q-beads {\it can} be created
in a collider, their signatures could be spectacular. For example, a soliton with both 
$B \ne 0$ and $L \ne 0$ would interact  as a massive leptoquark. Such objects
would  register in the passive detector system of the MoEDAL experiment.

\subsection{Strangelets} 

Strangelets are hypothetical baryonic objects 
consisting of approximately  equal  numbers of $u,d$ and $s$ quarks, which may be stable or 
metastable. The reason for this stability is that the Fermi energy for a 
large enough system with a fixed number of $u$ and $d$ quarks is higher than the 
corresponding system of $u, d$ and $s$ quarks due to the Pauli Principle~\cite{bodmer,witten}.
Calculations with the MIT bag model confirm  that strangelets 
could be stable~\cite{bagmodel,bagmodel1}.

The use of ultra-relativistic heavy ion collisions - for instance at the LHC -  is the only known way of creating
a QGP for laboratory studies and the goal has been for several years to create and
detect the QGP. More generally the motivation for these experiments is to study the thermodynamics
of strongly interacting matter. 
Strangelets could be produced as cooled remnants of a quark-gluon plasma (QGP) through the strangeness
distillation mechanism proposed in~\cite{sdistilling,sdistilling1}. Thus the detection of strangelets could be a signal for the
formation of a QGP. The basic idea in the distillation mechanism is that ${\bar s}$ quarks
produced in the interaction bind with the $u$  and $d$ quarks from the
initial-state nuclei, forming $K^{+}$ and $K^{0}$ mesons. When these mesons evaporate from the
system, they carry away anti-strangeness and entropy. The remaining baryon-rich system
is strongly enhanced with $s$ quarks, which could favour the formation of strangelets.

Coalescence models \cite{COALESCENCE,COALESCENCE1}  provide another production mechanism for strangelets, whereby hypernuclei
produced in heavy-ion interactions could decay to strangelets.
However, it should be pointed out that  the good agreement of measurements 
of particle production at RHIC  with simple thermodynamic models severely constrains  the production of 
strangelets in heavy-ion  collisions at the LHC  \cite{ellisstrange}.

The cross-sections and lifetimes of strangelets have recently been calculated using
the MIT bag model \cite{MITbagmodel}. The estimated lifetimes range from $10^{-5}$ - 10$^{-10}$~s for short-lived
ones to 10$^{-4}$ - 10$^{-5}$ s for long-lived ones. According to this model, charged strangelets are
expected to have a high negative charge  and to be rather massive. For example, interesting  candidates
 for longÐlived strangelets are lying in a valley of stability which starts at the quark alpha (6u6d6s) and continues by adding 
one unit of negative charge, i.e. (A, Z) = (8,-2),(9,-3), (10,-4), (11,-5), etc. 
 However, scenarios that include the production of negatively charged strangelets  at the LHC  been been
strongly  criticized  \cite{ellisstrange}.  Alternatively, the possibility that (meta-)stable strangelets with small positive charge could
exist has been discussed in a number of references \cite{PLUSCHARGE, PLUSCHARGE1,PLUSCHARGE2}. Massive, multiply charged
 strangelets could  be sufficiently  ionizing to reach the detection threshold of MoEDAL.

\subsection{Quirks}

Quirks are particles appearing~\cite{okun,kangandluty} in extensions of the SM that include heavy particles 
that are charged under both the SM Group and additional non-Abelian asymptotically-free gauge groups (``infracolour'' (IC)),
which may have masses reachable at the LHC. Quirks are analogous to the quarks of QCD,
but the confining gauge group is specifically \emph{not } the colour SU(3) of QCD.  The IC group may be of SU(N) type, 
and the quirks are assumed to be in its fundamental representation. 
The new group is assumed to become strong (confining) at a scale $\Lambda \ll m_Q$. 
Thus, in contrast to the familiar QCD scale, the confining scale of the new group is much lower than the fermion mass, 
hence the name infracolour. The phenomenologically interesting range of the quirk mass, $m_Q$, is 
$$  100~{\rm GeV} \le m_Q \le \mathcal{O}(10) ~{\rm TeV}.$$
Thus, such models are of relevance to LHC physics, including the MoEDAL experiment. 
The confinement scale $\Lambda$ can be as low as 100 eV~\cite{kangandluty}.

\begin{figure}[htb]
\begin{center}
\includegraphics[width=24pc]{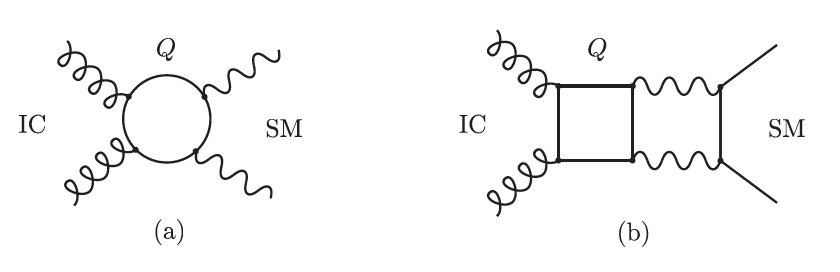}
\caption{ Loop diagrams that contribute to the coupling of the Infracolour (IC) gluons (helical structures) to the SM
gauge bosons (wavy lines)  or fermions (continuous external lines). Quirks are represented as 
internal-loop continuous lines.}
\label{Fig:IC}
\end{center}
\end{figure} 

Since the SM is uncharged under the infracolour sector, couplings of SM matter to the infracolour sector can arise
only through quirk-loop processes of the form depicted in Fig.~\ref{Fig:IC}. The leading coupling is provided by the 
diagram of Fig.~\ref{Fig:IC}(a), which leads  
to terms in the effective Lagrangian that mix the field strengths of the IC and SM gauge groups, of the 
form~\footnote{The reader should notice that the two-loop processes of Fig.~\ref{Fig:IC}(b), 
which couple the IC gluons to the fermionic SM sector suffer, in addition to the loop suppression, 
an additional helicity suppression, as compared to the diagram of Fig,~\ref{Fig:IC}(a), and are therefore non-leading contributions.}:
\begin{equation}\label{mixing}
\mathcal{L}_{\rm eff} \sim \dfrac{g^2 \, {g^\prime}^2 }{16\, \pi^2 \, m_Q^2}\, F_{\mu\nu}^2 \, {F^\prime}_{\rho\sigma}^2,
\end{equation}
where $g$ and $g^\prime$ denote the SM and IC gauge couplings respectively. The operator (\ref{mixing}) mediates
the decay of an IC gluon into photons and/or ordinary colour gluons with a life-time that depends crucially on the 
magnitude of the confining scale $\Lambda$. It can be shown that~\cite{kangandluty} for $\Lambda \ge 50~$GeV 
the IC glueball can decay within a particle detector, while its life-time becomes longer than the age of the 
Universe for $\Lambda \le 50~$MeV. 

\begin{figure}[htb]
\begin{center}
\includegraphics[width=26pc]{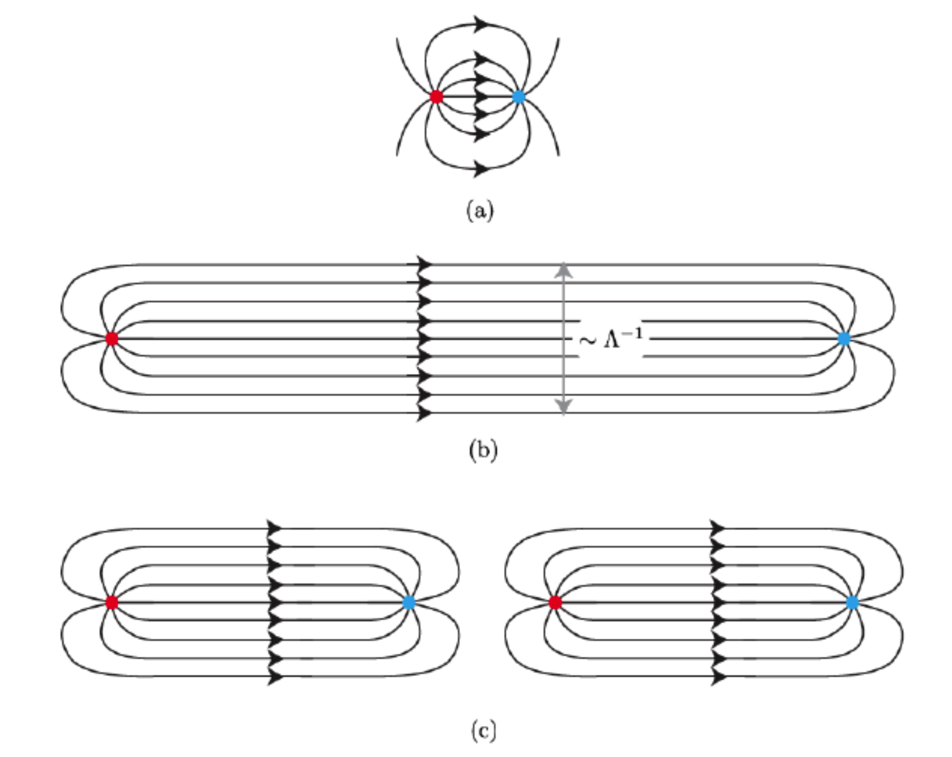}
\caption{Quirk-Antiquirk pairs are connected by IC flux strings which have the shape of tubes:
(a) tube for quirk-antiquirk separation $r \ll \Lambda^{-1} $, (b) tube for quirk-antiquirk separation $r \gg \Lambda^{-1}$.
(c) the breaking of strings is exponentially suppressed, requiring energy $2m_Q \gg \Lambda$. }
\label{Fig:quirkflux}
\end{center}
\end{figure} 

When quirks are pair-produced at the LHC they are bound together by a flux tube but, unlike QCD, 
there is no kinematical possibility of the pair hadronizing into jets, as the tension of the flux-tube is too low 
compared with the Quirk mass (\emph{cf.} fig.~\ref{Fig:quirkflux}). 
The quirk-antiquirk pair stays connected by the IC string like a ``rubber band'' that can stretch up to macroscopic lengths, 
depending on the magnitude of the confining scale $\Lambda$. One would expect the size of the confining flux tube 
to be~\cite{kangandluty}
\begin{equation}\label{tube}
L\sim\dfrac{m_Q}{\Lambda^2}\sim {\rm 10 m}\left(\dfrac{m_Q}{\rm TeV}\right)\left(\dfrac{\Lambda}{\rm 100 eV}\right)^2 \,
\end{equation}
where $m_Q$ is the mass of the quirks at the end of the confining fluxtubes.  
Technically speaking, for an energy $E \sim \pi \Lambda^2$ and area $A$, 
string breaking is exponentially suppressed, since the relevant life-time $\tau$ for a string of length $L \sim m_Q/\sigma $, 
with the string tension 
$\sigma \sim E^2 \, A$, can be estimated to be (by analogy with the Schwinger mechanism 
of pair creation of charged particles by a weak external electric field~\cite{kangandluty}):
$ \tau \sim \dfrac{4\pi^3}{m_Q}e^{m_Q^2/\Lambda^2} $,
and in general $m_Q \gg \Lambda$. This is already longer than the age of the Universe for $m_Q \ge 100$~GeV 
and $\Lambda / m_Q \le 0.1$.

\begin{figure}[htb]
\begin{center}
\includegraphics[width=20pc]{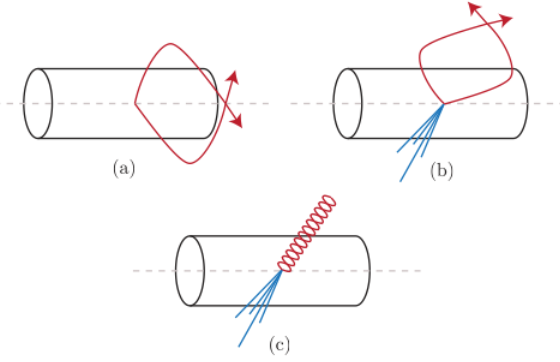}
\caption{Anomalous tracks from quirks with macroscopic strings }
\label{Fig:macroquirks}
\end{center}
\end{figure}

Assuming that the quirk mass $m$  is in the phenomenologically interesting range of 100 GeV to a few TeV, and that the new
gauge group gets strong at a scale $\Lambda < m$ and thus the  breaking of strings is exponentially suppressed, quirk production
 results in strings that are long compared to $\Lambda^{-1}$. The existence of these long stable strings
 would lead to highly exotic events at  the LHC. 
 For 100 eV $ \lesssim  \Lambda \lesssim$ keV  the strings are macroscopic. In this case one would observe
 events with two separated quirk tracks with
measurable curvature toward each other due to the string interaction, 
as shown in Fig.~\ref{Fig:macroquirks}~\cite{kangandluty}. 

The difficulty  in detecting quirks with macroscopic strings is that the  triggers and track reconstruction
 algorithms of conventional LHC detectors are designed for conventional tracks, and will likely
miss these events altogether. However, the MoEDAL detector  does not suffer from these drawbacks. Indeed,
the signature in the NTDs for two massive quirks separated by a  macroscopically string would be quite spectacular. 

For 10 keV $\lesssim \Lambda \lesssim $ MeV and $m_{Q}  \sim$ 1 TeV, the typical strings are too 
small to resolve (mesoscopic)  in the detector but  large compared to atomic scales. In this case the bound 
state appear in conventional LHC detector  as  as a single particle with the invariant
mass of a quirk pair.
 
For mesoscopic strings, we can no longer assume that matter interactions will randomize
the angular momentum and prevent the quirks from annihilating. In
other words, we need to know whether the bound state lives long enough to appear
in the detector. A very good vacuum  is maintained within the LHC beam pipe reducing
 matter interactions to a  negligible rate within the  pipe. The decay length of 
coloured quirks within the beam pipe has been estimated to be~\cite{kangandluty}:
\begin{equation}
c\tau \sim 1cm\left( \dfrac{\Lambda}{100 keV} \right)^{-2}\left( \dfrac{m_{Q}}{TeV} \right)^{2},
\end{equation}
while for uncoloured quirks:
\begin{equation}
c\tau \sim 10 cm \left(\dfrac{\Lambda}{MeV}\right)^{-3} \left( \dfrac{m_{Q}}{TeV}\right)^{2}, 
\end{equation}
one can see that heavier quirks with small $\Lambda$ ($\sim$ 10 KeV) would have
lifetimes sufficient to exit the beam-pipe.
However the efficiency of the mechanism of energy loss considered \cite{kangandluty}  
is uncertain, particularly for  the infracolor energy loss. The decay lengths
may be significantly longer than these estimates.
Once the quirk bound state reaches the beam pipe, interactions with matter  efficiently 
randomize the angular momentum and prevent annihilation. 

Thus, the  quirk  bound state could  appear  as collider stable  highly-ionizing
particle. In MoEDAL the latent tracks  resulting  from the passage of a highly ionizing 
particles are only of the order of  10nm  in diameter,  with resulting etch pits around 5-10 $\mu$m in 
diameter.  Thus,  even in the case of ``mesoscopic''  strings MoEDAL may  still be 
 able to resolve the ``di-quirk'' nature of the event. If not, MoEDAL will be
 able to detect the quirk state as a highly-ionizing, doubly-charged track. 

If the strings are microscopic, with length of the order of an Angstrom, corresponding
to $MeV \lesssim \Lambda \lesssim m_{Q}/few$,  the quirks annihilate promptly within the
detector. For coloured quirks, this can lead to hadronic  fireball events with $\sim$ 10$^{3}$ hadrons
each with energy of the order of a GeV, a spectacular signature that however is not detectable by MoEDAL.

\section{Stopped Stable and Metastable Particles} 

As we have seen, massive stable and metastable  particles are common in many models of new physics at the TeV scale. 
The discovery of new long-lived particles at the LHC would
provide fundamentally significant insights into the nature of dark matter, the presence
 of new symmetries of nature or extra dimensions and many other BSM scenarios.
 
 If such particles are charged and/or coloured,  a reasonable fraction of those produced at the LHC 
 will stop in the LHC detectors and the surrounding material, and  give observable out-of-time decays. 
 This is particularly true of very-highly-ionizing particles that lose energy quickly in matter. 
 Many such scenarios have been discussed in the preceding text.
 
 Particle metastability due to high-scale physics is well motivated. Also, as has been discussed above
 a new physics particle may also be long-lived because of its small couplings, as in $R$-violating
supersymmetry~\cite{RV}, or TeV-scale seesaw models~\cite{SEESAWS}.  A particle may also be 
 long-lived due to kinematics, e.g., if  the Massive Metastable Charge Particle (MMCP)  is nearly degenerate with the final state 
 into which it decays. There have been searches for slow-moving MMCP's at LEP \cite{LEPMMCPlims,LEPMMCPlims1,LEPMMCPlims2}, the
Tevatron \cite{TEVATRONMMCPlims, TEVATRONMMCPlims1}, and the LHC \cite{LHCMMCPlims,
LHCMMCPlims1,LHCMMCPlims2,LHCMMCPlims3}  that place bounds on their production.
 
 While much can be learned by studying the production and  propagation of  particles at the LHC,
  there is interesting physics that can only be  accessed by studying their decays. These measurements 
  can reveal the properties of  the decaying TeV-sector particles. For
example, the Lorentz structure of the decay and the branching fractions to different SM
particles can constrain the high-scale physics giving rise to the decay. 

It is often the case that MMCPs will be stopped in the collider detectors themselves.
 However, observing decays within the detector is experimentally challenging because the designs of the
standard   LHC detectors were optimized to measure particles  emanating from a central interaction point; 
moving near to the speed of light ($\beta \gtrsim$  0.5), and  not highly-ionizing enough to be absorbed
within the detector.  

Despite the experimental  difficulties, collider detectors  \cite{colliderdecays,colliderdecays1,colliderdecays2}
have managed to  perform  searches for decays  of stopped particles, demonstrating that tricky 
experimental issues such as triggering   can be solved in some cases~\cite{STOPPED}. 
By studying decays occurring out-of-time from beam crossings, using empty bunch crossings,  one can eliminate
 backgrounds from competing processes in collisions and  event-by-event background from pile-up or 
the underlying event. Nevertheless, for MMCPs produced at low rates, cosmic rays are still an  important source of 
statistical background.
  
 In addition,  there have been a number of  proposals to study decays of MMCPs that include searching for 
 decays from the surrounding rock~\cite{ROCK},  in the detectors   themselves \cite{SIBLEY}   and building 
 new detectors to capture MMCPs \cite{CAPTURE1, CAPTURE2,  CAPTURE3}.   The  MoEDAL experiment
 plans to deploy trapping detectors within roughly a metre of IP8, adjacent to the MoEDAL NTD sector array, in
 the MoEDAL/LHCb-VELO cavern. This ``MMT'' detector system has been described above.
 
 After the MMT detectors have been scanned for the presence of magnetic charge in a SQUID magnetometer 
 facility, e.g., at ETH Zurich, they will be transported to an underground laboratory, e.g., SNOLAB, to be monitored with a
 dedicated detector system over a long period for the decay of any massive metastable charged and/or coloured 
 particles that may have stopped within them.   The presence of trapped new physics particles also raises the possibility 
 that the particle can be freed   for direct study in the laboratory.  
 
 The  topologies and kinematic distributions of long-lived particle  decays can realistically be measured using 
 the MoEDAL approach. Due to the nature of its design, in this approach backgrounds from particle interactions 
 are eliminated without the need for restricting the search for out-of-time decays.
  Also, the fact that the decays are monitored in a deep underground laboratory means that
 cosmic-ray backgrounds are reduced to a minimum.

\section{Detecting Highly-Ionizing Particles at the LHC}

 The sub-detectors of general-purpose 
 experiments are designed to detect minimum-ionizing particles moving near to the speed of light. 
 Effects arising from the particleÕs low velocity and the high density of the energy deposition, 
 such as electronics saturation, light quenching in scintillators and adjacent hits from delta electrons,
are extremely challenging to deal with. Indeed, in some cases it may be impossible to make an accurate
measurement of the effective charge of the particle.  For example,  the resulting dead time as a result of 
electronics saturation may be of the order of the bunch crossing time. An example of this
is provided by  the effect of highly-ionizing particles on the CMS silicon strip tracker that was studied in~\cite{sensitivity,sensitivity1}. 
In addition, highly-ionizing particles  will be absorbed very quickly within 
 the mass of the standard collider detector.  Indeed, extremely-ionizing particles
may be absorbed before they penetrate far into the inner tracking detectors
 
 In order for stable or long lived massive particles  to be detected in general purpose collider detectors
 they need to be detected and  triggered on  in a sub-detector system, or group of sub-detector systems, 
 and be associated to the correct bunch crossing. This detection and triggering must happen  within $\Delta$t ns  - where
 $\Delta$t is the time between bunch crossings (nominally 25ns at the LHC with E$_{cm}$=14 TeV)  - 
 after the default arrival time of a particle travelling at the speed of light \cite{kraan,hauser}. Later arrival would 
 require  detection and triggering within the next crossing time window. The typically large size of the 
 general purpose collider detectors (the central ATLAS and CMS muon chambers extend to 10 and 7 m, 
 respectively) results in  this being an important source of inefficiency in detecting SMPs. For example,
it is only possible to reconstruct the track of a slowly-moving long lived massive particle
 in the ATLAS central muon chambers  within the correct bunch crossing window if $\beta  > $ 0. 5 \cite{kraan}.
 
But even if a long-lived massive particle travels through the  sub-detector systems within the timing window in
which it was created, additional problems may arise due its relatively slow speed. 
Naturally, the  time sampling and reconstruction software of collider detectors is optimized 
assuming all particles are  travelling near to the speed of light. Thus,  it is quite possible 
that the quality of the  read-out signal or reconstructed track or cluster will be degraded for
a collider stable slow moving mass long-lived particle , especially for sub-systems far away from the
 interaction point. However, if one relies on detector simulations it seems to be  possible to trigger and measure 
slowly-moving particles at, for example,  ATLAS and CMS~\cite{zalewski,kraan,BIRK}.  Of course, this is an area which
must continue to be studied as the simulation programs are further developed and
the detectors better understood.
  
The response of each of the general-purpose experiments'  subdetectors to highly-ionizing particles 
 cannot be calibrated directly in situ, and  consequently signal efficiency determination
 relies heavily on simulations. This point is exemplified by an  ATLAS search for which
  the dominant  source of uncertainty arises from the modelling of the effect of electron-ion
   recombination   in the liquid-argon calorimeter in the case of a high energy loss \cite{EIONRECOMB}.
 Last, but not least, the extremely high background from Standard Model particles at, say,
 the High Luminosity LHC in a detector with non-optimal granularity can give rise to backgrounds
 from, for example, multi-particle occupancy.
 
 \begin{figure}[htb]
\begin{center}
\includegraphics[width=30pc]{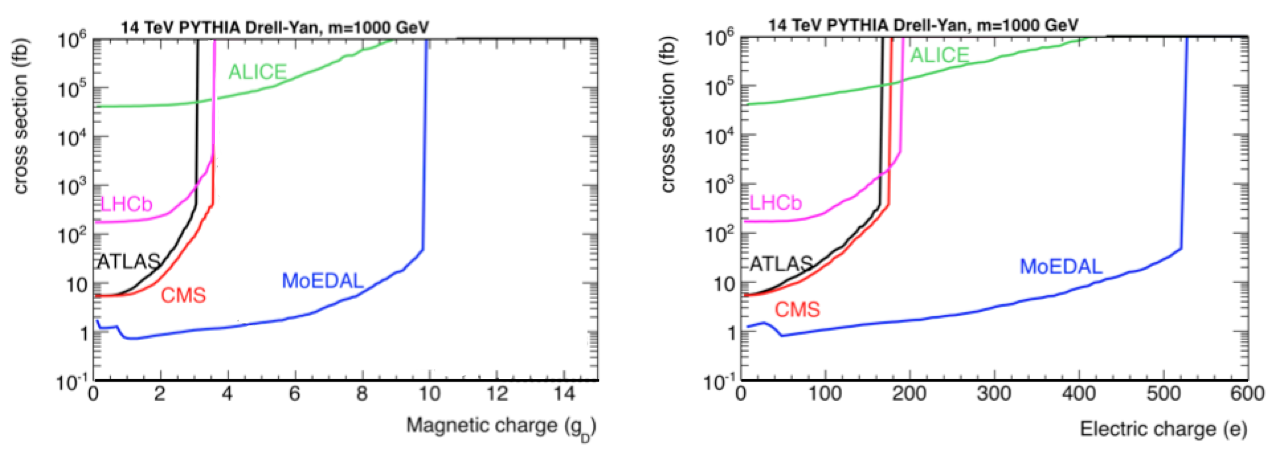}
\caption{ A comparison of the sensitivity of various LHC detectors for highly-ionizing particles.}
\label{Fig:sensitivity}
\end{center}
\end{figure} 

The MoEDAL-LHC detector is  designed to optimize the search for highly-ionizing 
particles and magnetic charge in a way that is  complementary to the reach of the general-purpose 
LHC detectors, and largely overcomes the experimental difficulties for
general purpose collider detectors mentioned above. The MoEDAL NTD detector system  is light, 
thin - around $\sim$1 cm thick - and with a detector density around 1.2 g/cm$^{3}$. Thus, 
 little material is added  to the relatively  small amount  of existing material comprising the
 LHCb vertex detector (VELO)  around LHC intersection Point IP8. The measurement of 
 the passage of highly-ionizing particles -  with  Z/$\beta \gtrsim$ 5 - is accomplished using 
 a plastic NTD system that can be easily calibrated  using heavy-ion beams.
 
In addition,  one can also  stop highly-ionizing particles in MoEDAL's dedicated  trapping 
detectors and then make delayed detection of their  presence in a remote  detector system.  
  In the case of monopoles,  one can use a SQUID magnetometer to
 detect trapped magnetic charge. The SQUID magnetometer is  calibrated
 for  a monopole's magnetic field using a very long solenoidal coil.   One can  monitor the decay of stopped  
 long-lived  massive  electrically charged particles in underground facilities 
 such as SNOLAB. 
 
 As has been discussed above Standard Model backgrounds  that can simulate the signals of 
 highly-ionizing particles arising from new physics scenarios  in MoEDAL are non-existent.  Thus only one event is 
enough to establish the presence of new  physics. However, in the case of general-purpose
 collider detectors, the experimental shortcomings and the backgrounds from the very large flux of SM particles
 described above require significantly more signal particles to be recorded before a 
 discovery can be claimed. 
 
 We estimate that in standard collider detectors between 10 and 100
 highly-ionizing particles will need to be registered before a discovery can be claimed. A comparison of
 the sensitivity of MoEDAL to highly-Ionizing particles with  that of the other LHC 
 detectors - assuming that ATLAS and CMS need to detect a conservative minimum of 100 events to
 claim discovery whereas MoEDAL only one -  is  shown in Fig.~\ref{Fig:sensitivity}. This
 figure is based on an initial study of the detection of highly-ionizing particles at the LHC \cite{mermod}.

\section{Summary and Conclusions}

The primary motivation for the MoEDAL experiment is to look for fundamental
magnetic monopoles. The discovery of such a particle would revolutionize
our understanding of electrodynamics, electroweak theory, the Standard Model
and scenarios for grand unification. Following the pioneering proposal by Dirac,
it was realized by 't Hooft and Polyakov that monopoles appear in generic unified
theories. In particular, they may well appear close to the electroweak scale and hence
be accessible at the LHC. Importantly, such a discovery would prove insights into the topology of the 
underlying theory at that electroweak scale.

As we have also seen, there are many scenarios for new physics in which new,
long-lived singly-charged particles may appear. Those produced with low velocities would
be highly-ionizing and therefore also detectable by the MoEDAL experiment.
There are also scenarios predicting the existence of (meta-)stable particles with
multiple electric charges, which would be even more highly-ionizing.
We have also discussed in this report several more exotic scenarios for new physics at the
LHC that might be detectable with MoEDAL.

The MoEDAL detector is dedicated to the search for the highly ionizing messengers of new
physics. It is  designed to be clearly superior to  the existing LHC detectors in this challenging
arena. Importantly,  MoEDAL extends the discovery reach of the LHC in a complementary way. 

MoEDAL's contribution is invaluable even  in those scenarios  where we can expect a
 considerable overlap between the physics  reach of MoEDAL and the other LHC experiments. 
Indeed, MoEDAL is a totally different kind of LHC detector with very different,  systematics and 
sensitivity. Also, any  new particle detected by the  MoEDAL would provide a permanent record and in some
cases the particle itself would be captured. Such considerations would be crucial in the verification
and understanding of any revolutionary discovery of beyond the Standard Model physics at the LHC.

We conclude that  the MoEDAL experiment, dedicated to the detection 
of new physics, would reveal unprecedented insights into such fundamental questions as:
does magnetic charge exist; are there new symmetries of nature; are there extra dimensions;
what is the nature of dark matter; and,  how did the universe unfurl at the earliest times.
 In short,   MoEDAL  has   a revolutionary physics potential that will significantly
  enhance the discovery horizon of the LHC.

\section*{Acknowledgements}

We thank the editors of International Journal of Modern Physics A for the invitation to write this review.
It is a  pleasure to thank David Milstead for valuable discussions. This work was partially supported by:  the London Centre
for Terauniverse Studies (LCTS), using funding from the European Research Council via the Advanced Investigator
Grant 267352, the Science and Technology  Facilities Council  (STFC, UK), a Natural Science and Engineering Research
 Council of Canada  project grant; the V-P Research of the University of Alberta;  the Provost of the University of Alberta; and,
 UEFISCDI (Romania).

\theendnotes


\begin{thebibliography}{999}


\bibitem{moedal-web} For general information on the MoEDAL experiment, see: \url{http://moedal.web.cern.ch/}

\bibitem{moedal-tdr} MoEDAL Collaboration, Technical Design Report of the MoEDAL Experiment {\it CERN Preprint} CERN-LHC-2009-006, MoEDAL-TDR-1.1 (2009).

\bibitem{LHC} 
  L.~Evans and P.~Bryant,
  JINST {\bf 3}, S08001 (2008).

\bibitem{Dirac1931kp} P. A. M. Dirac,
Proc. Roy. Soc. Lond. {\bf A 133}, 60-72 (1931). 

\bibitem{Diracs_idea} P. A. M. Dirac, Phys. Rev. {\bf 74}, 817 (1948).

\bibitem{tHooft1974qc}  G.  't Hooft, Nucl. Phys. {\bf  B 79}, 276-284 (1974).

\bibitem{Polyakov1974ek} A.  M.  Polyakov, JETP  Lett.  {\bf 20}, 194-195 (1974).

\bibitem{Julia1975ff}  B. Julia and A.  Zee, 
Phys. Rev.  {\bf D 11}, 2227 (1975). 

\bibitem{N} Y Nambu, Nucl. Phys. {\bf B 130}, 505 (1977).

\bibitem{Witten1979ey} E. Witten,
 ``Phys. Lett. {\bf B 86}, 283, (1979). 
 
 \bibitem{Lazarides1980cc} G. Lazarides, M.  Magg  and  Q. Shafi,
Phys. Lett. {\bf B 97}, 87 (1980).

 \bibitem{Sorkin1983ns} R. D. Sorkin,
Phys. Rev. Lett. {\bf 51}, 87-90 (1983).

\bibitem{Gross1983hb} D. J. Gross,  and  M. J. Perry, 
Nucl.Phys. {\bf B 226}, 29 (1983).

\bibitem{Cho1996qd}  Y. M. Cho and D. Maison,
Phys. Lett. {\bf B 391}, 360 (1997);  W. S. Bae and Y. M. Cho, JKPS {\bf 46}, 791 (2005).

\bibitem{Schwinger1969ib} J. S. Schwinger,
Science {\bf 165}, 757-761 (1969). 

\bibitem{Preskill1984gd}  J. Preskill,
Ann. Rev. Nucl. Part. Sci. 34, 461-530 (1984). 

 \bibitem{AV} A. Achucarro and T. Vachaspati, Physics Reports {\bf 327}, 347  (2000).
 
 \bibitem{Kephart2006zd}  T. W. Kephart, C-A Lee  and Q. Shafi,
 JHEP {\bf 0701}, 088 (2007). 
 
\bibitem{CHO-KIM-YOON} Y. M. Cho, K.  Kim and J. H. Yoon,  arXiv:1212.3885v5 [hep-ph] (2012).

 
\bibitem{ZHANG}   D.~G.~Pak, P.~M.~Zhang and L.~P.~Zou,
  arXiv:1311.7567 [hep-th], (2013).

\bibitem{Rajantie2005hi} A. Rajantie, 
JHEP {\bf 0601}, 088 (2006). 

\bibitem{Fairbairn07} M.~Fairbairn, A.~C.~Kraan, D.~A.~Milstead, T.~Sjostrand, P.~Z.~Skands and T.~Sloan,
  Phys.\ Rept.\  {\bf 438}, 1 (2007), [hep-ph/0611040] (2006).
 
\bibitem{SUSY}
H.~P.~Nilles,
  Phys.\ Rept.\  {\bf 110}, 1 (1984).
  
  \bibitem{MSSM}
H.~E.~Haber and G.~L.~Kane,
  Phys.\ Rept.\  {\bf 117}, 75 (1985).
 
\bibitem{CMSSM} M.~Drees and M.~M.~Nojiri,
Phys.\ Rev.\ D {\bf 47}, 376  (1993).
\bibitem{CMSSM1}G.~L.~Kane, C.~F.~Kolda, L.~Roszkowski and J.~D.~Wells,
  Phys.\ Rev.\  D {\bf 49}, 6173  (1994).
\bibitem{CMSSM2}  J.~R.~Ellis, T.~Falk, K.~A.~Olive and M.~Schmitt,
Phys.\ Lett.\ B {\bf 388}, 97 (1996).
\bibitem{CMSSM3} Phys.\ Lett.\ B {\bf 413}, 355 (1997).
\bibitem{CMSSM4}  H.~Baer and M.~Brhlik,
Phys.\ Rev.\ D {\bf 53}, 597 (1996).
\bibitem{CMSSM5}  Phys.\ Rev.\  D {\bf 57}, 567  (1998).
\bibitem{CMSSM6} V.~D.~Barger and C.~Kao,
Phys.\ Rev.\ D {\bf 57}, 3131 (1998).
\bibitem{CMSSM7}  H.~Baer, M.~Brhlik, M.~A.~Diaz, J.~Ferrandis, P.~Mercadante, P.~Quintana and X.~Tata,
    Phys.\ Rev.\  D {\bf 63}, 0105007 (2001).
 
\bibitem{RV}
R.~Barbieri, C.~Berat, M.~Besancon, M.~Chemtob, A.~Deandrea, E.~Dudas, P.~Fayet, S.~Lavignac {\it et al.},
  Phys.\ Rept.\  {\bf 420}, 1 (2005).
  
 \bibitem{FATHIGGS}  	
R. Harnik, Graham D. Kribs, D. T. Larson and H.  Murayama, 
Phys. Rev. {\bf D70}, 015002 (2004).
  
\bibitem{xyon}  Y. Nomura and B. Tweedie, Phys. Rev.  {\bf D72}, 015006  (2005).  

\bibitem{ADD} N. Arkani-Hamed, S. Dimopoulos and G. R. Dvali, Phys. Lett.  {\bf B 429}, 263 (1998).
\bibitem{ADD1}  I. Antoniadis,  N. Arkani-Hamed, S. Dimopoulos and G. R. Dvali, Phys. Lett.  {\bf B436} , 257 (1998).
\bibitem{ADD2} N. Arkani-Hamed, S. Dimopoulos, G. R. Dvali, Phys. Rev. {\bf D 59}, 086004 (1999).

\bibitem{Randall} L. Randall and R. Sundrum, Phys. Rev. Lett.  {\bf 83}, 4690 (1999).

\bibitem{TEV-1} I. Antoniadis,  Phys. Lett. {\bf B 377}, 246 (1990).
\bibitem{TEV-1a}  I. Antoniadis and K. Benakli,  Phys. Lett. {\bf B 326}, 
69 (1994).
\bibitem{TEV-1b}  I. Antoniadis, K. Benakli and M. Quiros, Phys. Lett.  {\bf B 331}, 313 (1994). 

\bibitem{UED} T. Appelquist, H.-Ch. Cheng and B. A.  Dobrescu, Phys. Rev.
{\bf D 64}, 035002 (2001). 

 \bibitem{westmuckett} J.~R.~Ellis, N.~E.~Mavromatos and D.~V.~Nanopoulos,
  Gen.\ Rel.\ Grav.\  {\bf 32}, 943 (2000).
 \bibitem{westmuckett1} J.~R.~Ellis, N.~E.~Mavromatos and D.~V.~Nanopoulos, Phys.\ Lett.\ B {\bf 665}, 412 (2008).
 \bibitem{westmuckett2} J.~R.~Ellis, N.~E.~Mavromatos and M.~Westmuckett, 
Phys.\ Rev.\ D \textbf{70}, 044036 (2004).
 \bibitem{westmuckett3} J.~R.~Ellis, N.~E.~Mavromatos and M.~Westmuckett, Phys. Rev. D \textbf{71}, 106006 (2005).
   
\bibitem{shiu}  
G.~Shiu and L.~-T.~Wang,
  Phys.\ Rev.\ D {\bf 69}, 126007 (2004).

\bibitem{mitsou} 
  N.~E.~Mavromatos, V.~A.~Mitsou, S.~Sarkar and A.~Vergou,
  Eur.\ Phys.\ J.\ C {\bf 72}, 1956 (2012).

\bibitem{vectorlike} D. E. Morrissey, T. Plehn and T. M. P. Tait,
arXiv:0912.3259 [hep-ph] (2009).

\bibitem{KILIC} 
C.  Kilic and T.  Okui, JHEP {\bf 1004}, 128  (2010).

\bibitem{Coleman:1985ki}
  S.~R.~Coleman,
  Nucl.\ Phys.\ B {\bf 262}, 263  (1985);
   [Erratum-ibid.\ B {\bf 269}, 744 (1986)].
  
 \bibitem{SUSYQballs} A. Kusenko, Phys. Lett.  {\bf B 405}, 108 (1997); Phys. Lett.  {\bf B 404}, 285 (1997);
  Phys. Lett.  {\bf B 406} , 26  (1997).

\bibitem{Rubakov1981rg}  V. A. Rubakov, 
JETP Lett.  {\bf 33}, 644 (1981).

\bibitem{Callan1982ah}    Curtis G. Callan Jr.
Phys. Rev. {\bf D 25}, 2141 (1982).

\bibitem{Fairbairn14} S. Burdin et al., Non-collider Searches for Stable MassiveParticle,
to be published in Phys. Rep. 2014.

\bibitem{LHCb-detector} 
  A.~A.~Alves, Jr. {\it et al.}  [LHCb Collaboration],
  JINST {\bf 3}, S08005 (2008).
  
  \bibitem{CR39} B. G Cartwright et al., Nucl. Instrum. Meth. {\bf 153}, 457 (1978)
  
  \bibitem{MAKROFOL} Z. Todorovic, Nucl. Tracks  Radiat. Meas. {\bf 17}, 23 (1990)
  
  \bibitem{lexan} R. C. Singh and H. S. Virk, Nucl. Instrum., Meth. {\bf 29}, 579 (1987)
  
  \bibitem{STEFANO} J. L. Pinfold, Rad. Meas. {\bf 44} 834 (2009).
  
  \bibitem{SQUIDS} A. de Roeck et al., Eur. Phys. J.  {\bf C 72}, 2212  (2012);  K. Bendtz {\it et al.}, 
  Phys. Rev. Lett. {\bf 110} ,  121803 (2013); K. Bendz et al., e-Print: arXiv:1311.6940 [physics.ins-det] (2013). 

\bibitem{TimePix} {\tt http://medipix.web.cern.ch/medipix/pages/medipix2/timepix.php}.

\bibitem{PDG} J.~Beringer {\it et al.}  [Particle Data Group Collaboration],
  Phys.\ Rev.\ D {\bf 86} (2012) 010001.

\bibitem{ICRU49} ``Stopping Powers and Ranges for Protons and Alpha Particles,''
ICRU Report No. 49 (1993); http://physics.nist.gov/PhysRefData/.

\bibitem{PDGHE} D.E. Groom, N.V. Mokhov, and S.I. Striganov, ``Muon stopping-power
and range tables: 10 MeV - 100 TeV'', Atomic Data and Nuclear Data Tables {\bf 78},
183 (2001).
\bibitem{PDGHE1} D. Yu. Ivanov {\it et al.}, Phys. Lett.  {\bf B 442}, 453 (1998).

\bibitem{NIEL} A. Chilingarov {\it et al.}, Nucl. Instrum. Meth.  {\bf A 449}, 277 (2000).

\bibitem{lewin} J. Lewin, Rutherford Laboratory Internal Note - RL-77 126/A (1977). 

\bibitem{sternheimer} R.M. Sternheimer, M.J. Berger and S.M. Seltzer, Atomic Data and Nuclear Data Tables {\bf 30},  261 (1984).

\bibitem{barkas} W.H. Barkas, W. Birnbaum and F.M. Smith, Phys. Rev. {\bf 101}, 778 (1956).



\bibitem{lindhard} J . Lindhard, V. Nielsen, M. Scharff and P. Thomsen,  Mat. Fys. Medd. Dan. Vid. Selsk. {\bf 33}, No. 10,  1
(1963).

\bibitem{AANDZ} H.H. Andersen and J.F. Ziegler, {\it Hydrogen: Stopping Powers and Ranges in All
Elements}, Vol. 3 of The Stopping and Ranges of Ions in Matter (Pergamon Press, 1977).

\bibitem{ahlen} S.P. Ahlen, Phys. Rev. {\bf D 17}, 229 (1978).

\bibitem{sternheimerD} R.M. Sternheimer, Phys. Rev.  {\bf 93}, 351 (1953).

\bibitem{monopole-range} H1, A. Aktas {\it et al.}, Eur. Phys. J.  {\bf C 41}, 133 (2005).

\bibitem{GEANT} G. Bauer {\it et al.}, Nucl. Instrum. Meth. {\bf A 545} , 503  (2005).

\bibitem{AHLENKINOSH} S. P Ahlen and K. Kinohsita, Phys. Rev. {\bf  D 26}, 2347 (1982).

\bibitem{APPROXLEEL} J. M. Barkov {\it et al}, CERN/EP 83-194 (1983).

\bibitem{GIORGIO} G. Giacomelli. ``Magnetic Monopoles''. La Rivistra del Nuovo Cimento, 1984.

\bibitem{BRACCI} L. Bracci,  G. Fiorentini, R. Tripiccione, Nucl. Phys. {\bf  B 238}, 167 (1984).

\bibitem{DERK} J. Derkaoui al., Astropart. Phys. {\bf 9}, 173 (1988).

\bibitem{AHLEN2} S.P. Ahlen, Phys. Rev. {\bf D 14}, 2935 (1976).

\bibitem{RITSON} D.M. Ritson, {\it Fermi-Teller theory of low velocity ionisation
 losses applied to monopoles}, SLAC-PUB 2950 (1982).
 
\bibitem{NAKAMURA} S. Nakamura, ``Search for supermassive relics by large area
 plastic track detectors'', PhD Thesis,  UT-ICEPP-88-04, University of Tokyo (1988).
 
\bibitem{GIORGIO2} G. Giacomelli et al., Nucl. Instrum. Meth.  A 411, 441-45 (1998).
\bibitem{GIORGIO2a}S.Cecchini {\it et al.},  Nuovo. Cim. {\bf A 109}, 1119 (1996).
\bibitem{GIORGIO2b}S. Balestra {\it et al.}, Nucl. Instrum. Meth. {\bf B 254}, 254 (2007).

\bibitem{LATENT} N. Yasuda, Rad. Measure. {\bf 34}, Issue 1-6, 45 (2001).

\bibitem{CHARGERES} H. Dekhissi {\it et al.}, Nucl. Phys. {\bf A 662}, 207 (2000).
\bibitem{CHARGERES1}S. Cecchini {\it et al.}, Astropart. Phys. {\bf 1}, 369 (1993).

\bibitem{BRACCIBOUND} 	
L. Bracci and  G. Fiorentini, Nucl. Phys. {\bf B 232}, 236  (1984).

\bibitem{MALKUS} W. V. R. Malkus, Phys. Rev. {\bf 83}, 899 (1951).

\bibitem{GAMBERG} 	
L. P. Gamberg, G. R. Kalbfleisch and  K. A. Milton, Found. Phys. {\bf 30}, 543-565  (2000).

\bibitem{OLAUSSEN}  K. Olaussen and R. Sollie, Nucl. Phys. {\bf B 255}, 465 (1985).

\bibitem{sivers} D. Sivers, Phys. Rev. {\bf  D 2}, 2048 (1970).
 
\bibitem{GOEBEL} C. J. Goebel, in {\it Monopole Õ83}, ed. J. L. Stone (Plenum, New York, 1984), p. 333.

\bibitem{GOTO}  E. Goto, H. Kolm, and K. Ford, Phys. Rev. {\bf 132}, 387 (1963).

\bibitem{Milton:2006cp}   K. A. Milton,
  Rept.  Prog. Phys.  {\bf 69}, 1637 (2006).
   
\bibitem{Rajantie:2012xh} 
  A. Rajantie,
  Contemp.  Phys. {\bf  53}, 195 (2012).  

\bibitem{ShnirBook} Y.M. Shnir, ``Magnetic Monopoles'', (Springer, 2005).

\bibitem{WeinbergBook} E.J.~Weinberg
``Classical solutions in quantum field theory : Solitons and Instantons in High 
Energy Physics'', (Cambridge University Press, 2012)

\bibitem{Wu:1975es} 
  T. T. Wu and C. N. Yang,
  Phys. Rev. {\bf D 12}, 3845 (1975).
  
\bibitem{Georgi1974sy} H. Georgi, H. and S. L.  Glashow,
Phys. Rev. Lett.  {\bf 32}, 438  (1974). 

 \bibitem{Pati1974yy} J. C. Pati and A. Salam, 
 Phys. Rev. {\bf D 10}, 275 (1974) [Erratum-ibid D 11, 703 (1975)].
 
\bibitem{Higgs-like} 
G.~Aad {\it et al.}  [ATLAS Collaboration],
  Phys.\ Lett.\ B {\bf 716}, 1 (2012)
  arXiv:1207.7214 [hep-ex] (2012).
 \bibitem{Higgs-like1}  S.~Chatrchyan {\it et al.}  [CMS Collaboration],
  Phys.\ Lett.\ B {\bf 716}, 30  (2012)
  arXiv:1207.7235 [hep-ex] (2012).

\bibitem{SMHiggs}   G.~Aad {\it et al.}  [ATLAS Collaboration],
  Phys.\ Lett.\ B {\bf 726}, 120 (2013), 
  arXiv:1307.1432 [hep-ex] (2013).
 \bibitem{SMHiggs1}   G.~Aad {\it et al.}  [ATLAS Collaboration],
  Phys.\ Lett.\ B {\bf 726}, 88 (2013),
  arXiv:1307.1427 [hep-ex] (2013).
\bibitem{SMHiggs2}    S.~Chatrchyan {\it et al.}  [CMS Collaboration],
  Phys.\ Rev.\ Lett.\  {\bf 110}, 081803 (2013),
  arXiv:1212.6639 [hep-ex] (2012),.
\bibitem{SMHiggs3}  S.~Chatrchyan {\it et al.}  [ CMS Collaboration],
  arXiv:1401.6527 [hep-ex] (2014).


\bibitem{Schwinger1966nj} J. S. Schwinger, Phys. Rev {\bf B 144}, 1087 (1966).

\bibitem{Zwanziger1970hk}
  D.~Zwanziger,
 Phys.  Rev. {\bf D 3}, 880  (1971).

\bibitem{Blagojevic1985sh}
  M.~Blagojevic and P.~Senjanovic,
  Phys.\ Rept.\  {\bf 157}, 233  (1988).
  
  \bibitem{derrick} G.H. Derrick, 
J.\ Math.\ Phys. \textbf{5}, 1252 (1964). 

\bibitem{Cho2014}Y.~M.~Cho \emph{et.~al.}, to be published.
  
 
 \bibitem{Buican:2006sn} 
  M.~Buican, D.~Malyshev, D.~R.~Morrison, H.~Verlinde and M.~Wijnholt,
  JHEP {\bf 0701}, 107 (2007), 
  [hep-th/0610007] (2005)
 
\bibitem{Verlinde:2006bc} 
  H.~Verlinde,
  hep-th/0611069 (2006).



\bibitem{JPV}M James, L Perivolaropoulos and T Vachaspati, Phys. Rev.
{\bf D 46}, R 5232 (1992).

\bibitem{Georgi1972cj}
  H.~Georgi and S.~L.~Glashow,
  Phys.  Rev. Lett.  {\bf 28} (1972) 1494.

\bibitem{Vento2013jua}
  V.~Vento and V.~S.~Mantovani,
  arXiv:1306.4213 [hep-ph] (2013).

\bibitem{khlopov} Y. B. Zeldovich and M. Y. Khlopov, Phys. 
Lett.  {\bf B 79},  239 (1978).

\bibitem{Monopolium} C. T. Hill, Nucl. Phys. B 224 469 (1983). 
\bibitem{Monopolium1} V. K. Dubrovich, Grav. Cosmol. Suppl. {\bf 8 N1},  122  (2002).

\bibitem{Epele} L. N. Epele, H. Fanchiotti, C. A. Garcia Canal and V. Vento, Eur. Phys. J. C 56,  87  (2008).
\bibitem{Epelea} L. N. Epele, H. Fanchiotti, C. A. G. Canal and V. Vento, Eur. Phys. J. C 62, 587 (2009).     

\bibitem{Epele1} L. N. Epele, H. Fanchiotti, C. A. Garcia Canal, V. A. Mitsou and V. Vento, Eur. Phys. J. Plus {\bf 127}, 60
(2012).


\bibitem{Zwanziger} D. Zwanziger, Phys. Rev. {\bf  D 3}, 880 (1971).

\bibitem{Gamberg}  L. P. Gamberg and K. A. Milton, Phys. Rev.  {\bf D 61}, 075013  (2000).

\bibitem{Mulhearn} M. J. Mulhearn, ``A Direct Search for Dirac Magnetic Monopoles'', Ph.D. Thesis, Massachusetts
Institue of Technology 2004, FERMILAB-THESIS-2004-51.

\bibitem{Dougall} T. Dougall and S. D. Wick, Eur. Phys. J.  {\bf A 39}, 213  (2009).

\bibitem{Kalbfleisch} G. R. Kalbfleisch, K. A. Milton, M. G. Strauss, L. P. Gamberg, E. H. Smith and W. Luo, Phys. Rev.
Lett. {\bf 85}, 5292 (2000).

\bibitem{Jauch} J.M. Jauch and F. Rorhlich, ``The theory of electrons and photons'', (Text and Monographs in Physics
Springer-Verlag, p533 1975).

\bibitem{Peskin} M.E. Peskin and D.V. Schroeder, ``An introduction to quantum field theory'', (HarperCollins, 1995).

\bibitem{chopin} Y.~M.~Cho and J.~L.~Pinfold,
  arXiv:1307.8390 [hep-ph] (2013).

\bibitem{Kalbfleisch2003yt}
G. R. Kalbfleisch, W.  Luo, K. A. Milton,  E. H. Smith and M. G. Strauss, 
Phys. Rev.  {\bf D 69},  052002 (2004). 

 \bibitem{MODAL} K. Kinoshita, G. Giacomelli, L. Patrizii, F. Predieri, P. Serra, M. Spurio, J.L. Pinfold, 
 Phys. Rev. {\bf D 46}, 881 (1992).
 
 \bibitem{OPALDEDICATED} J.L. Pinfold  {\it et al.}., Nucl. Instrum. Meth.  {\bf A 302} , 434 (1991).
 
 \bibitem{LEP} J.L. Pinfold, R. Du, K. Kinoshita, B. Lorazo, M. Regimbald, B. Price, Phys. Lett.  {\bf B 316}  407 (1993).
 
\bibitem{Abbiendi2007ab} G. Abbiendi {\it et al.} [OPAL Collaboration],  Phys. Lett.  {\bf B 663},
 37 (2008).
 
 \bibitem{CDF} A. Abulencia {\it et al.} [CDF Collaboration], Phys. Rev. Lett. {\bf 96}, 201801 (2006). 
 
 
\bibitem{Aad2012qi}
G.  Aad {\it et al.} [ATLAS Collaboration], Phys. Rev. Lett. 109, 261803 (2012).

\bibitem{Drukier1981fq} 
   A. K. Drukier  and  S. Nussinov,  Phys. Rev. Lett. {\bf 49}, 102 (1982).
     
\bibitem{Dobbins1993vc}
T. Dobbins L. E.  and Roberts,  Nuovo Cim.  {\bf A 106}, 1295  (1993). 
    
 \bibitem{Demidov2011dk}
   S. V. Demidov and  D. G. Levkov, Phys. Rev. Lett. {\bf 107},  071601 (2011).
    
\bibitem{Demidov2011eu},
 S. V. Demidov  and D. G.  Levkov, JHEP {\bf 1106} , 016 (2011).     

\bibitem{Raklev:2009mg} A.~R.~Raklev,
  Mod.\ Phys.\ Lett.\ A {\bf 24}, 1955 (2009), arXiv:0908.0315 [hep-ph] (2009).
  
\bibitem{NUHM}
D.~Matalliotakis and H.~P.~Nilles,
  Nucl.\ Phys.\  B {\bf 435}, 115  (1995),
  arXiv:hep-ph/9407251 (1994);
  M.~Olechowski and S.~Pokorski,
  Phys.\ Lett.\ B {\bf 344}, 201 (1995),
  arXiv:hep-ph/9407404 (1997);
V.~Berezinsky, A.~Bottino, J.~Ellis, N.~Fornengo,  G.~Mignola and S.~Scopel,
 Astropart.\ Phys.  {\bf 5}, 1 (1996),   hep-ph/9508249;
  M.~Drees, M.~Nojiri, D.~Roy and Y.~Yamada,
  Phys.\ Rev. {\bf D 56}, 276  (1997), 
  [Erratum-ibid.\ {\bf D 64}, 039901 (1997)], 
  hep-ph/9701219;
 P.~Nath and R.~Arnowitt,
 Phys.\ Rev. {\bf D 56}, 2820  (1997), 
  hep-ph/9701301 (1997);
   J.~R.~Ellis, T.~Falk, G.~Ganis, K.~A.~Olive and M.~Schmitt,
  Phys.\ Rev.\ D {\bf 58}, 095002 (1998),
  arXiv:hep-ph/9801445 (1998);
J.~Ellis, K.~Olive and Y.~Santoso,
Phys.\ Lett.\  B~{\bf 539}, 107 (2002),
arXiv:hep-ph/0204192 (2002);
H.~Baer, A.~Mustafayev, S.~Profumo, A.~Belyaev and X.~Tata,
  Phys.\ Rev.\  D {\bf 71}, 095008 (2005),
  arXiv:hep-ph/0412059 (2004);
  J.~R.~Ellis, K.~A.~Olive and P.~Sandick,
  Phys.\ Rev.\  D {\bf 78}, 075012 (2008),
  arXiv:0805.2343 [hep-ph] (2008).

\bibitem{Fayet}
P.~Fayet,
  Phys.\ Lett.\ B {\bf 69} (1977) 489.

\bibitem{EHNOS}
H.~Goldberg,
                Phys.\ Rev.\ Lett.\ {\bf 50}, 1419  (1983);
                J.~Ellis, J.~Hagelin, D.~Nanopoulos, K.~Olive and M.~Srednicki,
                Nucl.\ Phys.\ B {\bf 238}, 453 (1984).

\bibitem{stauNLSP}
J.~R.~Ellis, K.~A.~Olive, Y.~Santoso and V.~C.~Spanos,
Phys.\ Lett.\ B {\bf 565}, 176 (2003), 
arXiv:hep-ph/0303043 (2003);

\bibitem{sleptonNLSP}
J.~R.~Ellis, K.~A.~Olive and Y.~Santoso,
  JHEP {\bf 0810}, 005  (2008), 
  arXiv:0807.3736 [hep-ph] (2008).

\bibitem{stopNLSP}
J.~L.~Diaz-Cruz, J.~R.~Ellis, K.~A.~Olive and Y.~Santoso,
  JHEP {\bf 0705}, 003  (2007), 
  [hep-ph/0701229 [HEP-PH]].
 
\bibitem{SusyBBN}
M.~H.~Reno and D.~Seckel,
Phys.\ Rev.\ D {\bf 37}, 3441 (1988)
\bibitem{SusyBBN1} K.~Jedamzik,
  \bibitem{SusyBBN2} Phys.\ Rev.\ D {\bf 70}, 063524 (2004),
  arXiv:astro-ph/040234 (2004);
  M.~Kawasaki, K.~Kohri and T.~Moroi,
  Phys.\ Lett.\ B {\bf 625}, 7 (2005), 
  arXiv:astro-ph/0402490 (2004). 
  \bibitem{SusyBBN3} K.~Jedamzik, K.~-Y.~Choi, L.~Roszkowski and R.~Ruiz de Austri,
  JCAP {\bf 0607}, 7 (2006),
  [hep-ph/0512044] (2005).
 \bibitem{SusyBBN4} D.~Cumberbatch, K.~Ichikawa, M.~Kawasaki, K.~Kohri, J.~Silk and G.~D.~Starkman,
  Phys.\ Rev.\ D {\bf 76}, 123005 (2007), 
 arXiv:0708.0095 [astro-ph] (2007).
\bibitem{SusyBBN5}    K.~Jedamzik and M.~Pospelov,
  New J.\ Phys.\  {\bf 11}, 105028 (2009),
  arXiv:0906.2087 [hep-ph] (2009).
\bibitem{SusyBBN6} R.~H.~Cyburt, J.~Ellis, B.~D.~Fields, F.~Luo, K.~A.~Olive and V.~C.~Spanos,
  JCAP {\bf 0910}, 021 (2009),
  arXiv:0907.5003 [astro-ph.CO] (2009).
\bibitem{SusyBBN7}R.~H.~Cyburt, J.~Ellis, B.~D.~Fields, F.~Luo, K.~A.~Olive and V.~C.~Spanos,
  JCAP {\bf 1010}, 032  (2010), 
  arXiv:1007.4173 [astro-ph.CO] (2010).
   \bibitem{SusyBBN8} M.~Pospelov and J.~Pradler,
  Phys.\ Rev.\ Lett.\  {\bf 106}, 121305 (2011),
  arXiv:1010.4079 [astro-ph.CO] (2010).
 \bibitem{SusyBBN9}  M.~Kawasaki and M.~Kusakabe,
  Phys.\ Rev.\ D {\bf 83}, 055011 (2011)
  arXiv:1012.0435 [hep-ph] (2010).
\bibitem{SusyBBN10}  D.~Albornoz Vasquez, A.~Belikov, A.~Coc, J.~Silk and E.~Vangioni,
  Phys.\ Rev.\ D {\bf 86}, 063501 (2012),
  arXiv:1208.0443 [astro-ph.CO] (2012).

\bibitem{MC8}
O.~Buchmueller, R.~Cavanaugh, M.~Citron, A.~De Roeck, 
M.~J.~Dolan, J.~R.~Ellis, H.~Flacher and S.~Heinemeyer {\it et al.},
  Eur.\ Phys.\ J.\ C {\bf 72}, 2243  (2012), 
  [arXiv:1207.7315 [hep-ph]].
\bibitem{MC8a} O.~Buchmueller, R.~Cavanaugh, A.~De Roeck, M.~J.~Dolan, J.~R.~Ellis, H.~Flacher, S.~Heinemeyer
 and G.~Isidori {\it et al.},
  arXiv:1312.5250 [hep-ph].

\bibitem{Sato}
 T.~Jittoh, J.~Sato, T.~Shimomura and M.~Yamanaka,
  Phys.\ Rev.\ D {\bf 73}, 055009 (2006),
  [hep-ph/0512197] (2005).
  
\bibitem{CELMOV}
M.~Citron, J.~Ellis, F.~Luo, J.~Marrouche, K.~A.~Olive and K.~J.~de Vries,
  Phys.\ Rev.\ D {\bf 87}, 036012 (2013), arXiv:1212.2886 [hep-ph] (2013).
  
 \bibitem{CAPTURE3} K.~Hamaguchi, M.~M.~Nojiri and A.~de Roeck,
  JHEP {\bf 0703}, 046 (2007) [hep-ph/0612060].

\bibitem{splitSUSY}
N. Arkani-Hamed and S. Dimopoulos, JHEP 0506, 073 (2005).
\bibitem{splitSUSY1} G. F. Giudice and A. Romanino,  
Nucl. Phys. {\bf B 699}, 65 (2004) [ Erratum-ibid. B706, 65 (2005)].

\bibitem{GMSB} M. Dine {\it et al.} , Phys. Rev. {\bf D 53}, 2658  (1996).
\bibitem{GMSB1} M. Dine,  A.E. Nelson and Y. Shirman, Phys. Rev.{ \bf  D51}, 1362 (1995). 

\bibitem{GMSBFISCHLER1} M. Dine and W. Fischler,  Phys. Lett. B110, 227  (1982).

\bibitem{GMSBFISCHLER2} M.  Dine and W. Fischler,  Nucl. Phys. B204, 346  (1982).

\bibitem{GMSBFISCHLER3}  C. R.  Nappi, B. A. Ovrut,  Phys. Lett. B113, 175 (1982).

\bibitem{GMSBFISCHLER4} L. Alvarez-Gaume, M. Claudson and M. B. Wise, Nucl. Phys. B207, 96 (1982).

\bibitem{GMSBREVIEW} G.F. Giudice and R. Rattazzi, Phys. Rept. {\bf 322}, 419  (1999). 

\bibitem{AMSB1} L. Randall and R. Sundrum, Nucl. Phys. {\bf B 557}, 79  (1999). 

\bibitem{AMSB2}  G.F. Giudice {\it et al.}, JHEP {\bf 12}, 27  (1998).
 \bibitem{AMSB2a} J.L. Feng and T. Moroi, Phys. Rev. {\bf D 61}, 095004  (2000). 

\bibitem{G2-MSSM1}
  B.~S.~Acharya, K.~Bobkov, G.~L.~Kane, J.~Shao and P.~Kumar,
  Phys.\ Rev.\ D {\bf 78}, 065038 (2008),
  arXiv:0801.0478 [hep-ph]  (2008).

\bibitem{G2-MSSM2}
  B.~S.~Acharya, G.~Kane and P.~Kumar,
  Int.\ J.\ Mod.\ Phys.\ A {\bf 27}, 1230012  (2012), 
  arXiv:1204.2795 [hep-ph] (2012).
  
 \bibitem{2loopAMSB}
  M.~Ibe, S.~Matsumoto and R.~Sato,
  Phys.\ Lett.\ B {\bf 721}, 252  (2013), 
  arXiv:1212.5989 [hep-ph] (2013).

\bibitem{4top}
  B.~S.~Acharya, P.~Grajek, G.~L.~Kane, E.~Kuflik, K.~Suruliz and L.~-T.~Wang,
  arXiv:0901.3367 [hep-ph] (2009).

\bibitem{G2Wino}
  G.~Kane, R.~Lu and B.~Zheng,
  arXiv:1202.4448 [hep-ph] (2012).

\bibitem{INOSMB} H. Baer, S. de Alwis, K. Givens, S.  Rajagopalan, H. Summy,  JHEP {\bf 1005}, 069 (2010).

\bibitem{Higgsino}
  S.~Bobrovskyi, J.~Hajer and S.~Rydbeck,
  JHEP {\bf 1302} (2013) 133
  [arXiv:1211.5584 [hep-ph]].

\bibitem{SUSYLHP} 	
R.  Barbieri and  A.  Strumia, 
e-Print: hep-ph/0007265 (2000).
\bibitem{NEWFATHIGGS}  	
S. Chang, C. Kilic, R. Mahbubani,  Phys. Rev.  D71, 015003 (2005). 

\bibitem{FATHIGGSFATTOP} 	A. Delgado and T. M. P. Tait 
JHEP 0507, 023 (2005).

\bibitem{Ambrosanio} 
  S.~Ambrosanio, B.~Mele, A.~Nisati, S.~Petrarca, G.~Polesello, A.~Rimoldi and G.~Salvini,
  Rend.\ Lincei Sci.\ Fis.\ Nat.\  {\bf 12}, 5 (2001),  [hep-ph/0012192] (2001).

\bibitem{atlas-det} ATLAS Collab.\ (G.~Aad {\it et al}.), 
  {\it JINST} {\bf 3}, S08003 (2008).

\bibitem{cms-det} CMS Collab.\ (R.~Adolphi {\it et al}.),
  {\it JINST} {\bf 3}, S08004 (2008).
\bibitem{atlas-susy-results} \url{https://twiki.cern.ch/twiki/bin/view/AtlasPublic/SupersymmetryPublicResults}  
\bibitem{cms-susy-results} \url{https://twiki.cern.ch/twiki/bin/view/CMSPublic/PhysicsResultsSUS}
\bibitem{atlas-dv} 

G.~Aad {\it et al.} [ATLAS collaboration],
ATLAS-CONF-2013-092 (2013).

\bibitem{atlas-kink} 
  G.~Aad {\it et al.}  [ATLAS Collaboration],
   Phys.  Rev. {\bf D 88}, 112006 (2013),  arXiv:1310.3675 [hep-ex] (2012). 

\bibitem{atlas-slepton} 
G. Aad {\it et al.}, ATLAS collaboration,
 ATLAS-CONF-2013-058 (2013).

\bibitem{cms-slepton} 
  S.~Chatrchyan {\it et al.}  [CMS Collaboration],
  {\it JHEP} {\bf 07}, 112  (2013),  arXiv:1305.0491 [hep-ex] (2013).  
  
\bibitem{atlas-stopped} 
  G.~Aad {\it et al.}  [ATLAS Collaboration],
Phys. Rev.  {\bf D  88}, 112003 (2013), arXiv:1310.6584 [hep-ex] (2013).

\bibitem{GUIDICE} G.F. Giudice, R. Rattazzi, J.D.Wells, 
Nucl. Phys.  {\bf B 544}, 3 (1999), arXiv: hep-ph/9811291.

\bibitem{ED1} P.C. Argyres, S. Dimopoulos and J. March-Russell, Phys.
Lett. {\bf B 441}, 96 (1998). 
\bibitem{ED1a} S. Dimopoulos and G. L. Lands-
berg, Phys. Rev. Lett.  {\bf 87}, 161602 (2001).
\bibitem{ED1b} G.L. Alberghi, R. Casadio and
A. Tronconi, J. Phys. {\bf  G 34}, 767 (2007). 
\bibitem{ED1c} M. Cavaglia,
R. Godang, L. Cremaldi and D. Summers, Comput.
Phys. Commun. {\bf 177} , 506 (2007).
\bibitem{ED1d} D.C. Dai, G. Starkman, D. Stojkovic, C. Issever, E. Rizvi and J. Tseng,
Phys. Rev.  {\bf D 77}, 076007 (2008).


\bibitem{bhevaporation} S. B Giddings and S. Thomas, Phys. Rev. {\bf D 65}, 056010 (2002).

\bibitem{fischler1} W. Fischler,  arXiv:hep-th/9906038, (1999)

\bibitem{CHARYBDIS} C. M. Harris, P. Richardson and B. R. Webber, JHEP {\bf 0308}, 033 (2003).

\bibitem{ED2} R. Casadio and B. Harms, Int. J. Mod. Phys. {\bf A 17} , 4635
(2002)

\bibitem{ED3} M. Cavaglia, Int. J. Mod. Phys. {\bf A 18}, 1843 (2003).
\bibitem{ED3a} P. Kanti, Int. J. Mod. Phys.  {\bf A 19}, 4899 (2004).

\bibitem{hawkingradiation}  S.W. Hawking, Commun. Math. Phys. {\bf 43}, 199 (1975).
\bibitem{hawkingradiation1} Phys. Rev.{\bf  D 14}, 2460 (1976).

\bibitem{RSBH} R. Gregory, Lect. Notes Phys. {\bf 769}, 259 (2009),
arXiv:0804.2595 [hep-th] (2008)].
\bibitem{RSBH1}    P. Kanti, J. Phys. Conf. Serv. {\bf 189}, 012020 (2009), 
arXiv:0903.2147v1 [hep-th] (2009).

\bibitem{tidalcharge} N. Dadhich, R. Maartens, P. Papadopoulos, and V. Rezania,  Phys.Lett. B487, 1-6  (2000).

\bibitem{microcanonical} 
 A. Schelpe,
 arXiv:0809.2353v1 [hep-th] (2008).
 \bibitem{microcanonical1} R. Casadio, S. Fabi, and B. Harms, Phys. Rev. {\bf D 80}, 084036 (2009);
R. Casadio, S. Fabi, B. Harms, and O. Micu, 
 arXiv:0911.1884v1 [hep-th] (2009).
 
 \bibitem{gingrich-tidalcharge} 	 D.  M. Gingrich,  Phys. Rev. {\bf D 81}  057702 (2010). 

 \bibitem{infoloss} I.D. Novikov and V.P. Frolov, Black hole physics, Kluver Academic Publishers, Dordrecht
1998. 
\bibitem{infoloss1} S. Hawking, 
Commun. Math. Phys. {\bf 87}, 395  (1982).
\bibitem{infoloss2} J. Preskill, 
hep-th/9209058 (1992).

\bibitem{GUP} R.J. Adler, P. Chen and D.I. Santiago, 
Gen. Rel. Grav. 33, 2101  (2001),  [gr-qc/0106080]. 
\bibitem{GUP1}  K. Nozari and S.H. Mehdipour, Electron. J. Theor. Phys.  {\bf 3N11}. 151 (2006), 
gr-qc/0504099 (2005).

\bibitem{ARGSFORBHREMS}  Y. Aharonov, A. Casher and S. Nussinov, Phys. Lett. 
{\bf 191 B}, 51 (1987); 
\bibitem{ARGSFORBHREMS1}T. Banks, A. Dabholkar, M. R. Douglas and M. OÕLoughlin, Phys. 
Rev.{\bf  D 45},  3607 (1992). 
\bibitem{ARGSFORBHREMS2}T. Banks and M. OÕLoughlin, Phys. Rev. {\bf D 47}, 540 (1993). 
\bibitem{ARGSFORBHREMS3}T. Banks, M. OÕLoughlin and A. Strominger, Phys. Rev. {\bf D 47}, 4476 (1993).
\bibitem{ARGSFORBHREMS4}S. B.  Giddings, Phys. Rev. {\bf D 49}, 947 (1994). 
\bibitem{ARGSFORBHREMS5} M. D. Maia, Int. J. Mod. Phys. {\bf D 14}, 2251 (2005), 
[arXiv:gr-qc/0505119] (2005).  
\bibitem{ARGSFORBHREMS6}V. Husain and O. Winkler, Int. J. Mod. Phys. {\bf D 14}, 2233 (2005),
[arXiv:gr-qc/0505153] 2005).

\bibitem{NAIVEXS} M. B. Voloshin, Phys. Lett. {\bf B 518}, 137 (2001); Phys. Lett.  {\bf B 524}, 376 (2002);
\bibitem{NAIVEXS1} S. B. Giddings, in  {\it Proc. of the APS/DPF/DPB Summer Study on the Future of Particle Physics}
(Snowmass 2001) ed. N. Graf, eConf C010630, P328 (2001).

\bibitem{NAIVEXSGOODTO} S. N. Solodukhin, Phys. Lett. {\bf B 533}, 153 (2002).
\bibitem{NAIVEXSGOODTO1} A. Jevicki and J. Thaler, Phys. Rev.{\bf  D 66}, 024041 (2002).
\bibitem{NAIVEXSGOODTO2} T. G. Rizzo, in {\it Proc. of the APS/DPF/DPB Summer
Study on the Future of Particle Physics} (Snowmass 2001) ed. N. Graf, eConf C010630, 339 (2001).
\bibitem{NAIVEXSGOODTO3} D. M. Eardley and S. B. Giddings, Phys. Rev. {\bf D 66}, 044011 (2002).

\bibitem{STRINGSOK} G. T. Horowitz and J. Polchinski, Phys. Rev. D 66 103512 (2002).

\bibitem{BHMULT} B.  Koch, M. Bleicher and H.  Stoecker, J. Phys. G34, S535 (2007).

\bibitem{BELLAGAMBA}  L. Bellagamba, R. Casadi,, R. Di Sipio   and V. Viventi, [arXiv:1201.3208v2] (2012).

\bibitem{noncommutative} D. M. Gingrich,  arXiv:1003.1798v1 [hep-ph] (2010).

  
\bibitem{Kaluza1921tu}
T. Kaluza,
Sitzungsber. Preuss. Akad. Wiss. Berlin (Math.Phys.), {bf 1921}, 966 (1921).

\bibitem{Klein1926tv}
O. Klein, 
 Z.Phys. {\bf 37}, 895  (1926).

\bibitem{Cheng2002iz}
H-C Cheng, K. T.  Matchev  and  M. Schmaltz,
Phys. Rev. {\bf  D 66}, 036005 (2002).

\bibitem{Cembranos2006gt}
J. A. R. Cembranos, J. L.Feng  and  L. E. Strigari,
Phys. Rev. {\bf D 75}, 036004 (2007).
   
\bibitem{Servant2002aq}
G. Servant and T. M. P.  Tait, 
Nucl. Phys.  {\bf B 650}, 391(2003).
  
\bibitem{SEAWATER}
 P. Verkerk  {\it et al.}, Phys. Rev. Lett. {\bf 68}, 1116 (1992); 
\bibitem{SEAWATER1}  T. Hemmick {\it  et al.}, Phys. Rev.
{\bf D 41}, 2074 (1990).   

 \bibitem{Flacke2008ne}
T. Flacke, A.  Menon and  D. J. Phalen,
Phys. Rev. {\bf D 79}, 056009 (2009).

\bibitem{Witten2002wb} 
See, e.g.,
E.~Witten,
arXiv:hep-th/0212247 (2002), and references therein.

\bibitem{kolbmaverick} M.~Beltran, D.~Hooper, E.~W.~Kolb, Z.~A.~C.~Krusberg and T.~M.~P.~Tait,
  JHEP {\bf 1009}, 037 (2010),
  arXiv:1002.4137 [hep-ph] (2010).

\bibitem{lhcmaverick}  A.~Rajaraman, W.~Shepherd, T.~M.~P.~Tait and A.~M.~Wijangco,
  Phys.\ Rev.\ D {\bf 84}, 095013 (2011),
  [arXiv:1108.1196 [hep-ph]].
  
\bibitem{atlas-dm2014} 
  G.~Aad {\it et al.}  [ATLAS Collaboration],
  Phys.\ Rev.\ Lett.\  {\bf 112}, 041802 (2014),  [arXiv:1309.4017 [hep-ex]].

\bibitem{frampton} P. H. Frampton, P. Hung and M. Sher, Phys. Rept. {\bf 330}, 263 (2000).

\bibitem{eberhardt}  O. Eberhardt, G. Herbert, H. Lacker, A. Lenz A. Menzel,
{\it et al.}, Phys. Rev. {\bf  D86}, 013011 (2012).

\bibitem{dawson} S. Dawson and E. Furlan, Phys. Rev. {\bf D86}, 015021 (2012).

\bibitem{aguilla} F. del Aguila, M. Perez-Victoria and J. Santiago, JHEP
{\bf 0009}, 011 (2000).

\bibitem{campbelletal} J. M. Campbell, R. Frederix, F. Maltoni, and F. Tramontano, 
JHEP {\bf 0910}, 042 (2009).   
\bibitem{campbelletal1} A. Atre, G. Azuelos, M. Carena, T. Han, E. Ozcan, {\it et al.},
JHEP {\bf 1108}, 080 (2011).





\bibitem{buchkremer}     M. Buchkremer, J.-M. Gerard, and F. Maltoni, JHEP
{\bf 1206}, 135 (2012);  

\bibitem{buchkremeretal} 
A. Dighe, D. Ghosh, R. M. Godbole and A. Prasath,
Phys. Rev. {\bf D 85}, 114035 (2012).   
\bibitem{buchkremeretal1} S. Bar-Shalom, S. Nandi and A. Soni, Phys. Rev. {\bf D 84},
053009 (2011). 
\bibitem{buchkremeretal2}  E. Asilar, E. Cavlan, O. Dogangun, S. Kefeli, V. Erkcan Ozcan, {\it et al.}, Eur. Phys.
J. {\bf C 72}, 1966 (2012).

\bibitem{mackeprang} R. Mackeprang and D. Milstead, Eur. Phys. J. {\bf  C 66}, 493 (2010).

\bibitem{buchkremer1}  M.  Buchkremer and  A.  Schmidt, 
Adv. High  Energy Phys. {\bf  2013}, 690254  (2013).

\bibitem{ATLASHQ} G. Aad {\it et al.} [ATLAS Collaboration], Phys. Lett,  {\bf B 719}, 280 (2013).
     
\bibitem{CMSHQ} S. Chatrchyan {\it et al.} [CMS Collaboration], Phys.  Lett.  {\bf B 713}, 408 (2012).    

\bibitem{johansen}  M. Johansen, J. Edsjo, S. Hellman and D. Milstead,
JHEP {\bf 1008}, 005 (2010).

\bibitem{SANINNO} M. T. Frandsen, I. Masina and F. Saninno, Phys. Rev. {\bf D 81}, 035010 (2010).

\bibitem{MWT} F. Sannino and K. Tuominen, Phys. Rev. {\bf  D 71}, 051901 (2005).

\bibitem{SINISTER} Sheldon L. Glashow, e-Print: hep-ph/0504287 (2005).

 \bibitem{RAJASEKARAN} G. Rajasekaran, arXiv:1105.5213v2 [physics.gen-ph] ( 2011).

\bibitem{fchamp}   P.~Langacker and G.~Steigman,
  Phys.\ Rev.\ D {\bf 84}, 065040 (2011), 
  arXiv:1107.3131 [hep-ph] (20111).

\bibitem{LRSM} R. N. Mohapatra and J. C. Pati, Phys. Rev. {\bf D11}, 566 (1975).
\bibitem{LRSM1}G. Senjanovic and R. N. Mohapatra, Phys. Rev.  {\bf D 12}, 1502(1975).  
\bibitem{LRSM2}R. N. Mohapatra and G. Senjanovic, Phys. Rev. {\bf D 23}, 165 (1981). 

\bibitem{seesaw}  T. P. Cheng and L. -F. Li, Phys. Rev. {\bf D 22}, 2860 (1980).   
\bibitem{seesaw1} G. B. Gelmini and M. Roncadelli, Phys. Lett. {\bf B 99},  411(1981). 
\bibitem{seesaw2} A. Zee, Phys. Lett.  {\bf B 93}, 389 (1980),[Erratum-ibid. {\bf  B 95}, 461(1980)].
\bibitem{seesaw3}  T. Han, H. E. Logan, B. Mukhopadhyaya and R. Srikanth, Phys. Rev. {\bf D 72}, 053007 (2005).  
\bibitem{seesaw4}  J. Y. Lee, JHEP {\bf 0506}, 060 (2005);   I. Picek and B. Radovcic, Phys. Lett.  {\bf B 687}, 338 (2010). 
\bibitem{seesaw5}  S. K. Majee and N. Sahu, Phys. Rev. {\bf D 82}, 053007 (2010).  
\bibitem{seesaw6} M. Aoki, S. Kanemura and K. Yagyu, Phys. Lett. {\bf  B 702}, 355 (2011),  [Erratum-ibid. B 706, 495 (2012)]. 
 \bibitem{seesaw7} M. -C. Chen and J. Huang, Mod. Phys. Lett. {\bf A 26},  1147 (2011).  
 \bibitem{seesaw8} K. Kumericki, I. Picek and B. Radovcic, Phys. Rev. {\bf D 84},  093002 (2011).  
\bibitem{seesaw9}  M. Aoki, S. Kanemura  and K. Yagyu, Phys. Rev. {\bf D 85}, 055007 (2012). 
 \bibitem{seesaw10} C. -W. Chiang, T. Nomura and K. Tsumura, Phys. Rev. {\bf D 85}, 095023 (2012).  
 \bibitem{seesaw11}  K. Kumericki, I. Picek and B. Radovcic, Phys. Rev. {\bf  D 86}, 013006 (2012). 
\bibitem{seesaw12} H. Sugiyama, K. Tsumura and H. Yokoya, Phys. Lett.  {\bf B 717}, 229 (2012). 
\bibitem{seesaw13}  I. Picek and B. Radovcic, Phys. Lett.  {\bf B 719}, 404 (2013). 
\bibitem{seesaw14}  S. Kanemura, K. Yagyu and H. Yokoya, arXiv:1305.2383 [hep-ph] (2013).

 
 \bibitem{strings} M. Cvetic, J. Halverson and P. Langacker, JHEP {\bf 11}, 058 (2011), arXiv:1108.5187[hep-ph] (2011).
 
 \bibitem{susylr} D. A. Demir, M. Frank, K. Huitu, S. K. Rai and I. Turan, Phys. Rev. {\bf D 78}, 035013 (2008). 
 \bibitem{susylr1}M. Frank, K. Huitu and S. K. Rai, Phys. Rev. {\bf D 77}, 015006 (2008).  
  \bibitem{susylr2}K. S. Babu, A. Patra and S. K. Rai, Phys. Rev. {\bf D 88},  055006 (2013), arXiv:1306.2066 [hep-ph] (2013).  
  \bibitem{susylr3}R. Franceschini and R. N. Mohapatra, arXiv:1306.6108 [hep-ph]  (2013).  

\bibitem{extendedgg} P. H. Frampton and B. -H. Lee, Phys. Rev. Lett.{\bf  64}, 619 (1990).
\bibitem{extendedgg1} P. B. Pal, Phys. Rev. {\bf D 43}, 236 (1991). 
\bibitem{extendedgg2} F. Pisano and V. Pleitez, Phys. Rev. {\bf D 46}, 410 (1992). 
\bibitem{extendedgg3} P. H. Frampton, Mod. Phys. Lett. {\bf A 7}, 2017 (1992).  
\bibitem{extendedgg4} P. H. Frampton, Phys. Rev. Lett. {\bf  69}, 2889 (1992).

\bibitem{noncommtc} E. Farhi and L. Susskind, Phys. Rept. {\bf 74}, 277 (1981).   
\bibitem{noncommtc1}H. Harari, Phys. Rept. {\bf 104}, 159  (1984).
\bibitem{noncommtc2}N. Cabibbo, L. Maiani and Y. Srivastava, Phys. Lett. {\bf B 139}, 459 (1984).
\bibitem{noncommtc3} E. Eichten, I. Hinchliffe,  K. D. Lane and C. Quigg, Rev. Mod. Phys. {\bf 56}, 579  (1984) 579 [Addendum-ibid. {\bf 58}, 1065 (1986)].
\bibitem{noncommtc4} G. Pancheri and Y. N. Srivastava, Phys. Lett. {\bf B 146}, 87 (1984). 
\bibitem{noncommtc5}  W. Buchmuller, Acta Phys. Austriaca Suppl. {\bf 27}, 517 (1985);   C. A. Stephan, J. Phys. {\bf A 39}, 9657 (2006). 
\bibitem{noncommtc6}S. Biondini, O. Panella, G. Pancheri, Y. N. Srivastava and L. Fano, Phys. Rev. 
{\bf D 85}, 095018 (2012). 

\bibitem{WTCEF} S. B. Gudnason, C. Kouvaris and F. Sannino, 
Phys. Rev. {\bf D 73}, 115003 (2006).

\bibitem{Alloul:2013raa} 
  A.~Alloul, M.~Frank, B.~Fuks and M.~R.~de Traubenberg,
  Phys.\ Rev.\ D {\bf 88}, 075004 (2013),
  arXiv:1307.1711 [hep-ph] (2013).

\bibitem{xygaugino1} W.D. Goldberger, Y. Nomura and D.R. Smith, Phys. Rev. {\bf  D 67} 075021 (2003).

\bibitem{xygaugino2} Y. Nomura and D.R. Smith, Phys. Rev. {\bf D 68}  075003 (2003). 
\bibitem{xygaugino2a}  Y. Nomura, D. Tucker-Smith and  B. Tweedie, Phys. Rev. {\bf D 71}  075004 (2005).
 
\bibitem{xygaugino3} Y. Nomura and D. Tucker-Smith, Nucl. Phys. {\bf  B 698}  92 (2004). 

\bibitem{DE}
S.~Dimopoulos and J.~R.~Ellis,
  Nucl.\ Phys.\ B {\bf 182}, 505 (1981).
  
\bibitem{hole}
J.~R.~Ellis, M.~K.~Gaillard, D.~V.~Nanopoulos and P.~Sikivie,
  Nucl.\ Phys.\ B {\bf 182} (1981) 529.

\bibitem{WTC} D. K. Hong, S. D. H. Hsu and F. Sannino, Phys. Lett. B 597, 89 (2004);  D. D. Dietrich, F. Sannino and K. Tuominen, 
Phys. Rev. D 72, 055001 (2005).  
\bibitem{WTC1} D. D. Dietrich, F. Sannino and K. Tuominen, 
 Phys. Rev. D73 037701 (2006). 
\bibitem{WTC2} S. B. Gudnason, C. Kouvaris and F. Sannino, Phys. Rev. D74, 095008 (2006).

\bibitem{NUOSC}  Y. Ashie {\it et al.} [Super-Kamiokande Collaboration], Phys. Rev. Lett. {\bf 93}, 101801 (2004) and references therein.

\bibitem{SMALLNUMASS} R. N. Mohapatra and G. Senjanovic, Phys. Rev. Lett. {\bf 44}, 912 (1980).

\bibitem{LRSUSY0} C. S. Aulakh, A. Melfo, and G. Senjanovic, Phys. Rev. {\bf D 57}, 4174 (1998); 
\bibitem{LRSUSY1}Z. Chacko and R. N. Mohapatra, Phys. Rev. {\bf D 58}, 015003 (1998).
\bibitem{LRSUSY2}C. S. Aulakh, K. Benakli and
 G. Senjanovic, Phys. Rev. Lett. {\bf 79}, 2188 (1997).

\bibitem{DCHYUKAWACOUPS} K.Huitu, J.Maalampi, A.Pietila and M.Raidal, Nucl. Phys.  {\bf B 487}, 27
 (1997).
\bibitem{DCHYUKAWACOUPS1} M.L. Swartz, Phys. Rev.  {\bf D 40}, 1521 (1989).
 
\bibitem{Kuchimanchi:1993jg} 
  R.~Kuchimanchi and R.~N.~Mohapatra,
  Phys.\ Rev.\ D {\bf 48}, 4352 (1993),  [hep-ph/9306290] (1993).
 
\bibitem{Mohapatra:1996vg} 
  R.~N.~Mohapatra and A.~Rasin,
  Phys.\ Rev.\ D {\bf 54}, 5835 (1996), [hep-ph/9604445] (1996).
\bibitem{Mohapatra:1995xd} 
   R.~N.~Mohapatra and A.~Rasin,
  Phys.\ Rev.\ Lett.\  {\bf 76}, 3490 (1996), [hep-ph/9511391] (1995).
 
  \bibitem{CONNES} A. Connes,  Noncommutative Geometry (Academic Press, London and San Diego, 1994). 
 
\bibitem{ACmodel} M. Y. Khlopov, arXiv:astro-ph/0607048 (2006); 
C. A. Stephan, arXiv:hep-th/0509213 (2005). 
\bibitem{ACmodel1}  D. Fargion et al, Class. Quantum Grav. {\bf 23}, 7305 (2006); 
\bibitem{ACmodel2}  M. Y. Khlopov and C. A. Stephan, arXiv:astro-ph/0603187 
 (2006).
 
\bibitem{ASSUME100} D. Fargion, M. Khlopov and C. A. Stephan, 
 arXiv:astro-ph/0511789 (2005).
\bibitem{ASSUME100a} M. Y. Khlopov and C. A. Stephan, 
 arXiv:astro-ph /0603187 (2006).

 
 \bibitem{StableQballs} A. Kusenko and M. Shaposhnikov, Phys. Lett.  {\bf B 417}, 99 (1998).
 \bibitem{StableQballs1} A. Kusenko, V.A. Kuzmin, M. Shaposhnikov and P.G. Tinyakov, Phys.
Rev. Lett. 80, 3185 (1998).
 
  \bibitem{kusenkosolitons} 	
A.  Kusenko, Phys. Lett. {\bf  B 405}, 108  (1997).

  
  \bibitem{bodmer}
  A.~R.~Bodmer,
  Phys.\ Rev.\ D {\bf 4},  1601 (1971).

\bibitem{witten}
  E.~Witten,
  Phys.\ Rev.\ D {\bf 30}, 272  (1984).
  
\bibitem{bagmodel} S. A. Chin and A. K. Kerman, 
Phys. Rev. Lett.  {\bf 43}, 1292 (1979). 
\bibitem{bagmodel1} J. D. Bjorken and L. D. McLerran, 
Phys. Rev. {\bf D 20}, 2353 (1979).

\bibitem{sdistilling} H.-C. Liu and G. L. Shaw,
Phys. Rev. {\bf D 30}, 1137 (1984); 
\bibitem{sdistilling1} C. Greiner, Peter Koch and Horst Stocker, 
Phys. Rev. Lett. {\bf 58},  1825 (1987).

\bibitem{COALESCENCE} 	
Z. Arvay, J. Zimanyi, T. Csorgo, Carl B. Dover, Ulrich W. Heinz,  Z. Phys. {\bf A 348}, 201 (1994).
\bibitem{COALESCENCE1} C.  B. Dover, A.J. Baltz, Y. Pang, T.J. Schlagel, S.H. Kahana, 
``Production of strange clusters in relativistic heavy ion collisions'', Feb 1993. 17 pp.
BNL-48594, C93-01-13 

\bibitem{ellisstrange}
  J.~R.~Ellis, G.~Giudice, M.~L.~Mangano, I.~Tkachev and U.~Wiedemann,
  J.\ Phys.\ G G {\bf 35} (2008) 115004, 
  arXiv:0806.3414 [hep-ph] (2008).
  
    \bibitem{MITbagmodel} J. Schaffner-Bielich {\it et al.}, 
Phys. Rev. {\bf C 55}, 3038 (1997).
  
  \bibitem{PLUSCHARGE} 	
J.  Madsen, Phys. Rev. Lett. {\bf 85}, 4687  (2000).
  \bibitem{PLUSCHARGE1} E. P. Gilson and R. L. Jaffe,
Phys.Rev. Lett. {\bf 71}, 332 (1993). 
  \bibitem{PLUSCHARGE2} M. G. Mustafa and A. Ansari, Phys. Rev.{\bf  D 53}, 5136 (1996).
  
 \bibitem{okun}
  L.~B.~Okun,
  JETP Lett.\  {\bf 31}, 144 (1980) 
   [Pisma Zh.\ Eksp.\ Teor.\ Fiz.\  {\bf 31}, 156 (1979)].


\bibitem{kangandluty} 	
J.  Kang and  M,  A. Luty,  JHEP {\bf 0911}, 065 (2009). 

\bibitem{SEESAWS}
C. Cheung, L. J. Hall and D. Pinner, arXiv:1103.3520 [hep-ph] (2011).

\bibitem{LEPMMCPlims} A. Heister {\it et al.} [ ALEPH Collaboration ], Phys. Lett. {\bf B 537}, 5 (2002).
\bibitem{LEPMMCPlims1}A. Heister {\it et al.} [ ALEPH Collaboration ], Eur. Phys. J. {\bf C 31}, 327 (2003). 
\bibitem{LEPMMCPlims2} ALEPH, DELPHI, L3, and OPAL, LEP2 SUSY Working Group (2002).

\bibitem{TEVATRONMMCPlims} V. M. Abazov {\it et al.} [ D0 Collaboration ], Phys. Rev. Lett. 102, 161802 (2009).
 \bibitem{TEVATRONMMCPlims1}T. Aaltonen {\it et al.} [ CDF Collaboration ], Phys. Rev. Lett. {\bf 103}, 021802 (2009).

\bibitem{LHCMMCPlims} S. Chatrchyan {\it et al.} [CMS Collaboration], arXiv:1205.0272 [hep-ex].
\bibitem{LHCMMCPlims1} V. Khachatryan {\it et al.} [ CMS Collaboration ], JHEP {\bf 1103}, 024 (2011).
\bibitem{LHCMMCPlims2} G. Aad {\it et al.} [ ATLAS Collaboration ], Phys. Lett. {\bf B 701}, 1 (2011).
\bibitem{LHCMMCPlims3} G. Aad {\it et al.} [ ATLAS Collaboration ], Phys. Lett. {\bf B 703}, 428 (2011).

\bibitem{colliderdecays} V. M. Abazov {\it et al.} [ D0 Collaboration ], Phys. Rev. Lett. {\bf 99}, 131801 (2007).
\bibitem{colliderdecays1} V. Khachatryan {\it et al.} [CMS Collaboration], Phys. Rev. Lett. {\bf 106}, 011801 (2011). 
\bibitem{colliderdecays2} G.  Add {\it et al.} [ ATLAS Collaboration] Eur. Phys. J. {\bf C 72}, 1965 (2012). 

\bibitem{STOPPED} 	 P.  W. Graham, Kiel Howe, S.  Rajendran and D.  Stolarski, Phys. Rev. 
 {\bf D 86}, 034020  (2012).

\bibitem{ROCK} A. De Roeck, J. R. Ellis, F. Gianotti, F. Moortgat, K. A. Olive and L. Pape, Eur. Phys. J.
{\bf C 49}, 1041 (2007). 

\bibitem{SIBLEY} J. L. Pinfold and L. Sibley, Phys. Rev. {\bf D 83}, 035021 (2011). 

 \bibitem{CAPTURE1} K. Hamaguchi, Y. Kuno, T. Nakaya and M. M. Nojiri, Phys. Rev. {\bf  D 70}, 115007 (2004).


\bibitem{CAPTURE2}  J. L. Feng and B. T. Smith, Phys. Rev.{\bf  D 71}, 015004 (2005) [Erratum-ibid. D 71, 019904
(2005).

 

\bibitem{sensitivity} W. Adam {\it et al.}, Nucl. Instrum. Meth. {\bf A 543}, 463  (2005).
\bibitem{sensitivity1} R. Bainbridge, ``Prepared for 8th Workshop on Electronics for LHC Experiments'',
Colmar, France, 9-13 Sept. 2002.

\bibitem{kraan} A.C. Kraan, J.B. Hansen and P. Nevski, Eur. Phys. J. {\bf C 49}, 623 (2007),   hep-ex/0511014 (2005).

\bibitem{hauser}  R. Hauser, Eur. Phys. J. {\bf  C 34} , s173  (2004).

\bibitem{zalewski} P. Zalewski, CMS Conference Note, CMS-CR-1999-019 (1999).

\bibitem{BIRK} S.  Burdin, M.  Horbatsch, W.  Taylor, Nucl. Instrum. Meth. {\bf  A 664}, 111 (2012).

\bibitem{EIONRECOMB} G. Aad {\it et al.} [ATLAS Collaboration],  Phys. Lett. {\bf B 698}, 353  (2011).

\bibitem{mermod}  A. De Roeck,  A. Katre,  P. Mermod, D. Milstead and T. Sloan,  Eur. Phys. J. {\bf  C 72}, 1985  (2012).


\end{thebibliography}
\end{document}